\documentclass[11pt]{article}
\usepackage{setspace}
\usepackage{amsmath}

\usepackage{graphicx}
\usepackage{fullpage}
\usepackage{microtype}
\usepackage{amsmath,amsthm,amssymb}
\usepackage{color}
\usepackage{verbatim}
\usepackage{float}
\usepackage{framed}
\usepackage{tikz-cd}
\usetikzlibrary{arrows}
\usepackage[normalem]{ulem}

\usepackage{comment}
\usepackage{algorithm}
\usepackage{algpseudocode}
\usepackage{url}
\usepackage{hyperref}
\usepackage{chngcntr}
\usepackage{apptools}

\definecolor{darkred}{rgb}{0.8, 0.0, 0.0}
\newcommand{\facundo}[1]                {{ \textcolor{darkred} {#1}}}

\definecolor{darkgreen}{rgb}{0,0.8,0}

\begin{document}

\title{Sketching and Clustering Metric Measure Spaces}

\author{
Facundo M\'emoli\thanks{Department of Mathematics, The Ohio State University, \texttt{memoli@math.osu,edu}.}
\and
Anastasios Sidiropoulos\thanks{Department of Computer Science, University of Illinois at Chicago, \texttt{sidiropo@uic.edu}.}
\and
Kritika Singhal\thanks{Department of Mathematics, The Ohio State University, \texttt{singhal.53@osu.edu}.}
}

\newtheorem{theorem}{Theorem}[section]
\newtheorem{lemma}[theorem]{Lemma}
\newtheorem{proposition}[theorem]{Proposition}
\newtheorem{claim}[theorem]{Claim}
\newtheorem{corollary}[theorem]{Corollary}
\newtheorem{definition}[theorem]{Definition}
\newtheorem{observation}[theorem]{Observation}
\newtheorem{fact}[theorem]{Fact}
\newtheorem{property}{Property}
\newtheorem{Remark}{Remark}[section]
\newtheorem{notation}{Notation}[section]
\newtheorem{example}{Example}[section]
\newtheorem{conjecture}{Conjecture}
\newtheorem{question}[conjecture]{Question}

\newcommand{\Cl}{\mathrm{Cl}}
\newcommand{\Co}{\mathrm{Co}}
\newcommand{\Cov}{\mathrm{Cov}}
\newcommand{\dis}{\mathrm{dis}}
\newcommand{\diam}{\mathrm{diam}}
\newcommand{\rad}{\mathrm{rad}}
\newcommand{\partk}{\mathrm{Part}_{k}}
\newcommand{\eps}{\epsilon}
\newcommand{\len}{\mathrm{len}}

\newcommand{\dgh}{d_\mathrm{GH}}
\newcommand{\dgwp}{d_{\mathrm{GW}_p}}
\newcommand{\dha}{d_{\mathrm{H}}}
\newcommand{\dwp}{d_{\mathrm{W}_p}}

\newcommand{\shatter}{\mathbf{Shatter}}
\newcommand{\sketch}{\mathbf{Sketch}}
\AtAppendix{\counterwithin{theorem}{section}}

\clearpage

\date{}
\maketitle

\thispagestyle{empty}
\begin{abstract}
Two of the most fundamental data analysis tasks are clustering and simplification (sketching) of data. Clustering refers to partitioning a dataset, according to some rule, into sets of smaller size with the aim of extracting important information from the data. Sketching, or simplification of data, refers to approximating the input data with another dataset of much smaller size in such a way that properties of the input dataset are retained by the smaller dataset. In this sense, sketching facilitates understanding of the data. 

There are many clustering methods for metric spaces (mm spaces) already present in literature, such as $k$-center clustering, $k$-median clustering, $k$-means clustering,  etc. A natural method for obtaining a $k$-sketch of a metric space (mm space) is by viewing the space of all metric spaces (mm space) as a metric under Gromov-Hausdorff (Gromov-Wasserstein) distance, and then determining, under this distance, the $k$ point metric space (mm space) closest to the input metric space (mm space).

These two problems of sketching and clustering, a priori, look completely unrelated. However, in this paper, we establish a duality i.e.~an equivalence between these notions of sketching and clustering. In particular, we obtain the following results. For metric spaces, we consider the case where the clustering objective is minimizing the maximum cluster diameter. We show that the ratio between the sketching and clustering objectives is  constant over compact metric spaces. 

We extend these results to the setting of metric measure spaces where we prove that the ratio of sketching to clustering objectives is bounded both above and below by some universal constants. In this setting, the clustering objective involves minimizing various notions of the \emph{$\ell_p$-diameters} of the clusters. We consider two competing notions of sketching for metric measure spaces, with one of them being more demanding than the other. These notions arise from two different definitions of $p$-Gromov-Wasserstein distance that have appeared in the literature. We then prove that whereas the gap between these can be arbitrarily large, in the case of doubling metric spaces the resulting sketching objectives are polynomially related.

We also identify procedures/maps that transform a solution of the sketching problem to a solution of the clustering problem, and vice-versa. These maps give rise to algorithms for performing these transformations and, by virtue of these algorithms, we are able to obtain an approximation to the $k$-sketch of a metric measure space (metric space) using known approximation algorithms for the respective clustering objectives. In the end, we show that obtaining a duality between sketching and clustering objectives is non-trivial. This is done by presenting some natural clustering objectives that do not admit any dual sketching objectives.

\end{abstract}

\pagebreak{}
\setcounter{page}{1}
\tableofcontents

\section{Introduction and overview of our results}
With the advent of ever more efficient methods for the acquisition of data from a variety of domains, scientists are increasingly confronted with having to analyze large and complex datasets. In recent years this has led to an accelerated development of data analysis methods by mathematicians, statisticians, computer scientists, etc. Naturally, theoretical understanding of these methods is often achieved progressively after their inception. A natural question that arises is whether any two such methods are related or even \emph{equivalent}: the precise notion of equivalence being determined in terms of (1) the type of output they produce, and/or (2) the computational complexity they incur. This question is important because  equivalence relations based of either type help in classifying methods, and therefore, provide theoretical as well as computational guarantees for a particular method based on the theoretical and computational guarantees of its related methods. 
The mantra is that determining equivalence relations between data analysis methods permits widening our understanding of  extant data analysis tools.

%%%%%%%%%%%%%%%%%%%%%%%%%%%
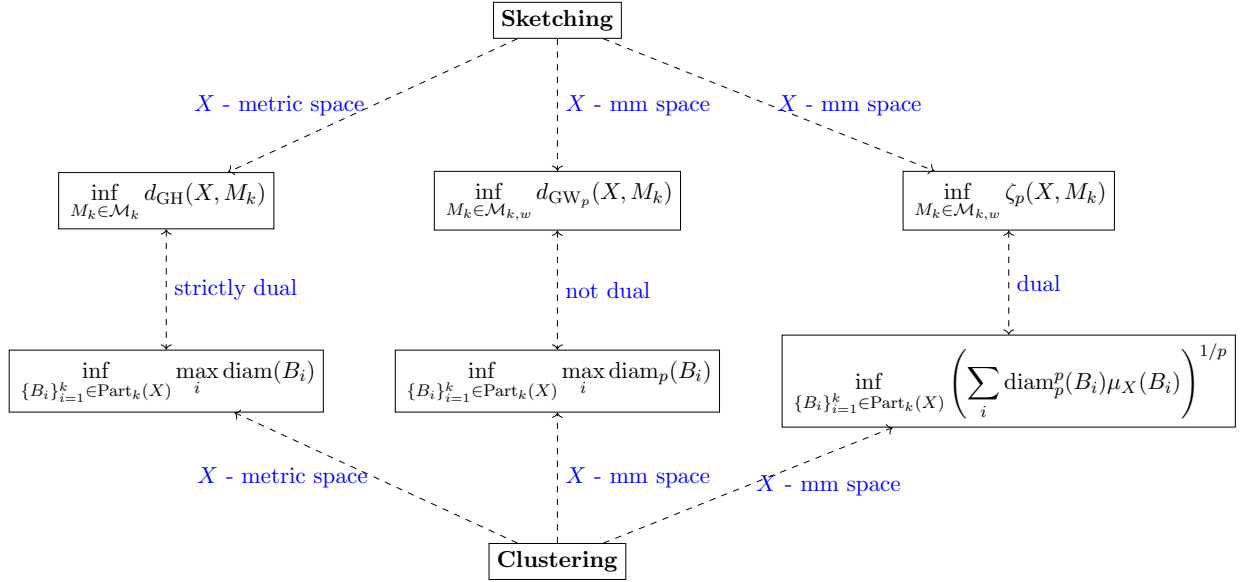
\begin{figure}
\centering
\scalebox{.8}{
\begin{tikzpicture}
\node[rectangle,draw] (a) at (0,0) {\textbf{Sketching}};
\node[rectangle,draw] (e) at (-6.5,-3) {$\begin{aligned} \inf_{M_k \in \mathcal{M}_k} \dgh(X,M_k) \end{aligned}$};
\node[rectangle,draw] (g) at (0,-3) {$\begin{aligned} \inf_{M_k \in \mathcal{M}_{k,w}} \dgwp(X,M_k) \end{aligned} $};
\node[rectangle,draw] (h) at (7.5,-3) {$\begin{aligned} \inf_{M_k \in \mathcal{M}_{k,w}} \zeta_p(X,M_k) \end{aligned} $};
\node[rectangle,draw] (f) at (-6.5,-6) {$\begin{aligned} \inf_{\{B_i \}_{i=1}^k \in \partk(X)} \max_i \diam(B_i) \end{aligned} $};
\node[rectangle,draw] (i) at (0, -6) {$\begin{aligned} \inf_{\{B_i\}_{i=1}^k \in \partk(X)} \max_i \diam_p(B_i) \end{aligned} $};
\node[rectangle,draw] (j) at (7.5,-6) {$\begin{aligned} \inf_{\{B_i \}_{i=1}^k \in \partk(X)} \left( \sum_i \diam_p^p(B_i) \mu_X(B_i) \right)^{1/p} \end{aligned} $};
\node[rectangle,draw] (b) at (0,-9) {\textbf{Clustering}};
\path[dashed,<->] (e) edge node[right,blue]{strictly dual} (f);
\path[dashed,<->] (g) edge node[right,blue]{not dual} (i);
\path[dashed,<->] (h) edge node[right,blue]{dual} (j);
\path[dashed,->] (a) edge node[left,blue]{$X$ - metric space} (e);
\path[dashed,->] (a) edge node[right,blue]{$X$ - mm space} (g);
\path[dashed,->] (a) edge node[right,blue]{$X$ - mm space} (h);
\path[dashed,->] (b) edge node[left,blue]{$X$ - metric space} (f);
\path[dashed,->] (b) edge node[right,blue]{$X$ - mm space} (i);
\path[dashed,->] (b) edge node[right,blue]{$X$ - mm space} (j);
\end{tikzpicture}
}
\caption{A summary of our duality results. See the text for details. \label{fig:pict}} 
\end{figure}

%%%%%%%%%%%%%%%%%%%%%%%%%%%

 In our work, we determine equivalence relations between two different families of data analysis methods: \textbf{Clustering} and \textbf{Sketching}.\footnote{We use  the term "sketching" to encompass data simplification methods.}  Informally speaking, clustering refers to partitioning data into ``meaningful" subsets, whereas sketching refers to approximating the dataset with a set of smaller cardinality which, crucially, retains some of the properties of the original dataset.  Clustering of data and sketching/simplification of data are two of the most fundamental data analysis methods with numerous applications in science and engineering \cite{lance67general,js71, jain1988algorithms, har2011geometric,hartigan-stat-clust,hartigan-book,ABSW07,GM09,EK03,BBK06,WSAB07,dhillon2001concept,forgy1965cluster}. As a consequence of this richness in applications, in the last 50 years various notions of sketching and clustering of data have been reported in the literature. 

\paragraph{Sketching.}
Sketching is, intuitively, the broad problem of obtaining a reasonably good ``summary'' of a dataset $X$.\footnote{We note that there is rich literature on \emph{random sampling methods}. We remark that in our paper we solely consider deterministic approximation/sampling methods.}  
We propose a general framework of sketching, and establish formal connections with clustering.

Before delving into our general treatment of sketching, let us discuss a motivating example.
In the applied literature, in particular in  machine learning, one form in which sketching/simplifications arise is when computing low rank approximations of kernel matrices, with the goal of reducing the computational cost of machine learning problems that depend on them \cite{hofmann2008kernel}. A common scenario is that given a dataset set $X$ one has a feature map $\varphi:X\rightarrow \mathcal{V}$ into some inner product space $(\mathcal{V},\langle\cdot,\cdot\rangle_{\mathcal{V}})$. This induces a kernel or Gram matrix $k_X$ on $X$: for $x,x'\in X$, $k_X(x,x') := \langle \varphi(x),\varphi(x')\rangle_\mathcal{V}.$  One of the basic tasks is approximating $k_X$ by a product of  matrices with lower rank. One such method  is the Nystr\"om method \cite{williams2001using}, which proceeds by making a choice of a small number $\ell$ of landmarks $L=\{p_1,\ldots,p_\ell\}$ from $X$, and then considers a certain low rank approximation $\widehat{k}_{X,L}$ to $k_X$. Many procedures for making the choice of $L$ are \emph{greedy} in that one adaptively chooses a new landmark based on the previous choices \cite{smola2000sparse,fine2001efficient,bach2002kernel,kumar}. 

Despite sharing the essential idea of approximating a given object (a kernel/Gram matrix) with a simpler object (a low rank approximation), 
our setting differs in that we represent data as a metric space, that is, instead of recording the inner products between different points in the dataset (which is the information contained in the Gram matrix), we represent the data via a distance matrix $d_X$. There is of course a connection between our metric representation of data, and the kernel representation: given the kernel matrix $k_X$, one can induce a distance matrix  $d_X$ on $X$ via
\begin{equation}\label{eq:kernel}d_X(x,x') = \big\langle \varphi(x)-\varphi(x'),\varphi(x)-\varphi(x')\big\rangle_{\mathcal{V}}^\frac{1}{2} = \big(k_X(x,x)+k_X(x',x')-2k_X(x,x')\big)^{1/2}.\end{equation} However, it is well known that there exist distance matrices that  do not admit an expression such as (\ref{eq:kernel}) -- the precise characterization of such distance matrices has been carried out in early work on distance geometry by Blumenthal and Schoenberg \cite{blumenthal1953theory,boor1988j}. The metric spaces one obtains in this way are a special strict subset of the collection of all metric spaces.

In this paper, we focus on datasets that can be represented as \emph{general} metric spaces -- not just those that admit a representation such as (\ref{eq:kernel}). Furthermore, we also adopt the point of view that a dataset is a point in some metric space $(D,\rho)$. This point of view has been considered in the literature; for example, $D$ could be the collection of all measures in the $2$-dimensional grid, and $\rho$ the earth-mover's (i.e. Wasserstein) distance (see \cite{andoni2009efficient, andoni2016impossibility}), which is a setting arising in image analysis and pattern recognition \cite{guibas-emd}. Another example is when ${D}$ is a set of strings, and $\rho$ the edit distance (see \cite{andoni2013homomorphic}), which is a natural setting in string analysis and computational biology. Similarly, $D$ could be some $\ell_p$ space (see \cite{indyk2006stable, kane2010exact}), or some general normed space (see \cite{andoni2015sketching}), which is an important setting in streaming algorithms. Here, streaming algorithms are algorithms for processing data streams in which the input is presented as a sequence of items and can be examined in only a few passes. These algorithms have found applications in networking and in databases.

In the same way that certain greedy methods are used for approximating kernel matrices, greedy methods also exist and are used for approximating general metric spaces.  Currently, one of the most widely used deterministic method for obtaining an approximation, or a sketch of a metric space is the so called Farthest Point Sampling method (FPS). FPS is a greedy algorithm that starts by choosing an arbitrary point from the metric space, and adaptively determines the subsequent points by maximizing the distance to the set of chosen points. This method was first introduced by Gonzales \cite{GONZ85}, and since then has been used for sampling metric spaces for various applications \cite{ABSW07,GM09,EK03,MS04,BBK06,WSAB07,javaplex}. In most of these works, it has been shown experimentally that FPS is a good heuristic for approximating metric spaces, but none of these works establish any theoretical guarantees for FPS. In this paper, we study the problem of approximating (sketching) metric spaces by using the existing notions of distance between metric spaces, specifically the Gromov-Hausdorff distance, with the approximation being defined as the (simpler) dataset minimizing this distance. 

Precisely, we study a specific structure of the  sketching objective, which is natural for the case of summarizing metric spaces and metric measure spaces. We consider the case where $D$ is either the collection $\mathcal{M}$ of all compact  metric spaces, or $\mathcal{M}_w$, the collection of all metric measure spaces,\footnote{A metric measure space is a metric space endowed with a Borel probability measure -- see Section \ref{sec:mms-res}.} endowed with some distance $\rho$. Here, $\rho$ is either the Gromov-Hausdorff distance \cite{gromov} between metric spaces and Gromov-Wasserstein distance \cite{Sturm2006,Memoli2011,sturm-ss} between metric measure spaces.
For $k \in \mathbb{N}$, we also let ${\cal M}_k$ be the collection of all $k$-point metric spaces, or metric measure spaces.
Then, for some $M\in {\cal M}$, the \emph{$k$-point sketch} of $M$ is defined to be a nearest-neighbor of $M$ in ${\cal M}_k$; that is, we define
\begin{equation}\label{eq:sketch-strcture}\tag{\#}
s_k(M) :=\inf_{M'\in {\cal M}_k} \rho(M, M'),
\end{equation}
breaking ties arbitrarily. We refer to the problem of computing this mapping as \emph{projective sketching} (i.e.~we project ${\cal M}$ onto ${\cal M}_k$). In this work, projective sketching is shown to be \emph{equivalent} to certain existing notions of clustering, both for metric spaces and for metric measure spaces. We describe these notions of clustering in the next paragraph.

We also note that there are some deep results in the sketching literature that do not fit within our broad definition of sketching a point $x$ in some metric space $M$.
Such examples include sketching information divergences \cite{DBLP:journals/ml/GuhaIM08}, and 
certain quadratic forms \cite{andoni2016sketching}.

\paragraph{Clustering.}
Clustering, given $k \in \mathbb{N} $, broadly refers to the problem of partitioning a metric space into $k$ clusters (or blocks), where the eventual  choice of clusters is usually determined through minimizing some given cost function over the set of all partitions of the given metric space into $k$ clusters. Several methods have been considered for clustering metric spaces such as $k$-center clustering \cite{GONZ85,hochbaum1986unified}, $k$-median clustering \cite{jain1988algorithms,arya2004local,charikar2005improved,korupolu2000analysis,kanungo2002local,jain2001approximation,thorup2001quick}, $k$-means clustering \cite{forgy1965cluster,lloyd1982least,dhillon2001concept,arthur2006slow,vattani2011k,har2004coresets,har2005fast,DBLP:journals/dcg/Har-PeledK07,matouvsek2000approximate,mahajan2009planar,aloise2009np,dasgupta2008hardness,drineas2004clustering,kanungo2002local,arthur2007k,Mac67,HW79}, hierarchical clustering \cite{Def77,Sib73,ELL11,clust-um}, distribution based clustering \cite{FR06,BR93,XEKS98} and so on.

For metric spaces, we consider the clustering objective corresponding to minimizing the maximal diameter of the $k$ clusters. This objective has been well studied \cite{GONZ85}, and there is a $2$-approximation algorithm for obtaining such a clustering. For metric measure spaces, for $1 \leq p \leq \infty $, we consider the clustering objective of minimizing the weighted mean of $p$-diameters of the clusters. This clustering objective agrees with the objective for $\ell_p$-facility location problem \cite{GT08} up to a constant. There exist approximation algorithms for the $\ell_p$-facility location problem for all $p \in [1, \infty] $ \cite{arya2004local,charikar2005improved, GT08}. 
These approximation results for the clustering objectives are in contrast to the projective sketching objective in equation (\ref{eq:sketch-strcture}) for which no computational complexity results were previously known. In the next two paragraphs, we describe how our work implies such computational complexity guarantees.

\paragraph{Duality between sketching and clustering.}
We establish a \emph{duality} i.e.~an equivalence between these two, seemingly different problems of sketching and clustering, both for metric spaces and for metric measure spaces. By duality we mean the following type of phenomenon: we show that for both metric spaces and metric measure spaces, there exist universal constants $0<C_1\leq C_2$ such that for every metric space or metric measure space $X$, the ratio between the optimal values of the clustering and sketching objectives for $X$ is always bounded above by $C_2$ and bounded below by $C_1$. For metric spaces, the constants $C_1$ and $C_2$ are equal -- we call this phenomenon \emph{strong duality}. The strong duality for metric spaces immediately has the following two surprising consequences:

\begin{itemize}
\item \emph{Computational consequence (cf. Section \ref{sec:comp-cons}):} Whereas it is known that it is NP-hard to compute any $3$-approximation to the  Gromov-Hausdorff distance between two  finite metric spaces \cite{Schmiedl17}, we prove that for a given $k$, the problem of computing the \emph{optimal} Gromov-Hausdorff approximation to a given metric space by a $k$-point metric space admits a $2$-approximation computable in polynomial time. This result is obtained by invoking the above duality and results from \cite{GONZ85}.

\item \emph{Theoretical consequence (cf. Theorem \ref{thm:ls} and Corollary \ref{cor:sketchs1}):} By  the Lusternik-Schnirelmann, we are first able to 	establish that  for $k\leq n+1$, any clustering into $k$ blocks of the sphere $\mathbb{S}^n$ (with geodesic distance) must necessarily have a block with diameter $\pi$. Then, invoking our strong duality for metric spaces, we can conclude that for all $k\leq n+1$, the optimal $k$-sketching cost of $\mathbb{S}^n$ is no less than (and actually equal to) $\frac{\pi}{2}$. Separately, we show that for all $k \in \mathbb{N}$, every partition of $\mathbb{S}^1$ into $k$ blocks must necessarily have a block of diameter at least $\frac{2 \pi}{k}$. We again invoke our strong duality for metric spaces to obtain that for all $k \in \mathbb{N}$, the optimal $k$-sketching cost of $\mathbb{S}^1$ is $\frac{\pi}{k}$. This implies that for every $k \in \mathbb{N}$, one of the closest $k$-point metric spaces to $\mathbb{S}^1$ is obtained by choosing $k$ points on $\mathbb{S}^1$, such that distance between consecutive points is $\frac{2 \pi}{k}$.  
\end{itemize}

The proof strategy for our strong duality relies on identifying maps --- the Hausdorff and Voronoi maps --- that respectively map any optimal $k$-clustering into an optimal $k$-sketch, and vice-versa.

For  metric measure spaces, the strong duality that we obtained for metric spaces is lost, but  we however retain a relaxed notion of duality: we show that for a certain metric on metric measure spaces, the optimal values of sketching and clustering are within a constant factor of each other. In more detail, we consider two sketching objectives for metric measure spaces, which arise from the two definitions of Gromov-Wasserstein distance present in literature \cite{Sturm2006,Memoli2011}. We prove that only one of the definitions provides duality, while with the other definition, we show counter examples depicting non-equivalence. In addition, we compare these two sketching objectives, and provide examples of metric measure spaces where the ratio between these sketching objectives is unbounded. However, for metric measure spaces of bounded doubling dimension and bounded diameter, we obtain that one of the sketching objective is both upper and lower bounded by a suitable function of the other sketching objective.

Complementing the above results, we also show that it is in general non-trivial to obtain such a duality. This is done by exhibiting some additional natural clustering objectives that do not admit any dual sketching objectives. 

By virtue of our duality results, we obtain additional \emph{computational results} for the sketching objectives, both for metric spaces and for metric measure spaces. In particular, we show NP-hardness of the sketching objective for metric spaces, and obtain approximation algorithms for sketching objectives for both metric spaces and metric measure spaces. These approximation results are obtained using known approximation algorithms for the dual clustering problem. We are also able to provide theoretical guarantees for the Farthest Point Sampling method, using the duality for metric spaces.  We note that, prior to our work, no computational results were known for computing such a $k$-sketch, $k \in \mathbb{N}$, for either a metric space or a metric measure space. In addition, it is possible that for some of these sketching objectives, the approximation factors obtained are best possible.

We formally state our main results in the next section. A pictorial summary of our results is given in Figure \ref{fig:pict}.

%%%%%%%%%%%%
\singlespacing
%%%%%%%%%%%
\subsection{Duality between clustering and sketching}
In this section we give an overview of our results. Let $\mathcal{M}$ denote the collection of all compact metric spaces. For $k \in \mathbf{N}$, let $\mathcal{M}_k \subset \mathcal{M}$ denote the collection of all finite metric spaces of cardinality at most $k$. For $(X, d_{X}) \in \mathcal{M}$ and $k \in \mathbf{N}$, let $\mathrm{Part}_{k}(X)$ denote the set of all partitions of $X$ into $k$ sets. 

We now define sketching and clustering cost functions in general. Examples of these concepts are given immediately after, in Sections \ref{sec:ms-res} and \ref{sec:mms-res}. The sketching and clustering cost functions induce shatter functional and sketch functional respectively, and these functionals are also defined below.

\begin{definition}[\textbf{Clustering cost function}] \label{def:clustercost}
A clustering cost function is any function 
$$\Phi : \mathcal{M} \times \mathrm{Part}_{k}(\cdot) \rightarrow \mathbf{R}_{+}.$$ 
\end{definition}
The interpretation is that on input of some $(X,d_{X}) \in \mathcal{M}$ and some $P \in \mathrm{Part}_{k}(X)$, $\Phi$ returns the cost of partitioning the space $(X,d_{X})$ according to $P$.

\begin{definition}[\textbf{Shatter functional}] \label{def:clustfun}
Given a clustering cost function $\Phi$, we define the $k$-th Shatter functional induced by $\Phi$ $$\shatter_k^\Phi:\mathcal{M}\rightarrow \mathbb{R}_+$$ as follows: for  $(X,d_{X}) \in \mathcal{M}$,
$$\shatter_{k}^{\Phi}(X) := \inf_{P \in \mathrm{Part}_k(X)} \Phi(X,P). $$
\end{definition}

A partition $P \in \partk(X) $ is called an \emph{optimal clustering} of $X$ if it achieves the infimum in the above definition. 
 
\begin{definition}[\textbf{Sketching cost function}] \label{def:sketchcost}
A sketching cost function is any function 
$$\Psi: \mathcal{M} \times \mathcal{M}_{k} \rightarrow \mathbf{R}_{+}.$$
\end{definition}
 The interpretation is that on input of some $(X,d_{X}) \in \mathcal{M}$ and some $(M_{k},d_{k}) \in \mathcal{M}_k$, $\Psi$ returns the cost of sketching (approximating) $(X,d_{X})$ by $(M_{k},d_{k})$.
 
\begin{definition}[\textbf{Sketch functional}] \label{def:sketchfun}
Given a sketching cost function $\Psi$ and we define the $k$-th Sketch functional induced by $\Psi$ $$\sketch_k^\Psi:\mathcal{M}\rightarrow \mathbb{R}_+$$ as follows:  for $(X,d_X) \in \mathcal{M}$,
$$ \sketch_{k}^{\Psi}(X) := \inf_{M_k \in \mathcal{M}_k} \Psi(X,M_k). $$
\end{definition}

A $k$-sketch  of the compact metric space $X$ consists of any pair $(M_k,R)$ where $M_k\in\mathcal{M}_k$ and $R\in\mathcal{R}(X,M_k)$.

\begin{Remark}
Notice that in our definition of a $k$-sketch of $X$ we retain information not only about the approximating space $M_k$, but also about the \emph{way in which $M_k$ relates to $X$} (as given by the correspondence $R$). 
\end{Remark}

We say that a $k$-sketch $(M_k,R)$ of $X$ is an \emph{optimal $k$-sketch} for $X$ if it achieves the infimum in the above definition, that is $\frac{\dis(R)}{2} = \sketch_k^\Psi(X)$. Given $\lambda\geq 1$, we say that the $k$-sketch $(M_k,R)$ of $X$ is a $\lambda$-approximation of $\sketch_k^\Psi(X)$ if 
$$\frac{\dis(R)}{2}\leq \lambda \cdot \sketch_k^\Psi(X).$$

The notion of duality between sketching and clustering cost functions is defined via comparison of the sketching and shatter functionals that they induce.  

\begin{definition}[\textbf{Duality}] \label{def:duality}
A sketching cost function $\Psi$ and a clustering cost function $\Phi$ are called \emph{dual} if there exist constants $C_2 \geq  C_1 > 0$ such that 
\begin{equation}\label{eq:duality} C_1 \cdot \shatter_{k}^{\Phi}(X) \leq \sketch_{k}^{\Psi}(X) \leq C_2 \cdot \shatter_{k}^{\Phi}(X) \end{equation}

for every $X \in \mathcal{M}$ and $k \in \mathbf{N}$. The duality is \emph{strict} when $C_{1} = C_{2}$. Interchangeably, we will say that $\shatter_{k}^{\Phi}$ and $\shatter_{k}^{\Phi}$ are dual (resp. strictly dual) when the above conditions hold.
\end{definition}

We now describe the duality results that we obtain for metric spaces and for metric measure spaces.

\subsection{Our results in the case of metric spaces}\label{sec:ms-res}

Given $X \in \mathcal{M}$, $k \in \mathbf{N}$ and $M_k \in \mathcal{M}_k$, we consider the particular sketching cost function $$\Psi_{\mathcal{M}}(X,M_k) := \dgh(X,M_k).$$  Here $\dgh$ is the Gromov-Hausdorff distance between metric spaces, see Section \ref{sec:GH-details} for details. The resulting sketching objective is defined as 
\begin{equation}\label{eq:sketch-ms}
\sketch_{k}(X) := \sketch^{\Psi_{\mathcal{M}} }_k(X) = \inf_{(M_{k}, d_{k}) \in \mathcal{M}_{k}} \dgh(X, M_{k}). \end{equation}
 
For a partition  $P = \{B_{i} \}_{i=1}^{k} \in \partk(X)$, we consider the specific  clustering cost function $$\Phi_{\mathcal{M}} (X,P) := \max_{i} \diam(B_{i}),$$
where for any set $S \subseteq X $, its diameter is $\diam(S) := \mathrm{sup}_{x,x' \in S}d_{X}(x,x')$.  The resulting clustering objective is defined as
$$ \shatter_{k}(X) := \shatter_k^{\Phi_{\mathcal{M}}}(X) = \inf_{P = \{B_i \}_{i=1}^k \in \mathrm{Part}_{k}(X)} \max_i \diam(B_i).$$

\begin{Remark}
We observe that $\shatter_k(X)$ is within a factor of $2$ from the $k$-center objective \cite{har2011geometric}. Indeed, recall that the $k$-center problem for finite metric spaces is to find a subset $L$ of the input metric space $X$ such that the quantity $\max_{x \in X} \min_{l \in L}d_X(x,l)$ is minimized. Now, any subset $L$ of $X$ determines a clustering of $X$ by assigning every point in $X \setminus L$ to its closest point in $L$, and this clustering satisfies that the maximum diameter of any cluster is at most $2 \cdot \max_{x \in X} \min_{l \in l}d_X(x,l)$.  
\end{Remark}

\begin{proposition}\label{prop:shatter-s1}
For every $k \in \mathbf{N}$, $\shatter(\mathbb{S}^1) = \frac{2 \pi}{k}$.
\end{proposition}

\begin{proof}
We first show that for every $k \in \mathbf{N}$, we have $\shatter_k(\mathbb{S}^1) = \frac{2 \pi}{k}$. Clearly, $\shatter_k(\mathbb{S}^1) \leq \frac{2 \pi}{k}$, since the equipartition of $\mathbb{S}^1$ into $k$ blocks satisfies that the diameter of every block is $\frac{2 \pi}{k}$. Let $P = \{B_1, \ldots, B_k \}$ be a partition of $\mathbb{S}^1$ into $k$ blocks such that for every $i \in [k]$, $\diam(B_i) \leq \frac{2 \pi}{k}$. For every $i \in [k]$, we denote by $\overline{B_i}$ the closure of the convex hull of $B_i$ in $\mathbb{S}^1$. We observe that $\diam(\overline{B_i}) \leq \diam(B_i)$ for every $i \in [k]$. This is because if $\overline{B_i} = [\theta_1, \theta_2] $, where $\theta_1$ and $\theta_2$ are angles in the anticlockwise direction from the positive x-axis, then both $\theta_1, \theta_2 \in B_i$, and therefore $\diam(\overline{B_i}) = d_{\mathbb{S}^1}(\theta_1, \theta_2) \leq \diam(B_i) $. Now, we define $\hat{B_1} = \overline{B_1}, \hat{B_2} = \overline{B_2} \setminus \hat{B_1}, \ldots, \hat{B_k} = \overline{B_k} \setminus (\cup_{i=1}^{k-1} \hat{B_i})$. Since $B_1, \ldots, B_k$ was a partition of $\mathbb{S}^1$, we obtain that $\hat{B_1}, \hat{B_2}, \ldots, \hat{B_k}$ is also a partition of $\mathbb{S}^1$. In addition, we obtain that each $\hat{B_i}$ is a connected set. We use $\len(\hat{B_i})$ to denote the length of the arc $\hat{B_i}$. Thus, we have that $\len(\hat{B_1}) + \ldots + \len(\hat{B_k}) = 2 \pi$. This implies that there exists an $i \in [k]$ such that $\len(\hat{B_i}) \geq \frac{2 \pi}{k}$. We have proved that for every partition $\{B_i \}_{i=1}^k \in \partk(\mathbb{S}^1) $ satisfying $\max_i \diam(B_i) \leq \frac{2 \pi}{k}$, there exists a partition $\{\hat{B_i} \}_{i=1}^k$ such that each $\hat{B_i}$ is connected, $\diam(\hat{B_i}) \leq \diam(B_i)$ for every $i \in [k]$, and $\diam(\hat{B_i}) \geq \frac{2 \pi}{k}$ for some $i \in [k]$. This implies that $\max_i \diam(B_i) \geq \frac{2 \pi}{k}$, and we conclude that $\shatter_k(\mathbb{S}^1) \geq \frac{2 \pi}{k}$. Thus, we obtain that $\shatter_k(\mathbb{S}^1) = \frac{2 \pi}{k}$. 
\end{proof}

Our main result in this setting is the following theorem which establishes a strict duality between sketching and clustering:

\begin{theorem}[Strict duality for metric spaces] \label{thm:main1}
For every $(X, d_{X}) \in \mathcal{M}$ and every $k \in \mathbf{N}$, we have
$$ \sketch_{k}(X) = \frac{1}{2} \cdot \shatter_{k}(X).$$
\end{theorem}

Algorithmic applications of the strict duality theorem above are treated in Sections \ref{sec:sketch-np-hard} and  \ref{sec:poly-sketch}. 
From a different perspective, this strict duality permits obtaining  sharp results for the sketch functional of spheres via a topological argument. 

\paragraph{An application: $\sketch_k(\mathbb{S}^n)$.}

We now crucially invoke the duality established between $\sketch_k(X)$ and $\shatter_k(X)$, along with the Lusternik-Schnirelmann theorem to show that for $k,n \in \mathbb{N}$ with $k \leq n+1$, $\sketch_k(\mathbb{S}^n) = \frac{\pi}{2} $.\footnote{Note that the Lusternik-Schnirelmann theorem is a topological fact about spheres.} Here, we view spheres as metric spaces by endowing them with the geodesic distance. The informal interpretation of this fact is the following: 

\begin{quote}\emph{
Any finite metric space with at most $n+1$ points will provide a poor approximation to $\mathbb{S}^n$. Furthermore, this is so for topological reasons.} \end{quote}

\begin{theorem}\label{thm:ls}
For all $k,n \in \mathbb{N}$ with $k \leq n+1$, $\sketch_k(\mathbb{S}^n) = \frac{\pi}{2}$.
\end{theorem}

\begin{proof}
Fix $k \leq n+1$. Let $P = \{B_1, \ldots, B_k\}$ denote a partition of $\mathbb{S}^n $ into $k$ sets. We now invoke the Lusternik-Schnirelmann theorem \cite{bollobás_2006} that states the following: if the sphere $\mathbb{S}^n$ is covered by $n+1$ closed subsets, then one of these sets contains a pair of antipodal points. Since $\{ \overline{B_1}, \ldots, \overline{B_k} \} $ forms a closed cover of $\mathbb{S}^n$, we have that there exists $i_0 \in [k] $ such that $\overline{B_{i_0}} $ contains a pair of antipodal points. This implies that $\diam(\overline{B_{i_0}}) = \diam(B_{i_0}) = \pi $. Since $P$ was an arbitrary partition of $\mathbb{S}^n$ into $k$ blocks, we have that $\shatter_k(\mathbb{S}^n) = \pi$. Thus, by Theorem \ref{thm:main1}, we obtain $\sketch_k(\mathbb{S}^n) = \frac{1}{2} \cdot \shatter_k(\mathbb{S}^n) = \frac{\pi}{2} $.
\end{proof}

The above theorem stipulates that for $n\geq k+1$ the number $\sketch_k(\mathbb{S}^n)$ is large and independent of $k$. Now, for $n=1$ fixed, we characterize precisely how increasing $k$ translates into a decrease of $\sketch_k(\mathbb{S}^1).$ This is an immediate consequence of Proposition \ref{prop:shatter-s1} and Theorem \ref{thm:main1}:
\begin{corollary} \label{cor:sketchs1}
For all $k\in \mathbb{N}$, we have $\sketch_k(\mathbb{S}^1) = \frac{\pi}{k}.$
\end{corollary}

\subsection{Our results in the case of metric measure spaces}\label{sec:mms-res}
We now state the duality results that we obtain for metric measure spaces.  In this section we introduce clustering cost functions for mm-spaces (cf. Definition \ref{def:clustercost}) and from these derive a definition of Shatter functional (cf. Definition \ref{def:clustfun}) applicable to mm-spaces. Similarly, we also introduce sketching cost functions in the context of mm-spaces (cf. Definition \ref{def:sketchcost}), and also induce a corresponding definition of sketch functionals (cf. Definition \ref{def:sketchfun}) for mm-spaces.

\begin{definition}[\textbf{Metric measure space}]
A compact metric space $(X,d_{X}) \in \mathcal{M}$ is called a \emph{metric measure space} (mm-space for short) if $X$ is endowed with a probability measure $\mu_{X}$ such that its support is all of $X$: $\mathrm{supp}[\mu_{X}] = X$.
\end{definition}

We denote by $\mathcal{M}_w$ the collection of all mm-spaces and by $\mathcal{M}_{k,w}$ the collection of all metric measure spaces of cardinality at most $k$. We now define the sketching cost function for mm-spaces. The definition is analogous to that for metric spaces, but the Gromov-Hausdorff distance between metric spaces is replaced by Gromov-Wasserstein distance between mm-spaces \cite{Memoli2011}. Given $X \in \mathcal{M}_w$, $k \in \mathbf{N}$, $1 \leq p \leq \infty$ and $M_k \in \mathcal{M}_{k,w}$, we consider the sketching cost function $$\Psi_p(X,M_k) := \dgwp(X, M_{k}).$$ Here $\dgwp$ is the $p$-Gromov-Wasserstein distance, see Section \ref{sec:weak-GW-Defn} for details. The resulting sketching objective is defined as 
\begin{equation}\label{eq:sketch-mm} \sketch_{k,p}(X) := \sketch^{\Psi_p}_k(X) = \inf_{(M_{k}, d_{k}, \mu_{k})} \dgwp(X, M_{k}). 
\end{equation}

In order to define the clustering objective, we need to define the $p$-diameter of a metric measure space.

\begin{definition}[\textbf{$p$-diameter}]
Given $X \in \mathcal{M}_{w}$, $1 \leq p < \infty$ and $B \subseteq X$, define \emph{$p$-diameter} of $B$ as 
$$ \diam_{p}(B) := \left( \int_{B} \int_{B} d^{p}(x,x') \frac{d \mu_{X}(x) \,d \mu_{X}(x')}{(\mu_{X}(B))^2} \right)^{1/p}. $$
For $p = \infty$, define
$ \diam_{\infty}(B) := \sup_{x,x' \in B}d_{X}(x,x'). $
\end{definition}

\begin{example} 
For example, under Euclidean distance and Lebesgue measure, we have that for all $1 \leq p \leq \infty$, $$\diam_p([0,1]) = \left(\frac{2}{(p+1)(p+2)} \right)^{1/p}.$$ This is proved as follows: by definition,  $\diam_p^p([0,1]) = \int_{[0,1]} \int_{[0,1]} d^p(x,x') \,dx \,dx' $. Since for all $x,x' \in [0,1]$, $d(x,x') = |x-x'|$, we have that
$$ \diam_p^p([0,1]) = \int_0^1 \int_0^1 |x-x'|^p \, dx \, dx'= \int_0^1 \int_{x \leq x'} (x'-x)^p dx \, dx' + \int_0^1 \int_{x \geq x'} (x-x')^p \,dx \, dx'. $$
By symmetry, we have that $\int_0^1 \int_{x \leq x'} (x'-x)^p dx \,dx' = \int_0^1 \int_{x \geq x'} (x-x')^p \, dx \, dx' $. It is easy to check that $\int_0^1 \int_{x \leq x'} (x'-x)^p dx \, dx' = \frac{1}{(p+1)(p+2)} $. Thus, we obtain $\diam_p^p([0,1]) = \frac{2}{(p+1)(p+2)} $. 

A similar calculation which we omit shows that under spherical distance and normalized length measure, we have $$\diam_p(\mathbb{S}^1) = \frac{\pi}{(p+1)^{1/p}}.$$
\end{example}

We observe that our sketching objective for mm-spaces eq. (\ref{eq:sketch-mm}) is analogous to that for metric spaces eq. (\ref{eq:sketch-ms}), with the Gromov-Hausdorff distance between metric spaces being replaced by the $p$-Gromov-Wasserstein distance between mm-spaces. We want to define the clustering objective for mm-spaces similarly. Therefore, instead of considering the maximum diameter of the clustering, as we did for metric spaces, we consider the weighted sum of the $p$-diameters of the clusters for mm-spaces.
The formal definition is as follows: given $X \in \mathcal{M}_{w}$, $k \in \mathbf{N}$, $1 \leq p < \infty$, $1 \leq q < \infty$ and $P = \{B_{i} \}_{i=1}^{k} \in \partk(X)$, we consider the  clustering cost function $$  \Phi_{p,q}(X,P) := \left(\sum_{i} \diam_{p}^{q}(B_{i}) \,\mu_{X}(B_{i}) \right)^{1/q}.$$ The resulting clustering objective is defined as
$$ \shatter_{k,p,q}(X) 
:= \inf_{P \in \partk(X)} \Phi_{p,q}(X,P) = \inf_{P \in \partk(X)} \left(\sum_{i} \diam_{p}^{q}(B_{i}) \,\mu_{X}(B_{i}) \right)^{1/q} . $$
For $1 \leq p \leq \infty$, $q = \infty$ and any $P = \{B_{i} \}_{i=1}^{k} \in \partk(X)$, we consider $ \Phi_{p,\infty}(X,P) := \max_{i} \diam_{p}(B_{i})$ and define the clustering objective as  
$$ \shatter_{k,p,\infty}(X) 
:= \inf_{P \in \partk(X)} \Phi_{p, \infty}(X,P) = \inf_{P \in \partk(X)}\max_{i} \diam_{p}(B_{i}). $$

\begin{Remark}
We remark that 
$\shatter_{k,1,1}(X)$ is within a factor of $2$ from the  $k$-median objective (see Remark \ref{rem:kmed}),
and $\shatter_{k,2,2}(X)$ is within a factor of $2$ from the $k$-means objective (see Remark \ref{rem:kmed}).

\end{Remark}

We show the following bound.
\begin{theorem}\label{thm:main-weak}
$\sketch_{k,p}(X) \leq \shatter_{k,p, \infty}(X)$ for all $k\in \mathbf{N}$ and $1\leq p < \infty$.
\end{theorem}
We also show that, in general, $\sketch_{k,p}$ and $\shatter_{k,p,\infty}$ are \emph{not} dual (Theorem \ref{thm:counter-ex}).

The negative result of Theorem \ref{thm:main-weak} motivates the following definition of a stronger sketching objective for metric measure spaces.
Given $X \in \mathcal{M}_{w}$, $k \in \mathbf{N}$, $1 \leq p \leq \infty$ and $M_k \in \mathcal{M}_{k,w}$, we consider the sketching objective $\Psi^S_{p}(X,M_k) = \zeta_p(X,M_k)$. Here $\zeta_p(X,M_k)$ is the Sturm's version of $p$-Gromov-Wasserstein distance \cite{Sturm2006}, see Section \ref{sec:Sturmsdefn} for details. The resulting sketching objective is defined as
$$ \sketch^{S}_{k,p}(X) := \sketch_k^{\Psi^S_p}(X) =  \inf_{(M_{k}, d_{k}, \mu_{k})} \zeta_{p}(X,M_{k}). $$

Our main duality result for metric measure spaces is the following.

\begin{theorem}[Duality for mm-spaces] \label{thm:main2}
For all $(X,d_X,\mu_X) \in \mathcal{M}_{w}$, $k \in \mathbf{N}$ and $1 \leq p < \infty$, we have
$$\frac{1}{2} \cdot \shatter_{k,p,p}(X) \leq \sketch^{S}_{k,p}(X) \leq \shatter_{k,p,p}(X). $$ 
\end{theorem}

We also show a relation between the weak and strong sketching objectives for the case of mm-spaces of bounded doubling dimension and bounded diameter. In this case, $\sketch^S_{k,p}(X)$ is bounded by a function of $\sketch_{k,p}(X)$.

\begin{theorem} \label{thm:doubling}
Let $X \in \mathcal{M}_{w}$ with doubling constant $C>0$ and $p \in [1, \infty)$. Then,
$$ \delta\leq \sketch^{S}_{k,p}(X) < \left(8 \cdot \diam(X)\cdot \delta^{1/ (5\log_{2}C)} + \delta^{1/5} \right)^{1/p} \cdot M_X $$
whenever $ \sketch_{k,p}(X) = \delta < 2^{-5}$. Here $M_X = 2 \cdot \diam(X) + 45 $.  
\end{theorem}

\subsection{Computational considerations and results}\label{sec:comp-cons}
We also consider computational problems that arise in the context of optimizing the sketching objective, either exactly or approximately.
Interestingly, we obtain polynomial-time approximation algorithms for various sketching objectives, by combining the duality theorems with known approximation algorithms for clustering.
We obtain algorithms via this duality paradigm both for metric spaces (Section \ref{sec:approx-sketch}) and for mm-spaces (Section \ref{sec:sketch-s-approx}). We would like to emphasize here that computing Gromov-Hausdorff distances between finite metric spaces leads to solving an instance of the Quadratic Assignment Problem, which is NP-Hard. In addition, computing any $(1+\epsilon)$-approximation is also NP-Hard, for all $\epsilon > 0$ \cite{dgh-props,Schmiedl17,AgarwalFNSW15}. However, by virtue of the strong duality for metric spaces, we obtain a simple $2$-approximation algorithm for computing the Gromov-Hausdorff distance between a finite metric space $X$ and the $k$-point metric space closest to $X$ for $k < |X|$. In addition, we show that obtaining an approximation factor of $2 - \eps$ is NP-Hard.

The strong duality for metric spaces also enables us to analyze the Farthest Point Sampling (FPS) method. We show that the set of $k$ centers obtained using FPS is a $4$-approximation to the optimal set of $k$ centers, when the cost of picking a set of centers $K \subseteq X$ is $\dgh(X,K)$. Thus, we provide the first theoretical guarantees for FPS. We complement the above results with NP-hardness proof of the sketching objective for metric spaces. This is done by means of a reduction from the set cover decision problem (Section \ref{sec:sketch-np-hard}). 
\subsection{Impossibility results}

We finally show that there exist clustering objectives that do not admit any ``natural'' dual sketching objective. An example of such a clustering objective is maximizing the minimum inter-cluster distance. 
We define a notion of ``natural'' sketching objective (called \emph{admissible}), that follows from a few simple axioms.
We show that no natural sketching objective is dual to the above clustering objective. See Section \ref{sec:impossibility} for details.

\subsection{Organization}
This paper is organized as follows. In Section \ref{sec:GH-details}, we prove strict duality between $\sketch_k(X)$ and $\shatter_k(X)$, and present results on NP-hardness and $2$-approximation of $\sketch_k(X)$. In Section \ref{sec:mm-spaces}, we prove duality between $\sketch_{k,p}(X)$ and $\shatter_{k,p,\infty}(X)$ for some values of $k$ and $p$, but show that, in general, $\sketch_{k,p}(X)$ and $\shatter_{k,p,\infty}(X)$ are not dual to each other. We then prove the duality between $\sketch^S_{k,p}(X)$ and $\shatter_{k,p,p}(X)$ for all values of $k$ and $p$. In Section \ref{sec:doubling}, we prove a relation between $\sketch_{k,p}(X)$ and $\sketch^S_{k,p}(X)$ for doubling metric measure spaces. In Section \ref{sec:impossibility}, we present examples of clustering cost functions that do not admit a dual sketching cost function.

\section{The case of metric spaces} \label{sec:GH-details}
In this section, we first prove that for all $X \in \mathcal{M}$ and $k \in \mathbf{N}$, $\sketch_k(X)$ is strictly dual to $\shatter_k(X)$. We use this duality to compute $\sketch_k(\mathbb{S}^n) $ for $n \in \mathbb{N} $ and $k \leq n+1 $. Then, in Section \ref{sec:sketch-np-hard}, we show that, in general, computation of $\sketch_k(X)$ is NP-Hard. However, as a result of the duality, in Section \ref{sec:approx-sketch}, we obtain a polynomial time approximation of $\sketch_k(X)$. In Section \ref{sec:minimizers}, we show that an optimal sketch of a metric space may not be any of its subspaces.

We start by defining the notion of Gromov-Hausdorff distance between metric spaces. Informally speaking, the Gromov-Hausdorff distance between $(X,d_X) $ and $(Y, d_Y)$ measures the minimum cost incurred from associating elements of $X$ to elements of $Y$. This association is called a \emph{correspondence}. We now, mathematically define, a correspondence between metric spaces.

\begin{definition}[\textbf{Correspondence}]
Given $(X, d_{X}), (Y, d_{Y}) \in \mathcal{M} $, a \emph{correspondence} $R$ between $X$ and $Y$ is a subset $R \subset X \times Y$ such that $\pi_{1}(R) = X$ and $\pi_{2}(R) = Y$. Here $\pi_1$ and $\pi_2$ are the projection maps.
\end{definition} 
Let $\mathcal{R}(X,Y)$ denote the set of all correspondences between $X$ and $Y$. 
We define next the \emph{distortion} of a correspondence $R \in \mathcal{R}(X,Y)$, which is the cost associated with $R$.

\begin{definition}[\textbf{Distortion of a correspondence}]
Given a correspondence $R$ between $(X, d_{X})$ and $(Y, d_{Y})$, its \emph{distortion} is given by
$$\dis(R):=\sup_{(x,y),(x',y')\in R}\big|d_X(x,x')-d_Y(y,y')\big|.$$
\end{definition}

We are now ready to define the Gromov-Hausdorff distance between metric spaces. 
\begin{definition}[\textbf{Gromov-Hausdorff distance}]
The Gromov-Hausdorff distance between \\$(X, d_{X}), (Y, d_{Y}) \in \mathcal{M}$ is defined as follows:
$$\dgh(X,Y):=\frac{1}{2}\inf_{R \in \mathcal{R}(X,Y)} \dis(R).$$
\end{definition}

The Gromov-Hausdorff distance is a pseudo-metric on $\mathcal{M}$ and satisfies triangle inequality, symmetry, positivity, and $\dgh(X,Y)=0$ if and only if $X$ and $Y$ are isometric \cite{burago-book}. A fact is that in general the computation of GH distance leads to NP-hard problems. Furthermore, any $(1+\epsilon)$-approximation is also NP-hard, see  \cite{dgh-props,Schmiedl17,AgarwalFNSW15}. 

\begin{example}[$\sketch_1(\cdot)$]\label{ex:dgh-ast} The distance between any $(X,d_X)\in\mathcal{M}$ and the one point metric space $\ast$ satisfies $d_{\mathrm{GH}}(X,\ast)=\frac{\mathrm{diam}(X)}{2},$
see \cite[Chapter 7]{burago-book}.  This means that for any $(X,d_X)\in\mathcal{M}$, $\sketch_1(X)=\frac{\mathrm{diam}(X)}{2}.$
\end{example}

A correspondence $R \subseteq X \times Y $ is called \emph{optimal} if it achieves the infimum i.e.$$2 \cdot \dgh(X,Y) = \dis(R).$$ 

\begin{Remark}
It is shown in Proposition 1.1 of \cite{cm16} that the set of closed optimal correspondences between a pair of metric spaces is non-empty. This implies that given $(X,d_X), (Y,d_Y) \in \mathcal{M} $, there exists $R \in \mathcal{R}(X,Y)$ closed such that $2 \cdot \dgh(X,Y) = \dis(R) $.
\end{Remark}

The Gromov-Hausdorff distance is a generalization of the Hausdorff distance between metric spaces. We first define the Hausdorff distance and then state the relation between these two notions of distances on metric spaces.

\begin{definition}[\textbf{Hausdorff distance}]
Given $(X,d_X) \in \mathcal{M}$ and $C,D \subseteq X $, let $C^\eps = \{ x \in X~|~\min_{c \in C}d_X(x,c) \leq \eps \} $. Then, Hausdorff distance between $C$ and $D$ is defined as 
$$\dha^X(C,D) := \inf\{\epsilon>0~|~\,C\subset D^\epsilon\,\mbox{and}\, D\subset C^\epsilon\}.$$
\end{definition}

It follows from the definition that for a finite metric space $(X,d_X)$, the Hausdorff distance between a pair of its subsets is computable in time polynomial in the size of subsets. For $(X,d_X), (Y,d_Y) \in \mathcal{M}$, we have \cite[Theorem 7.3.25]{burago-book} 
$$ \dgh(X,Y) = \inf \{ \dha^Z(i(X), j(Y))~|~(Z,d_Z) \in \mathcal{M}~\mbox{and}~i:X \hookrightarrow Z, j:Y \hookrightarrow Z~\mathrm{are~ isometries} \}. $$

There is another definition of Gromov-Hausdorff distance using metric couplings between $(X,d_X), (Y,d_Y) \in \mathcal{M}$. 

\begin{definition}[\textbf{Metric Coupling}]
A metric coupling between $(X,d_X), (Y,d_Y) \in \mathcal{M}$ is a metric $d$ on the disjoint union $X \sqcup Y$, such that $d|_{X\times X} = d_{X}$ and $d|_{Y\times Y} = d_{Y}$. For $X, Y \in \mathcal{M}$, let $\mathcal{D}(d_X,d_Y)$ denote the set of all metric couplings between $X$ and $Y$.
\end{definition}

Given any correspondence $R\in\mathcal{R}(X,Y)$, we can define $d_R:X\sqcup Y \times X\sqcup Y\rightarrow \mathbb{R}_+$ by: $d_R=d_X$ on $X\times X$, $d = d_Y$ on $Y\times Y$, and for $x\in X, \,y\in Y$,
\begin{equation}\label{eq:dR-def}d_R(x,y):=\inf_{(x',y')\in R}\bigg(d_X(x,x')+d_Y(y,y')+\frac{\dis(R)}{2}\bigg).\end{equation}

It is easy to check that $d_{R}$ is a valid metric on $X \sqcup Y $ \cite[Theorem $7.3.25$]{burago-book}, and hence $d_R\in\mathcal{D}(d_X,d_Y)$.

A metric coupling $d \in \mathcal{D}(d_X,d_Y) $ is called \emph{optimal} if $\dgh(X,Y) = \dha^{(X \sqcup Y,d)}(X,Y) $. 

\begin{lemma}\cite[Proposition $2.1$]{Memoli2011} \label{lem:haus}
Let $(Z,d_Z)$ be a compact metric space. Then, the Hausdorff distance between any two subsets $A,B \subset Z$ can be expressed as
$$ \dha^Z(A,B) = \inf_{S \in \mathcal{R}(A,B)} \sup_{(a,b) \in S}d_Z(a,b). $$
\end{lemma}

\begin{corollary}\label{coro-dh-dR}
For any $X,Y\in\mathcal{M}$ and any $R\in\mathcal{R}(X,Y),$ we have that 
$$\dha^{(X\sqcup Y,d_R)}(X,Y)\leq \frac{\dis(R)}{2}.$$
\end{corollary}

\begin{proof}
By Lemma \ref{lem:haus}, we have that 
\begin{equation}\label{eq:dgh-bound} \dha^{(X \sqcup Y,d_{R})}(X,Y) = \inf_{S \in \mathcal{S}(X,Y)} \sup_{(x,y) \in S}d_{R}(x,y) \leq \sup_{(x,y) \in R}d_{R}(x,y). \end{equation}
For every $(x,y) \in R$, for $(x',y') = (x,y)$, we obtain $d_X(x,x') + d_Y(y,y') + \frac{\dis(R)}{2} = \frac{\dis(R)}{2}$, and since clearly $d_{R}(x,y)\geq \frac{\dis(R)}{2}$ for all $(x,y)\in X\times Y$, this means that $d_{R}(x,y)=\frac{\dis(R)}{2}$ for $(x,y)\in R$. 
\end{proof}

\begin{lemma}
For all $(X,d_X), (Y,d_Y) \in \mathcal{M}$, there exists an optimal metric coupling $d \in \mathcal{D}(d_X,d_Y)$.
\end{lemma}

\begin{proof}
Let $R^\ast \in \mathcal{R}(X,Y)$ be a closed optimal correspondence, and let $\dis(R^\ast) = 2r$. Then, since $R^\ast$ is a closed optimal correspondence between $(X,d_X)$ and $(Y,d_Y)$, we have that $\dgh(X,Y) = r $. We now consider $d_{R^\ast} \in \mathcal{D}(d_X,d_Y)$ given by (\ref{eq:dR-def}).

 Now, we show that $\dgh(X,Y) = \dha^{(X \sqcup Y,d_{R^\ast})}(X,Y)$. By definition of the Gromov-Hausdorff distance, we have that
$\dha^{(X \sqcup Y,d_{R^\ast})}(X,Y) \geq \dgh(X,Y)=r.$ By Corollary \ref{coro-dh-dR}, we have that
$\dha^{(X\sqcup Y,d_{R^\ast})}(X,Y)\leq r.$

Thus, we have shown that $\dha^{(X \sqcup Y,d_{R^\ast} )}(X,Y) = r =  \dgh(X,Y)$. We note that since $R^\ast$ is a closed subset of $X \times Y$,  the infimum in the definition of the metric $d_{R^\ast}$ is achieved. We conclude that the metric $d_{R^\ast}$ defined above is an optimal metric coupling.
\end{proof}

We now define the Voronoi partition of a metric space with respect to one of its finite subsets.

\begin{definition}[\bf Voronoi partition]
Given a metric space $(X,d_X) $, $k \in \mathbf{N}$ and an ordered finite subset of $L = \{l_1, \ldots, l_k \} \subseteq X $, the Voronoi partition $\{B_1,\ldots,B_k\}$ of $X$ with respect to $L$ is  defined as follows: 
\begin{itemize} 
\item First, for every $i \in [k]$ first consider the subset
$$ B'_{i} = \{ x \in X~|~d_X(x,l_i) \leq d_X(x,l_j)~ \forall~ j \in [k], j \neq i  \}. $$
\item  Since te sets $B_1',B_2',\ldots,B_k'$ so formed might have non-empty intersections, we disjointify them as follows. We define 
$$ B_1 = B'_1, B_2 = B'_2 \setminus B'_1, \ldots, B_k = B'_k \setminus (\cup_{i=1}^{k-1}B'_i). $$
\end{itemize}

We observe that a reordering of the elements of $L$ might result in a different Voronoi partition of $X$. 

\end{definition}

For each $(X,d_X) \in \mathcal{M}$ and $k \in \mathbf{N}$, we now define the \emph{Voronoi map} and the \emph{Hausdorff map}. The Voronoi map associated to $X$ returns a partition of $X$ into $k$ clusters, whereas the Hausdorff map associated to $X$ acts on a partition of $X$ into $k$ clusters and returns a $k$-point metric space. The definition of the Voronoi map will make use of the existence of optimal metric couplings.

\begin{definition}[\textbf{Voronoi map}] \label{def:Voronoi_map}
Let $(X,d_X) \in \mathcal{M}$ and $k \in \mathbf{N}$. Given $(M_k,d_k) \in \mathcal{M}_k $ with $M_k = \{1,2, \ldots, k \} $, we define
$$ \mathcal{V}_{X,k,M_k}: \mathcal{D}(d_X,d_k) \rightarrow \partk(X) $$
as follows: given $d \in \mathcal{D}(d_X,d_k)$, let $\{B_i \}_{i=1}^k $ denote the Voronoi partition of $(X \sqcup M_k,d)$ with respect to $M_k$. Then, we define
$$ \mathcal{V}_{X,k,M_k}(d) := \{B_{i} \setminus \{i \},\,i\in[k] \}.$$
\end{definition}

\begin{definition}[\textbf{Hausdorff Map}] \label{def:Hausdorff-map}
Let $(X,d_X) \in \mathcal{M}$ and $k \in \mathbf{N}$. We define a map 
$$\mathcal{H}_{X,k}: \partk(X) \rightarrow \mathcal{M}_{k}$$ as follows: for any $P = \{B_{i}\}_{i=1}^{k} \in \partk(X)$, we define a metric $d_{k}$ on $M_{k}:=\{1,2,\ldots,k\}$ by $ d_{k}(i,j) = \dha^{X}(B_{i}, B_{j})$ for $i,j\in\{1,2,\ldots,k\}$. We now define
$$ \mathcal{H}_{X,k}(P) := (M_{k},d_{k}).$$
\end{definition}

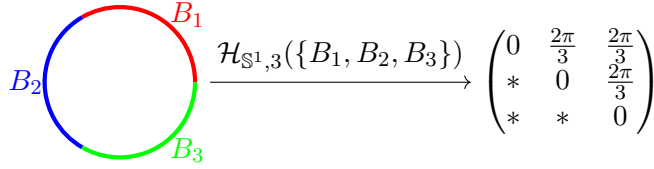
\begin{figure}
\centering
\begin{tikzpicture}
\draw [ultra thick,red] (1,0) arc (0:120:1cm);
\draw [ultra thick, blue](-1/2,1.732/2) arc (120:240:1cm);
\draw [ultra thick, green](-1/2,-1.732/2) arc (240:360:1cm);
\node [red] at (0.9,0.9) {$B_1$};
\node[blue] at (-1.25,0) {$B_2$};
\node[green] at (0.9,-0.9) {$B_3$};
\node(matrix) at (6,0) {$\begin{pmatrix}
0 & \frac{2 \pi}{3} & \frac{2 \pi}{3} \\
* & 0 & \frac{2 \pi}{3} \\
* & *& 0
\end{pmatrix} $};
\path[->] (1.2,0) edge node[above]{$\mathcal{H}_{\mathbb{S}^1,3}(\{B_1,B_2,B_3 \}) $} (matrix);
\end{tikzpicture}
\caption{The Hausdorff map applied to clusters $\{B_1,B_2,B_3 \} \in \mathrm{Part}_3(\mathbb{S}^1)$ outputs the $3$-point metric space on the right.} \label{fig:hausdorff_map_example}
\end{figure}
An example illustrating the Hausdorff map is depicted in Figure \ref{fig:hausdorff_map_example}. 

In the next section, we use these two maps to establish the strict duality between $\sketch_k(X)$ and $\shatter_k(X)$.

\subsection{The strict duality between $\sketch_{k}(X)$ and $\shatter_{k}(X)$} \label{sec:strict-duality-sketch-shatter}

In this section, we show that $\sketch_k(X)$ and $\shatter_k(X)$ are strictly dual to each other. In addition, we provide a scheme for obtaining an optimal sketch of a metric space from an optimal clustering of that metric space and vice-versa.

The strict duality between $\sketch_k(X)$ and $\shatter_k(X)$ is established by virtue of Lemma \ref{lem:Voronoi-map} and Lemma \ref{lem:Hausdorff-map}. The first lemma states that given a $k$-point metric space $(M_k,d_k)$, the cost of clustering $(X,d_X)$ according to the partition obtained from applying the Voronoi map, is at most twice the Gromov-Hausdorff distance between $X$ and $M_k$. This lemma is formally stated as follows.

\begin{lemma} \label{lem:Voronoi-map}
For every $(X,d_{X}) \in \mathcal{M}$, $k \in \mathbf{N}$, $(M_{k}, d_{k}) \in \mathcal{M}_{k}$ with $M_k = \{1,2,\ldots,k \}$ and $d \in \mathcal{D}(d_X,d_k)$, we have
$$ \Phi_{\mathcal{M}}(X, \mathcal{V}_{X,k,M_k}(d)) \leq \dha^{(X\sqcup M_k,d)}(X,M_k).$$
\end{lemma}

\begin{proof}
Given $(X, d_{X}) \in \mathcal{M}$, $k \in \mathbf{N}$, $(M_{k}, d_{k}) \in \mathcal{M}_{k}$ with $M_k = \{1,2,\ldots,k \}$ and $d \in \mathcal{D}(d_X,d_k)$, let $\dha^{(X \sqcup M_k,d)}(X,M_k) < \eta $. This implies that for every $x \in X$, there exists $i \in M_k $ such that $d(x,i) \leq \eta $.

We now consider the partition $P = \mathcal{V}_{X,k,M_k}(d) = \{ B_{1}, \ldots, B_{k} \}$. For every $i \in M_k$, we have 
\begin{align*}
\mathrm{diam}(B_{i}) &= \max_{x,x' \in B_{i}} d_{X}(x,x') \\
&\leq \max_{x,x' \in B_{i}} (d(x,i) + d(x',i))\\
&\leq 2 \eta.
\end{align*}
The first inequality is due to triangle inequality on $d$, and the second inequality is due to the fact that $\{B_i \cup \{i\} \}_{i=1}^k $ is a Voronoi partition of $(X \sqcup M_k,d)$ with respect to $M_k$.
Since the above inequalities hold for every $i \in M_k$, we have that $\max_{i} \mathrm{diam}(B_{i}) \leq 2 \eta$. Thus, we have shown that $\Phi_{\mathcal{M}} (X, \mathcal{V}_{X,k,M_k}(d)) \leq 2 \eta $. Hence, we conclude that $\Phi_{\mathcal{M}} (X, \mathcal{V}_{X,k,M_k}(d)) \leq 2 \cdot \dha^{(X \sqcup M_k,d)} (X, M_{k})$.
\end{proof}

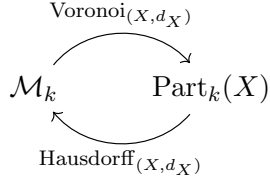
\begin{figure}
\centering
\begin{tikzcd}
\mathcal{M}_k \arrow[r, bend left = 50, "\mathrm{Voronoi}_{(X,d_X)}"]  
& \partk(X) \arrow[l, bend left = 50,  "\mathrm{Hausdorff}_{(X,d_X)}"] 
\end{tikzcd}
\caption{An illustration of the Voronoi and Hausdorff maps associated to $(X,d_X)$.}
\end{figure}

The next lemma states that given a partition $P$ of $(X,d_X)$, the cost of clustering $X$ according to $P$ is at least the Gromov-Hausdorff distance between $X$ and the $k$-point metric space obtained from applying the Hausdorff map to $P$.

\begin{lemma} \label{lem:Hausdorff-map}
For every $(X,d_{X}) \in \mathcal{M}$, for every $k \in \mathbf{N}$ and for every $P \in \mathrm{Part}_{k}(X)$, we have
$$ 2 \cdot \dgh(X, \mathcal{H}_{X,k}(P)) \leq \Phi_{\mathcal{M}} (X,P). $$
\end{lemma}

\begin{proof}
Given $(X,d_{X}) \in \mathcal{M}$, $k \in \mathbf{N}$ and $P = \{B_{i} \}_{i=1}^{k} \in \mathrm{Part}_{k}(X)$, let $\eta = \Phi_{\mathcal{M}} (X,P) = \max_{i \in [k]} \diam(B_{i})$. We define a metric $d_{k}$ on $M_{k}$ as follows: for $i,j \in M_k$,
$$ d_{k}(i,j) = \dha^{X}(B_{i}, B_{j}), $$
where $\dha^{X}(B_{i}, B_{j})$ is the Hausdorff distance in $X$ between sets $B_{i}$ and $B_{j}$. We define a map $\phi: X \rightarrow M_{k}$ as $\phi(x) = i$, where $i \in [k] $ is such that $x \in B_{i}$. This map is surjective and its graph $R(\phi) := \{(x, \phi(x)) \mid x \in X \}$ is a correspondence between $(X,d_{X})$ and $(M_{k}, d_{k})$. This is because every $x \in X$ belongs to some cluster $B_{i}, i \in [k] $ and all clusters are non-empty.

For $(x,i), (y,j) \in R(\phi)$ with $i = j$, we have
$$\big| d_{X}(x,y) - d_{k}(i,j) \big| = d_{X}(x,y) \leq \mathrm{diam}(B_{i}) \leq \eta.$$
For $ (x,i), (y,j) \in R(\phi) $ with $i \neq j$, we have  
$$\big| d_{X}(x,y) - d_{k}(i,j) \big| = \big| d_{X}(x,y) - \dha^{X}(B_{i}, B_{j}) \big|.$$

For $1 \leq i,j \leq k$, we define 
$$d_{X}(B_{i}, B_{j}) := \min_{x \in B_{i}} \min_{x' \in B_{j}} d_{X}(x,x').$$ Let $v_{i} \in B_{i}$ be such that $d_{X}(v_{i}, B_{j})$ is the smallest and $v_{j} \in B_{j}$ be such that $d_{X}(v_{j}, B_{i})$ is the smallest. Then, we have $d_{X}(v_{i}, B_{j}) = d_{X}(v_{j}, B_{i}) = d_{X}(B_{i}, B_{j})$. 

Let $u_{i} \in B_{i}$ be such that $d_{X}(u_{i}, B_{j})$ is the largest and $u_{j} \in B_{j}$ be such that $d_{X}(u_{j}, B_{i})$ is the largest. Then, we observe that
$$ \dha^{X}(B_{i}, B_{j}) = \mathrm{max} \{ d_{X}(u_{i}, B_{j}), d_{X}(u_{j}, B_{i}) \}. $$
By triangle inequality, we have
$$d_{X}(u_{i}, B_{j}) \leq d_{X}(u_{i}, v_{i}) + d_{X}(v_{i}, B_{j}) \leq \mathrm{diam}(B_{i}) + d_{X}(B_{i}, B_{j}) \leq \eta + d_{X}(B_{i}, B_{j}).$$
Similarly, $d_{X}(u_{j}, B_{i}) \leq \eta + d_{X}(B_{i}, B_{j}) $. This implies that $\dha^{X}(B_{i}, B_{j}) \leq \eta + d_{X}(B_{i}, B_{j})$. Since $ \dha^{X}(B_{i}, B_{j}) \geq d_{X}(B_{i}, B_{j})$, we have that for $ (x,i), (y,j) \in R(\phi) $ with $i \neq j$,
$$ \big| \dha^{X}(B_{i}, B_{j}) - d_{X}(x,y) \big| \leq \big| \eta + d_{X}(B_{i}, B_{j}) - d_{X}(x,y) \big| \leq \eta. $$ 
The second inequality holds because $$d_{X}(B_{i}, B_{j}) \leq d_{X}(x,y) \leq \eta + \dha^{X}(B_i,B_j).$$
The leftmost inequality holds because
$$d_{X}(x,y)\leq d_{X}(x,B_{j})+\mathrm{diam}(B_{j}) \leq d_{X}(u_{i}, B_{j}) + \eta \leq \dha^{X}(B_{i}, B_{j}) + \eta.$$

Thus, for the correspondence $R(\phi)$, we obtain $\dis(R(\phi)) \leq \eta$. This shows that $\dgh(X,M_{k}) \leq \frac{\eta}{2}$. Since $\mathcal{H}_{X,k}(P) = (M_{k}, d_{k})$, we have that $2 \cdot \dgh(X, \mathcal{H}_{X,k}(P)) \leq \Phi_{\mathcal{M}} (X,P).$
\end{proof}

We are now ready to prove Theorem \ref{thm:main1}.

\begin{proof}[Proof of Theorem \ref{thm:main1}]
Fix $(X, d_{X}) \in \mathcal{M}$ and $k \in \mathbf{N}$. Suppose $\sketch_{k}(X) < \eta$. Then, there exists $(M_{k}, d_{k}) \in \mathcal{M}_{k}, M_k = \{1,2,\ldots,k \}$ and $d \in \mathcal{D}(d_X,d_k) $ such that $\dha^{(X \sqcup Y,d)} (X, M_{k}) < \eta $. From Lemma \ref{lem:Voronoi-map}, we obtain that
$$\Phi_{\mathcal{M}} (X, \mathcal{V}_{X,k,M_k}( d)) \leq 2 \cdot \dha^{(X \sqcup Y,d)} (X, M_{k}) < 2 \eta.$$
Since $\mathcal{V}_{X,k,M_k} \left( d \right) \in \mathrm{Part}_{k}(X)$ and $\shatter_{k}(X) = \inf_{P \in \mathrm{Part}_{k}(X)} \Phi_{\mathcal{M}} (X,P)$, we obtain $\shatter_{k}(X) < 2 \eta.$ Thus, we have shown that whenever $\sketch_{k}(X) < \eta$, we have $\shatter_{k}(X) < 2 \eta$. We conclude that $\shatter_{k}(X) \leq 2 \cdot \sketch_{k}(X)$. 

We now prove the opposite direction of the last inequality. Suppose $\shatter_{k}(X) < \eta$. Then, there exists $P \in \mathrm{Part}_{k}(X)$ such that $\Phi_{\mathcal{M}} (X,P) < \eta$. From Lemma \ref{lem:Hausdorff-map}, we have that
$$ 2 \cdot \dgh(X, \mathcal{H}_{X,k}(P)) \leq \Phi_{\mathcal{M}} (X,P) < \eta. $$
Since $\mathcal{H}_{X,k}(P) \in \mathcal{M}_{k}$ and $\sketch_{k}(X) = \inf_{(M_{k},d_{k}) \in \mathcal{M}_{k}} \dgh(X,M_{k})$, we have that $2 \cdot \sketch_{k}(X) < \eta$. Thus, we have shown that whenever $\shatter_{k}(X) < \eta$, we have $2 \cdot \sketch_{k}(X) < \eta$. We conclude that $2 \cdot \sketch_k(X) \leq \shatter_k(X)$.
\end{proof}

The following corollary provides a procedure for converting optimal sketches into optimal clusterings, and vice-versa.
\begin{corollary}[From optimal sketch to optimal shatter] \label{cor:strong-duality}
We have that
\begin{itemize}
\item If $(M_{k}, d_{k}) \in \mathcal{M}_{k}, M_k = \{1,2,\ldots,k \}$ is s.t. $\sketch_{k}(X) = \dgh(X, M_{k})$, then for any optimal $d \in \mathcal{D}(d_X,d_k)$, we obtain $\shatter_{k}(X) = \Phi_{\mathcal{M}} (X,\mathcal{V}_{X,k,M_k}( d ))$.
\item If $P = \{B_{i} \}_{i=1}^{k} \in \mathrm{Part}_{k}(X)$ is s.t. $\shatter_{k}(X) = \Phi_{\mathcal{M}} (X,P)$, then $\sketch_{k}(X) = \dgh(X, \mathcal{H}_{X,k}(P))$.
\end{itemize}
\end{corollary}

\begin{Remark} \label{rem:strict-duality}
Note that Corollary \ref{cor:strong-duality} along with Lemmas \ref{lem:Voronoi-map} and \ref{lem:Hausdorff-map} tells us that for any $(X,d_X) \in \mathcal{M}$ and $k \in \mathbf{N}$, given an optimal $k$-sketch $(M_k,d_k)$ of $X$ and an optimal metric coupling $d \in \mathcal{D}(d_X,d_k)$, an optimal clustering of $X$ is computable via the Voronoi map. On the other hand, given an optimal clustering $P = \{B_i \}_{i=1}^d \in \partk(X)$, an optimal $k$-sketch of $X$ is obtained by computing the Hausdorff distance between sets $B_i$, $i \in [k]$. Thus, the procedures mentioned in the introduction, for transforming an optimal clustering of $(X,d_X) \in \mathcal{M} $ to an optimal $k$-sketch of $X$, and vice-versa, are provided by the Hausdorff and the Voronoi maps respectively.  
\end{Remark}

\subsection{The computation of $\sketch_{k}(X)$ is NP-hard}  \label{sec:sketch-np-hard}

 We show that computing $\sketch_k(X)$ is NP-hard\footnote{See \cite{Joh90} for background material on computational hardness concepts.}  by reducing, in polynomial time, an NP-hard problem to the problem of computing $\shatter_k(X)$.  The NP-Hard problem which we reduce the computation of  $\shatter_k(X)$ is the set cover decision problem \cite{Cor09}. Since $\sketch_{k}(X) = \frac{1}{2} \cdot \shatter_{k}(X)$, we obtain that the computation of $\sketch_{k}(X)$ is NP-hard. We start by defining the set cover decision problem.

\begin{definition}[\textbf{Set Cover Decision Problem}]
Given a universe of elements $U = \{u_{1}, \ldots, u_{n} \}$, a collection $S = \{S_{1}, \ldots, S_{m}\} $ of subsets of $U$ such that $\cup_{i=1}^{m} S_{i} = U$ and $k \in \mathbf{N}$, the set cover decision problem asks if there is a collection $\{S_{i_{1}}, \ldots, S_{i_{l}}\} \subseteq S$ such that $\cup_{j=1}^{l} S_{i_{j}} = U$ and $l \leq k$.   
\end{definition}

We now describe the reduction of an instance of the set cover decision problem to an instance of computing $\shatter_k(X)$. Given an instance of a set cover decision problem with input $U,S, |U|=n, |S|=m$ and $k \in \mathbf{N}$, we construct a graph $G$ as follows: for every $i \in [n] $, we add a vertex $u_i$ to $G$, and for every $i \in [m]$, we add a vertex $S_i$ to $G$. We add two more vertices $r$ and $r'$ to $G$. We now add edges between the vertices of $G$ in the following manner. We add an edge of length $1$ joining $r$ and $r'$. For every $i \in [m]$, we add an edge of length $1$ from $r$ to every $S_{i}$. For every $i \in [n]$ and $j \in [m]$, if $u_{i} \in S_{j}$, we add an edge of length $1$ between vertices $u_{i}$ and $S_{j}$ in $G$. Let $V(G)$ denote the set of vertices of $G$ and $E(G)$ denote the set of edges of $G$. We now have the following theorem.

\begin{theorem} \label{thm:sketch-np-hard} 
Given $k \in \mathbb{N}$, there is a set cover of $U$ of size $k$ if and only if there exists a partition of $G$ into $k+1$ blocks such that the diameter of every block is at most $2$. 
\end{theorem}

\begin{proof}
Fix $k \in \mathbb{N}$. Suppose there is a set cover of $U$ of size $k$. Let $S' = \{S_{i_{1}}, \ldots, S_{i_{k}} \}$ be such a set cover. Then $\cup_{j=1}^{k} S_{i_{j}} = U$. We construct a partition of $G$ as follows: for $j \in [k]$ and $S_{i_{j}} \in S'$, let $B_{j} = S_{i_{j}} \cup \{u \in U ~|~ u \in S_{i_{j}} \}$. Let $B_{k+1} = V(G) \setminus \cup_{j=1}^{k} B_{j}$. It is easy to check that the for every $j \in [k+1]$, $\diam(B_{j}) \leq 2$. We observe that $\shatter_{k+1}(G) = 2$. Thus, the partition $\{B_{1}, \ldots, B_{k+1} \}$ is an optimal partition.

Conversely, suppose that there exists a partition $P$ of $G$ into $k+1$ blocks such that the diameter of every block is at most $2$. We may assume that all blocks are connected. 

We first show how to modify this partition $P$ so as to ensure that no block of $P$ is a singleton containing some $u \in U$.

\begin{claim} \label{clm:partition}
There is no block in partition $P$ that is a singleton containing some $u \in U$.
\end{claim}
\begin{proof}
Suppose $P$ has a block $B$ such that $B = \{u \}$ for some $u \in U$. Then we consider any set $S_{i}$ such that $u \in S_{i}$. Let $B'$ be the block such that $S_{i} \in B'$. There are five cases:
\begin{enumerate}
\item $B' = S_{i}$. In this case, we merge $B$ and $B'$ into one block. 
\item $B'$ contains $u' \in U$, $u' \neq u$, but $B'$ neither contains $r$ nor any other $S_{j}$, $j \neq i$. In this case, we merge $B'$ and $B$ into a single block.
\item $B'$ contains both $r$ and $r'$. In this case, if $B'$ contains any $S_{j} \neq S_{i}$, then we remove $S_{i}$ from $B'$ and add $S_{i}$ to $B$. If $B'$ does not contain any other $S_{j}$, $j \neq i$ then we remove $S_i$ from $B'$ and add $S_i$ to $B$.
\item $B'$ does not contain $r$ but contains some $S_{j} \neq S_{i}$. This happens if there is a $u' \in U$ such that $u' \in B'$ and $u' \in S_{j} \cap S_{i}$. In this case, we remove $S_{i}$ from $B'$ and add it to $B$. 
\item $B'$ contains $r$ but not $r'$. This means that $r'$ is in a block by itself. We remove $r$ from $B$ and add it to the block of $r'$. Now we are in one of the above cases.
\end{enumerate}
We observe that in all the above cases, we do not change the maximum diameter of the partition. Therefore, we may assume that there is no block in $P$ that is a singleton containing some $u \in U$.
\end{proof}

We now show how to modify the partition $P$ such that vertices $r$ and $r'$ belong to the same block.

\begin{claim}
Vertices $r$ and $r'$ belong to the same block of partition $P$.
\end{claim}

\begin{proof}
Let $B_{r'}$ be the block containing the vertex $r'$. Suppose $B_{r'}$ does not contain $r$. Since the blocks are connected, we have $B_{r'} = \{r' \}$. Let $B_r$ be the block containing the vertex $r$. If $B_{r} = \{ r \}$, we merge $B_r$ and $B_{r'}$ to a single block. If $B_{r}$ contains some $u \in U$, then it also contains some $S_{i} \in S$. In this case, we remove $r$ from $B_{r}$ and add $r$ to $B_{r'}$. Note that removing $r$ from $B_{r}$ does not disconnect $B_{r}$. If $B_{r}$ does not contains any $u \in U$ but contains some $S_{i} \in S$, then we merge $B_{r}$ and $B_{r'}$ to a single cluster. We observe that any of the above steps do not change the maximum diameter of partition $P$.
\end{proof}
We now show that the $k$ blocks of $P$ obtained by excluding the block containing vertices $r$ and $r'$ form a set cover of $U$ size $k$. We observe that the block containing $r$ does not contain any $u \in U$ because this block also contains $r'$ and the diameter of the partition is at most $2$. Therefore, all $u \in U$ are covered by the remaining $k$ blocks. Since the blocks are connected and there is no block that is a singleton containing some $u \in U$, we have that every block contains some $S_{i} \in S$. Let $L$ be the set of blocks of $P$ containing exactly one $S_{i} \in S$ and $M$ be the set of blocks of $P$ containing more than one $S_{i}$. We observe that for every block $B \in M$, every $u \in U \cap B$ is contained in every $S_{i} \in B$. We define a set cover $H$ as follows: for every block $B \in L$, add the unique $S_{i} \in B$ to $H$. For every block $B \in M$, pick some $S_{i} \in B$ arbitrarily and add to $H$. The set $H$ contains $k$ elements of $S$ that cover $U$. Thus, we obtain a set cover of $U$ of size $k$.
\end{proof}

\subsection{A polynomial time approximation of $\sketch_k(X)$} \label{sec:poly-sketch}
In the previous section, we proved that computing $\sketch_k(X)$ is NP-Hard.  In this section, we present a polynomial time algorithm that computes a $2$-approximation of $\sketch_k(X)$. Furthermore, we establish that the $2$ factor is optimal.

\label{sec:approx-sketch}
\begin{theorem}
Given $X \in \mathcal{M}$ and $k \in \mathbf{N}$, there is a polynomial time algorithm that outputs a $2$-approximation of $\sketch_k(X)$. Furthermore, for any $\eps > 0$, obtaining a $(2 - \eps)$-approximation of $\sketch_k(X)$ is NP-Hard.
\end{theorem}

\begin{proof}
Computation of $\shatter_k(X)$ is same as solving the problem of clustering a metric space to minimize the maximum inter-cluster distance. The latter problem was first studied by Gonzales \cite{GONZ85}, and he obtained a $2$-approximation algorithm for this problem. Additionally, he showed that for any $\eps > 0$, obtaining a $(2- \eps)$-approximation for this problem is NP-Hard. This implies the existence of a $2$-approximation algorithm for computing $\shatter_k(X)$, as well as that obtaining a $(2 - \eps)$-approximation of $\shatter_k(X)$ is NP-Hard, for any $\eps > 0$. Let $P \in \partk(X)$ be the partition of $X$ obtained from the $2$-approximation algorithm of Gonzales \cite{GONZ85}. We apply the Hausdorff map (Definition \ref{def:Hausdorff-map}) to $P$ to obtain a $k$-point metric space, say $M_k$. From Lemma \ref{lem:Hausdorff-map}, we obtain that
$$ 2 \cdot \dgh(X,M_k) \leq \max_{\{B_i \}_{i=1}^k \in P} \diam(B_i) \leq 2 \cdot \shatter_k(X) = 4 \cdot \sketch_k(X). $$
Here, the last equality is due to the strong duality theorem (Theorem \ref{thm:main1}). Since $\sketch_k(X) \leq \dgh(X,M_k) $, we obtain that $M_k$ is a $2$-approximation of $\sketch_k(X)$. We note that computation of $M_k$ is polynomial time, since the computation involves calculating Hausdorff distance between blocks of partition $P$, and this can be done in polynomial time. 

We now prove that for any $\eps > 0$, obtaining a $(2 - \eps)$-approximation of $\sketch_k(X)$ is NP-Hard. We recall that for $\lambda > 0$, a $\lambda$-approximation of $\sketch_k(X)$ is a pair $(M_k,R)$, where $M_k$ is a $k$-point metric space and $R$ is a correspondence between $X$ and $M_k$ satisfying $\frac{\dis(R)}{2} \leq \lambda \cdot \sketch_k(X)$. We assume that there exists a polynomial time algorithm that outputs an $(M_k,d_k)$, $M_k = \{1,2, \ldots,k \}$, and $R \in \mathcal{R}(X,M_k) $ that is a $(2 - \eps)$-approximation of $\sketch_k(X)$, for some $\eps > 0$. Clearly, we have $\dis(R) \leq 2\,(2 -\eps) \cdot \sketch_k(X)$. By applying the Voronoi map (Definition \ref{def:Voronoi_map}) to $d_R \in \mathcal{D}(d_X,d_k)$, defined in (\ref{eq:dR-def}), we obtain a partition of $X$, say $\{B_i \}_{i=1}^k $ that satisfies 
\begin{align*}
\shatter_k(X) &\leq \max_i \diam(B_i) \leq 2 \cdot \dha^{(X \sqcup M_k,d_R)}(X,M_k) \leq \dis(R) \\
&\leq 2 \,(2 - \eps) \cdot \sketch_k(X) = (2 - \eps) \cdot \shatter_k(X).
\end{align*}
Here, the first inequality follows from the definition of $\shatter_k(X)$, the second is from Lemma \ref{lem:Voronoi-map}, the third is from Corollary \ref{coro-dh-dR}  and the equality in the end is due to the strict duality theorem (Theorem \ref{thm:main1}). Thus, we have a polynomial time algorithm that computes a $(2 - \eps)$-approximation of $\shatter_k(X)$. This contradicts the fact that obtaining a $(2 - \eps)$-approximation of $\shatter_k(X)$ is NP-Hard for all $\eps > 0$.
\end{proof}

\begin{Remark}
It is known that computing Gromov-Hausdorff distance between finite metric spaces leads to solving an instance of the Quadratic Assignment Problem, which is NP-Hard. In addition, computing any $(1+\epsilon)$-approximation is also NP-Hard, for all $\epsilon > 0$. However, the strong duality (Theorem \ref{thm:main1}) for metric spaces proves that there is a simple $2$-approximation algorithm for computing the Gromov-Hausdorff distance between a finite metric space $X$ and the $k$-point metric space closest to $X$ for $k < |X|$. 
\end{Remark}

\subsection{Theoretical guarantees for Farthest Point Sampling}
Farthest point sampling (FPS) method was first used by Gonzales \cite{GONZ85} to determine a $k$ clustering of a point cloud, $k \in \mathbb{N}$, that minimizes the maximum cluster diameter. The method did not produce an optimal clustering, but produced the best approximation to optimal clustering. It was later proved that finding a better approximation to optimal clustering is NP-Hard. Since then, FPS has been used in many applications such as computing approximate geodesic distances \cite{ABSW07, GM09}, for shape recognition \cite{EK03,MS04}, for efficiently computing point-to-point correspondence between surfaces \cite{BBK06,WSAB07}, and for subsampling finite metric spaces as a preprocessing step in computational topology \cite{javaplex} (where it is known as the maxmin algorithm). In all these works, it has been shown experimentally that FPS is a good heuristic for isometry-invariant surface processing tasks, but none of these works establish any theoretical guarantees behind using FPS. 

In \cite{KLMW16}, the stretch factor of approximate geodesics computed using FPS is defined as the maximum over all pairs of distinct vertices, of their approximated distance over their geodesic distance in the graph. The authors obtain a bound on this stretch factor in terms of the minimal stretch factor, and therefore provide a theoretical explanation for why FPS is a good heuristic for surface processing tasks. However, the bound obtained on the stretch factor depends on the intrinsic properties of the input graph, i.e.~the ratio of the lengths of the longest and shortest edges of the graph, and this number can be quite large. In addition, FPS has been analyzed for a specific application of approximating geodesics on a graph, with a fixed definition of approximation. 

In this work, we show that a $k$-point metric space $M_k$ that is closest to a point cloud $X$ satisfies 
$$\dgh(X,M_k) = \frac{1}{2} \min_{\{B_i \}_{i=1}^k \in \partk(X)} \max_i \diam(B_i) = \shatter_k(X) . $$
The objective of minimizing the maximum cluster diameter was studied by Gonzales \cite{GONZ85}, and he obtained a $2$-approximation for this objective using a Voronoi partition associated to the points sampled via FPS. He also proved that obtaining a $2 - \eps$ approximation is NP-Hard for any $\eps > 0$. Thus, the Voronoi partition of $X$ associated to its sample obtained via FPS provides the best possible approximation for $\shatter_k(X)$. In this work, we prove that if $\{B_i \}_{i=1}^k$ is a Voronoi partition of $X$ associated to its sample obtained via FPS, then $M_k$ with $d_k(i,j) = \dha^X(B_i, B_j)$ provides the best approximation to $\inf_{(M_k,d_k)} \dgh(X,M_k)$. The $k$-point approximation of $X$ produced by our algorithm is not necessarily a subset of $X$. However, we show that the set of $k$-points obtained using FPS provides a $4$-approximation to $\sketch_k(X)$. Precisely, let $F_k = \{a_1, \ldots, a_k \} $ be the set of $k$-points obtained using FPS and $\{B_i \}_{i=1}^k, a_i \in B_i$ be the partition obtained by applying the Voronoi map $\mathcal{V}_{X,k,F_k,d_X} $ on any Voronoi clustering of X with respect to $F_k$. Then we have that
\begin{align*}
\dgh(X,F_k) &\leq \dha^X(X,F_k) = \max_{1 \leq i \leq k} \max_{x \in B_i}d_X(x,a_i) \leq \max_{1 \leq i \leq k} \diam(B_i) \\
&\leq 2 \cdot \shatter_k(X) = 4 \cdot \sketch_k(X).
\end{align*}
The third inequality holds because $\{B_i \}_{i=1}^k $ provides a $2$-approximation to $\shatter_k(X)$, and the last equality is by our strong duality theorem for metric spaces (Theorem \ref{thm:main1}). Thus, we have proved that the set of $k$-points obtained using FPS provide a $4$-approximation to the $k$-point metric space closest to $X$. We remark that these are the first results providing theoretical guarantees for using FPS to sample point clouds.

\subsection{Minimizers of $M_k\mapsto \dgh(X, M_{k})$ are not always subsets} \label{sec:minimizers}

In this section, we show that for any $m \in \mathbf{N}$, there exists a metric space $(X, d_{X})$ such that minimizers of $M_{3m}\mapsto\dgh(X, M_{3m})$ are \emph{not} subsets of $X$. This is surprising since most of the sampling algorithms only sample subsets of the input metric space, while this result shows that there are metric spaces for which the closest metric space of smaller cardinality is not one of its subset. Here, being closest in terms of the Gromov-Hausdorff distance implies that this set retains crucial properties of the input metric space better than any subset of the input metric space.

In Proposition \ref{thm:tree-metric-spaces}, we first consider the case when $X$ is tree metric space, and in Proposition \ref{thm:euclidean-metric-spaces}, we deal with the case when $X$ is a metric space that can be isometrically embedded into the Euclidean plane.

\begin{definition}[\textbf{Tree metric space} \cite{SS03}]
A metric space $(X,d_X)$ is called a tree metric space if there exists a weighted tree $T = (V_T, E_T,w_T)$ and a map $\phi: X \rightarrow V_T$ such that for all $x,x' \in X$, $d_X(x,x') = d_T(\phi(x), \phi(x'))$. Here, $d_T$ is the metric on $V_T$ induced by $w_T$.
\end{definition}

\begin{figure} 
\begin{center}
\scalebox{0.5}{\includegraphics{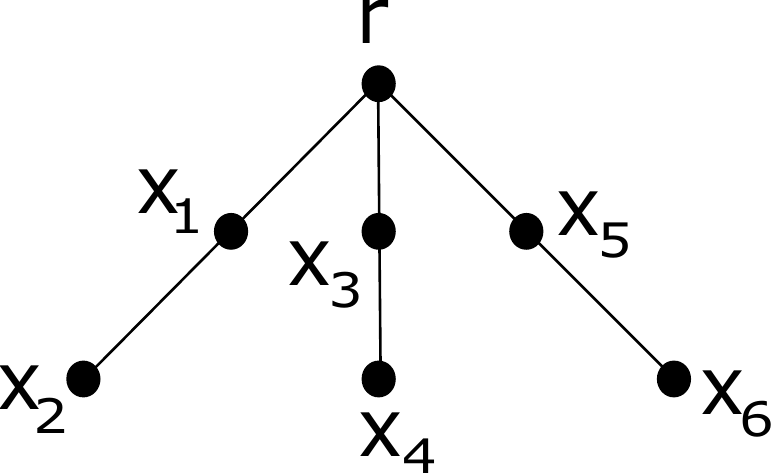}} \hspace{2cm}
\scalebox{0.5}{\includegraphics{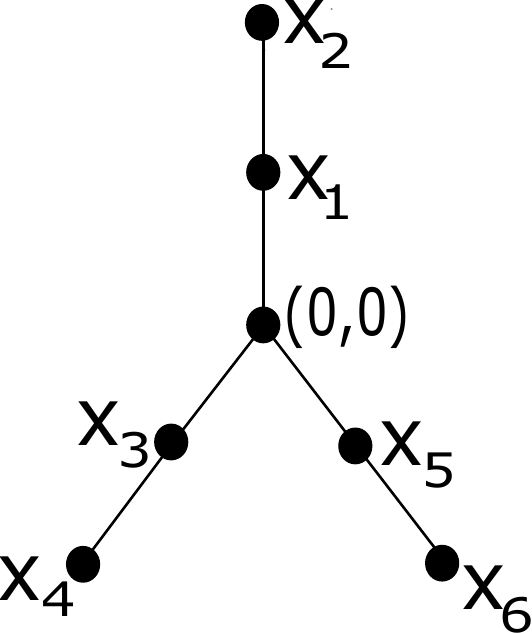}}
\caption{The tree metric space (left) and the Euclidean metric space (right) $X = \{x_1, \ldots, x_6 \}$ corresponding to $m=1$.} \label{fig:tree}
\end{center}
\end{figure}

\begin{proposition}[Tree metric spaces] \label{thm:tree-metric-spaces}
For all $m \in \mathbf{N}$ and $n=3m$, there exist tree metric spaces $(X, d_{X})$ and $(Y, d_{Y})$ such that (1) $|Y|=n$, and
 (2) for any $K \subset X$ with $|K| \leq n$, we have $\dgh(X, K) > \dgh(X, Y)$.
 \end{proposition}
 
The proof of the above proposition has been moved to the appendix. The tree metric space $X$ corresponding to $m=1$ is depicted in Figure \ref{fig:tree}. Now, we show that we have a similar behavior even for subsets of Euclidean space.

\begin{proposition}[Euclidean metric spaces] \label{thm:euclidean-metric-spaces}
For all $m \in \mathbf{N}$ and $n=3m$, there exist $X,Y \subset \mathbf{R}^{2} $ such that
(1) $|Y|=n$, and
(2) for any $K \subset X$ with $|K| \leq n$, we have $\dgh(X, K) > \dgh(X, Y)$.
\end{proposition}

The proof of the above proposition can also be found in the appendix. The euclidean metric space $X$ corresponding to $m=1$ is depicted in Figure \ref{fig:tree}. This completes our duality and computational results for metric spaces. We now start with metric measure spaces.

\section{The case of metric measure spaces} \label{sec:mm-spaces}

In this section, we first prove that for $k \in \mathbb{N}$ and $X \in \mathcal{M}_{k,w}$, $\sketch_{k,p}(X)$ is dual to $\shatter_{k,p,\infty}(X)$,  for $k = 1$ and $1 \leq p < \infty$, and for $p = \infty$ and any finite $k$. However, in \ref{sec:relationship-infty} we prove that, in general, these objectives are not dual to each other. Thus, in \ref{sec:Sturmsdefn}, we define Sturm's version of $p$-Gromov-Wasserstein distance and prove in Section \ref{sec:sketch-sturm-shatter} that for all $k \in \mathbb{N}$ and $1 \leq p \leq \infty$, $\sketch^S_{k,p}(X)$ is dual to $\shatter_{k,p,p}(X)$.

\subsection{The weak Gromov-Wasserstein distance} \label{sec:weak-GW-Defn}

In this section, we define the $p$-Gromov-Wasserstein distance between mm-spaces following \cite{Memoli2011}.

We say that two mm-spaces $X$ and $Y$ are isomorphic whenever there exists a map $\phi:X\rightarrow Y$ such that (1) $d_X(x,x') = d_Y(\phi(x),\phi(x'))$ for all $x,x'\in X$, and (2) $\mu_X(\phi^{-1}(B))=\mu_Y(B)$ for all $B$ measurable.

The analogue of correspondence for metric measure spaces is given by the notion of \emph{measure coupling}. 

\begin{definition}[\textbf{Measure coupling}]
Let $X,Y \in \mathcal{M}_{w}$. A probability measure $\mu$ on the product space $X \times Y$ is called a coupling of $\mu_{X}$ and $\mu_{Y}$ iff we have the following:
\begin{itemize}
\item For all measurable sets $A \subseteq X$, $\mu(A \times Y) = \mu_{X}(A)$ and,
\item For all measurable sets $B \subseteq Y$, $\mu(X \times B) = \mu_{Y}(B)$.
\end{itemize}
\end{definition}

For $X,Y \in \mathcal{M}_{w}$, let $\mathcal{U}(\mu_X,\mu_Y)$ denote the set of all couplings between $\mu_{X}$ and $\mu_{Y}$. For $\mu \in \mathcal{U}(\mu_X,\mu_Y)$, let $R(\mu) = \mathrm{supp}[\mu] = \{(x,y) \in X \times Y~|~\mu(x,y) \neq 0 \}$. The distortion of a measure coupling $\mu$ is  defined as follows.
 
\begin{definition}[\textbf{$p$-distortion}] 
Given $X,Y \in \mathcal{M}_{w}, p \in [1, \infty]$ and $\mu \in \mathcal{U}(X,Y)$, the $p$-distortion of $\mu$ is defined as follows:
for $1 \leq p < \infty$, we have
$$ \dis_{p}(\mu) := \left(\int_{X \times Y} \int_{X \times Y} |d_{X}(x,x') - d_{Y}(y,y')|^{p} d \mu(x,y) \, d \mu(x',y')\right)^{1/p}. $$
For $p = \infty$, we have $ \dis_{\infty}(\mu) := \sup_{x,x' \in X,~ y,y' \in Y,~(x,y),(x',y') \in R(\mu)} |d_{X}(x,x') - d_{Y}(y,y')|. $
\end{definition} 

We are now ready to define the $p$-Gromov-Wasserstein distance between metric measure spaces.

\begin{definition}[\textbf{$p$-Gromov-Wasserstein Distance}]
Given $X,Y \in \mathcal{M}_{w}$ and $1 \leq p \leq \infty$, the $p$-Gromov-Wasserstein distance between $X$ and $Y$ is defined as 
$$ \dgwp(X, Y) := \frac{1}{2} \inf_{\mu \in \mathcal{U}(\mu_X,\mu_Y)} \dis_{p}(\mu).$$
\end{definition}
It is known \cite{Memoli2011} that the $p$-Gromov-Wasserstein distance defines a proper metric on the collection of isomorphism classes of mm-spaces. The analogue of Hausdorff distance for metric measure spaces is the $p$-Wasserstein distance. 

\begin{definition}[\textbf{$p$-Wasserstein distance}, \cite{villani2003topics}]\label{def:dW}
Let $(X,d_X,\mu_X) \in \mathcal{M}_{w}$ and $A,B \subset X$ be compact subsets of $X$. Let $\mu_A, \mu_B$ be Borel probability measures with $\mathrm{supp}[\mu_A] = A$ and $\mathrm{supp}[\mu_B] = B$. The $p$-Wasserstein distance between $(A,\mu_A)$ and $(B,\mu_B)$ is defined as 
$$ \dwp^X(\mu_A,\mu_B) := \inf_{\mu \in \mathcal{U}(\mu_A, \mu_B)} \left( \int_{A \times B} d_X^p(a,b) d \mu(a,b) \right)^{1/p}, $$
for $1 \leq p < \infty$, and
$ d^X_{\mathrm{W}_\infty}(\mu_A,\mu_B) := \inf_{\mu \in \mathcal{U}(\mu_A, \mu_B)} \sup_{(a,b) \in R(\mu)} d_X(a,b). $
\end{definition}

\subsection{Relationship between $\sketch_{k,p}$ and $\shatter_{k,p,\infty}$}\label{sec:relationship-infty}

In this section, we show the relation between $\sketch_{k,p}$ and $\shatter_{k,p,\infty}$ for various values of $k \in \mathbf{N}$ and $1 \leq p \leq \infty$. It turns out that these are strictly dual to each other for certain values of $p$ and $k$:

\begin{theorem}[The case of $k=1$ and $p$ finite] \label{thm:equalitypfinite}
For all $X \in \mathcal{M}_{w}$ and $1 \leq p < \infty$, we have $$\sketch_{1,p}(X) = \frac{1}{2} \cdot \shatter_{1,p, \infty}(X).$$
\end{theorem}

\begin{proof}
We have $\sketch_{1,p}(X) = \mathrm{inf}_{(M_{1}, d_{1}, \mu_{1})} \dgwp(X,M_{1})$. There is a unique probability measure $\mu_{1}$ that is defined on a one point metric space and hence a unique coupling between $\mu_{X}$ and $\mu_{1}$. Thus, we obtain that $\sketch_{1,p}(X) = \dgwp(X,M_{1}) = \frac{1}{2} \cdot \diam_{p}(X)$. We have $\shatter_{1,p,\infty}(X) = \mathrm{diam}_{p}(X)$ and this proves the theorem.  
\end{proof}

\begin{theorem}[The case of $k$ finite and $p=\infty$] \label{thm:equalityforpinf} 
For all $X \in \mathcal{M}_{w}$ and $k \in \mathbf{N}$, we have
$$\sketch_{k,\infty}(X) = \frac{1}{2} \shatter_{k,\infty, \infty}(X).$$
\end{theorem}

\begin{proof}
We observe that $d_{\mathrm{GW}_\infty}(X,M_{k}) \geq \dgh(X,M_{k})$. Therefore, we have that
$$\sketch_{k,\infty}(X) = \inf_{(M_{k},d_{k},\mu_{k})} d_{\mathrm{GW}_\infty}(X,M_{k}) \geq \inf_{(M_{k}, d_{k})} \dgh(X, M_{k}) = \frac{1}{2} \cdot \shatter_{k}(X).$$
Since $\shatter_{k}(X) = \shatter_{k,\infty, \infty}(X)$, we obtain $\sketch_{k, \infty}(X) \geq \frac{1}{2} \cdot \shatter_{k, \infty, \infty}(X)$. We now prove the opposite inequality. Let $\shatter_{k,\infty, \infty}(X) \leq \eta$. Then, there exists a partition $\{ B_{1}, \cdots B_{k} \}$ of $X$ such that for all $i \in [k]$, $\mathrm{diam}(B_{i}) \leq \eta$. Let $(M_{k}, d_{k}, \mu_{k})$ be the $k$ point metric space such that for all $i,j \in [k]$, $d_{k}(i,j) = \dha^{X}(B_{i}, B_{j})$ and $\mu_{k}(i) = \mu_{X}(B_{i})$. Let $\gamma$ be the probability measure on $X\times M_k$  defined by $\gamma(A\times \{i\}) = \mu_X(A\cap B_i)$, for all measurable sets $A \subseteq X$ and $i \in M_k$. Clearly, $ \gamma(X \times \{i\}) = \mu_{X}(B_i) = \mu_k(i) ~ \forall ~ i \in M_{k}.$ Now, for all $A\subset X$ measurable, 
$ \gamma(A \times M_{k}) = \sum_{i\in[k]} \mu_X(A\cap B_i) = \mu_X(A)$. Thus, $\gamma$ is a coupling between $\mu_X$ and $\mu_k$.

Now, we have 
$$d_{\mathrm{GW}_\infty}(X,M_{k}) \leq \frac{1}{2} \sup_{\substack{x,x' ~ \in ~ X \\ i,j  ~\in ~ M_{k} \\ (x,i),(x',j) ~ \in ~ supp(\mu)}} \vert d_{X}(x,x') - d_{k}(i,j) \vert  \leq \frac{1}{2} \sup_{\substack{x ~ \in ~ B_{i} \\ x' ~ \in ~ B_{j}}} \vert d_{X}(x,x') - d_{k}(i,j) \vert. $$

We have the following two cases:
\begin{itemize}
\item \textbf{Case $1: i=j$}

Here we have $\mathrm{sup}_{x,x' \in B_{i}} d_{X}(x,x') = \mathrm{max}_{i} \mathrm{diam}(B_{i}) \leq \eta $.
\item \textbf{Case $2: i \neq j$}

Here we have 
$$\sup_{x \in B_{i}, y \in B_{j}} \vert d_{X}(x,y) - d_{k}(i,j) \vert = \sup_{x \in B_{i}, y \in B_{j}} \vert d_{X}(x,y) - \dha^{X}(B_{i}, B_{j}) \vert.$$
Now, $\dha^{X}(B_{i}, B_{j}) \leq \eta + d_{X}(B_{i}, B_{j}) $, where $d_{X}(B_{i}, B_{j}) = \mathrm{inf}_{x \in B_{i}, y \in B_{j}} d_{X}(x,y)$.  From the proof of Lemma \ref{lem:Hausdorff-map}, we know that for any $x \in B_{i}$ and $ y \in B_{j}$,
$$ d_{X}(B_{i}, B_{j}) \leq d_{X}(x,y) \leq \eta + \dha^{X}(B_{i}, B_{j}) $$
and 
$ d_{X}(B_{i}, B_{j}) \leq \dha^{X}(B_{i}, B_{j}) \leq \eta + d_{X}(B_{i}, B_{j}). $
This gives 
$ \vert d_{X}(x,y) - \dha^{X}(B_{i}, B_{j}) \vert \leq \eta. $
Thus, $d_{\mathrm{GW}_\infty}(X, M_{k}) \leq \dfrac{\eta}{2}$. So $\sketch_{k, \infty}(X) \leq \dfrac{\eta}{2}$. Thus, we obtain $\sketch_{k, \infty}(X) \leq \frac{1}{2} \shatter_{k, \infty, \infty}(X)$.
\end{itemize}
\end{proof}

We showed that for $k \in \mathbb{N}$ and $1 \leq p \leq \infty$ $\sketch_{k,p}$ and $\shatter_{k,p,p}$ are strictly dual for certain combinations of $k$ and $p$. We now prove Theorem \ref{thm:main-weak} which states that for all metric measure spaces $(X,d_X,\mu_X)$, $k \in \mathbb{N} $ and $1 \leq p \leq \infty $, we have $\sketch_{k,p}(X) \leq \shatter_{k,p,\infty}(X) $.

\begin{proof}[Proof of Theorem \ref{thm:main-weak}]
Let $\shatter_{k,p, \infty}(X) \leq \eta$. This means that there exists a partition $B_{1}, \ldots, B_{k}$ of $X$ such that for all $1 \leq i \leq k$, $\mathrm{diam}_{p}(B_{i}) \leq \eta$. Consider a set of $k$ points $M_{k}$. We define a metric $d_{k}$ on $M_{k}$ as follows:
$$ d_{k}(i,j) = \dwp^X(\mu_{B_{i}}, \mu_{B_{j}}) $$
where metric on $B_{i}$ is $d_{X}\vert_{B_{i}}$ and $\mu_{B_{i}} = \frac{\mu_{X}\vert_{B_{i}}}{\mu_X(B_i)}$. The measure $\mu_{k}$ on $M_{k}$ is defined as follows:
$ \mu_{k}(i) = \mu_{X}(B_{i})$ for each $i\in[k].$ Let $\gamma$ be the probability measure on $X\times M_k$ defined by: $\gamma(A\times \{i\}) = \mu_X(A\cap B_i)$ for all measurable sets $A \subseteq X$ and $i \in M_k$. Clearly, $ \gamma(X \times \{i\}) = \mu_{X}(B_i) = \mu_k(i) ~ \forall ~ i \in M_{k}.$ Now, for all $A\subset X$ measurable, 
$ \gamma(A \times M_{k}) = \sum_{i\in[k]} \mu_X(A\cap B_i) = \mu_X(A)$. Thus, $\gamma$ is a valid coupling between $\mu_X$ and $\mu_k$. Now, 
\begin{align*}
\dgwp(X, M_{k}) & \leq \frac{1}{2} \left( \iint \displaylimits_{(X \times M_{k}) \times (X \times M_k)} \vert d_{X}(x,y) - d_{k}(i,j) \vert^{p} d\gamma(x,i) d\gamma(y,j)\right)^{1/p} \\
& \leq \frac{1}{2} \left( \iint \displaylimits_{(X \times M_{k}) \times (X \times M_k)} \vert d_{X}(x,y) - \dwp^X(\mu_{B_{i}}, \mu_{B_{j}}) \vert^{p} d\gamma(x,i) d\gamma(y,j)\right)^{1/p} \\
& \leq \frac{1}{2} \left( \iint \displaylimits_{(X \times M_{k}) \times (X \times M_k)} \vert \dwp^X(\delta_{x}, \delta_{y}) - \dwp^X(\mu_{B_{i}}, \mu_{B_{j}})\vert^{p} d\gamma(x,i) d\gamma(y,j)\right)^{1/p} \\
& \leq \frac{1}{2} \left( \iint \displaylimits_{(X \times M_{k}) \times (X \times M_k)} \vert \dwp^X(\delta_{x}, \mu_{B_{i}}) + \dwp^X(\delta_{y}, \mu_{B_{j}}) \vert^{p} d\gamma(x,i) d\gamma(y,j)\right)^{1/p} \\
& \leq \frac{1}{2} \left( \iint \displaylimits_{(X \times M_{k}) \times (X \times M_k)} \vert \dwp^X(\delta_{x}, \mu_{B_{i}}) \vert^{p} d\gamma(x,i) d\gamma(y,j)\right)^{1/p} \\
&+ \frac{1}{2} \left( \iint \displaylimits_{(X \times M_{k}) \times (X \times M_k)} \vert \dwp^X(\delta_{y}, \mu_{B_{j}}) \vert^{p} d\gamma(x,i) d\gamma(y,j)\right)^{1/p} \\
& \leq \frac{1}{2} \left( \sum_{i} \int \displaylimits_{B_i} \vert \dwp^X(\delta_{x}, \mu_{B_{i}}) \vert^{p} d\mu_{X}(x) \right)^{1/p}  + \frac{1}{2} \left( \sum_{j} \int \displaylimits_{B_j} \vert \dwp^X(\delta_{y}, \mu_{B_{j}}) \vert^{p} d\mu_{X}(y) \right)^{1/p} \\
& = \left( \sum_{i} \int \displaylimits_{B_i} \vert \dwp^X(\delta_{x}, \mu_{B_{i}}) \vert^{p} d\mu_{X}(x) \right)^{1/p}.
\end{align*}
Here, the fourth inequality is due to the triangle inequality and the fifth inequality is due to Minkowski's inequality. Now, for all $x \in X$, $ \dwp^X(\delta_{x}, \mu_{B_{i}}) = \left(\int_{y \in B_{i}} d(x,y)^{p} \frac{d\mu_{X}(y)}{\mu_X(B_i)} \right)^{1/p} $. Therefore, we have
$$ \left( \sum_{i} \int \displaylimits_{B_i} \vert \dwp^X(\delta_{x}, \mu_{B_{i}}) \vert^{p} d\mu_{X}(x) \right)^{1/p} = \left(\sum_{i} \int_{x \in B_{i}} \int_{y \in B_{i}} d(x,y)^{p} d\mu_{X}(x) \frac{d\mu_{X}(y)}{\mu_X(B_i)} \right)^{1/p}. $$
Now, $\mathrm{diam}_{p}(B_{i}) = \left( \int_{x \in B_{i}} \int_{y \in B_i} d(x,y)^{p} \frac{d\mu_{X}(x)}{\mu_{X}(B_{i})} \frac{d\mu_{X}(y)}{\mu_{X}(B_{i})} \right)^{1/p}$. Thus, we obtain 
$$ \left(\sum_{i} \int_{x \in B_{i}} \int_{y \in B_{i}} d(x,y)^{p} d\mu_{X}(x) \frac{d\mu_{X}(y)}{\mu_X(B_i)} \right)^{1/p} = \left( \sum_{i} \mathrm{diam}_{p}^p(B_{i}) \cdot \mu_{X}(B_{i}) \right)^{1/p}. $$
We know that for every $i \in [k] $, $\mathrm{diam}_{p}(B_{i}) \leq \eta$ and $\sum_{i} \mu_{X}(B_{i}) = 1$. This implies that
$$ \left( \sum_{i} \mathrm{diam}_{p}^p(B_{i}) \cdot \mu_{X}(B_{i}) \right)^{1/p} \leq \eta.$$
We have shown that whenever $\shatter_{k,p, \infty}(X) \leq \eta$, we obtain $\sketch_{k,p}(X) \leq \eta$. Thus, $\sketch_{k,p}(X) \leq \shatter_{k,p, \infty}(X)$.
\end{proof}

In Theorem \ref{thm:main-weak}, we showed that $\sketch_{k,p}(X) \leq \shatter_{k,p,\infty}(X)$ for $X \in \mathcal{M}_w$. In general, however, these objectives are not dual to each other as we show next.

\subparagraph*{$\shatter_{k,p,\infty}$ is not dual to $\sketch_{k,p}.$}Let $\Delta_{m} = ([m], d_{m}, \mu_{m}) $ denote the metric measure space with $m$ points such that (1)for all $i,j \in [m]$, $d_{m}(i,j) = 1$ for $i \neq j$ and $d_{m}(i,j) = 0$ for $i=j$, and (2) for every $i \in [m] $, we have $\mu_{m}(i) = \frac{1}{m}$. The next theorem shows that for certain values of $m$, $\sketch_{k,p}(\Delta_m)$ and $\shatter_{k,p,\infty}(\Delta_m)$ are not within constant factor of each there. This implies that the functionals $\sketch_{k,p}$ and $\shatter_{k,p,\infty}$ are not dual to each other.

\begin{theorem} \label{thm:counter-ex}
For every constant $C_1 > 0$ and every $r \in \mathbf{N}$ such that $r \geq 2$, there exists $k \in \mathbf{N}$ such that for $m=rk$, we have $ C_1 \cdot \shatter_{k,p,\infty}(\Delta_m) > \sketch_{k,p}(\Delta_m),$
for every $1 \leq p < \infty$. 
\end{theorem}

\begin{proof}
Fix $p \in [1, \infty)$. We first show that for any $r,k \in \mathbf{N}$ and $m = rk$, we have 
$\shatter_{k,p, \infty}(\Delta_{m}) = \left( 1 - \frac{k}{m}\right)^{\frac{1}{p}}$. We know that for any $B \subseteq \Delta_{m}$ with $|B| = l$, we have $\mathrm{diam}_{p}(B) = \left(1-\frac{1}{l} \right)^{\frac{1}{p}} $. Therefore for any partition $\{B_{1}, \ldots, B_{k} \}$ of $\Delta_{m}$ with $|B_{i}| = l_{i}$, we have $\mathrm{max}_{i} \mathrm{diam}_{p}(B_{i}) = \mathrm{max}_{i} \left(1 - \frac{1}{l_{i}}\right)^{\frac{1}{p}} = \left(1 - \frac{1}{\mathrm{max}_{i} l_{i}} \right)^{\frac{1}{p}} $. Thus,
$$ \shatter_{k,p, \infty}(\Delta_{m}) = \min_{\substack{(l_{1}, \ldots, l_{k}) \\ \sum_{i = 1}^{k} l_{i} = m}} \max_{i} \left(1-\frac{1}{l_{i}} \right)^{\frac{1}{p}} = \left( 1-\frac{1}{ \mathrm{min}_{\substack{(l_{1}, \ldots, l_{k}) \\ \sum_{i = 1}^{k} l_{i} = m} }\mathrm{max}_{i} l_{i} } \right)^{\frac{1}{p}}. $$
Since $m = rk$, we have that $$\mathrm{min}_{\substack{(l_{1}, \ldots, l_{k}) \\ \sum_{i = 1}^{k} l_{i} = m}} \mathrm{max}_{i} l_{i} = \frac{m}{k} = r.$$ 
Thus, we obtain $\shatter_{k,p, \infty}(\Delta_{m}) = \left( 1 - \frac{k}{m}\right)^{\frac{1}{p}}$.

From the definition of $\sketch_{k,p}(X)$, we know that $\sketch_{k,p}(\Delta_{m}) \leq \dgwp(\Delta_{m}, \Delta_{k})$. We use claim $5.1$ from \cite{Memoli2011} to obtain that $\dgwp(\Delta_{k}, \Delta_{m}) \leq \frac{1}{2} \left( \frac{1}{k} + \frac{1}{m} \right)^{\frac{1}{p}}$. This implies that $\sketch_{k,p}(\Delta_m) \leq \frac{1}{2} \left( \frac{1}{k} + \frac{1}{m} \right)^{\frac{1}{p}}$.

Now, we fix a constant $C_1 > 0$ and $r \in \mathbf{N}$ such that $r \geq 2$. Let $k \in \mathbf{N}$ be such that $k > \dfrac{1+r^{-1}}{2^p \cdot C_1^p \cdot (1-r^{-1})}$. Then, we have that 
$$C_1 \cdot (1-r^{-1})^{1/p}  > \frac{1}{2 \cdot k^{1/p}} \cdot (1+r^{-1})^{1/p}. $$
This implies that $ C_1 \cdot \shatter_{k,p,\infty}(\Delta_m) > \sketch_{k,p}(\Delta_m)$. 
\end{proof}

From the definition of duality (equation (\ref{eq:duality})) and the above theorem, we conclude that $\shatter_{k,p,\infty}$ is not dual to $\sketch_{k,p}$.

\subsection{Using Sturm's version of the Gromov-Wasserstein distance} \label{sec:Sturmsdefn}
In this section, we define another sketching objective for metric measure spaces, $\sketch^S_{k,p}(X)$ using Sturm's definition of $p$-Gromov-Wasserstein distance \cite{Sturm2006}. We show that $\sketch^S_{k,p}$ is dual to the clustering objective $\shatter_{k,p,p}$, for all $k \in \mathbf{N}$ and $1 \leq p \leq \infty$. 

We recall that given $X,Y \in \mathcal{M}_w $, $\mathcal{D}(d_X,d_Y)$ denotes the set of all metric couplings between $d_X$ and $d_Y$. The Sturm's version of $p$-Gromov-Wasserstein distance, for $1 \leq p \leq \infty$, is defined as follows.

\begin{definition}[\textbf{Sturm's version of $p$-Gromov-Wasserstein distance}] \cite{Sturm2006}
Given $X,Y \in \mathcal{M}_{w}$ and $1 \leq p < \infty$, define
$$ \zeta_{p}(X,Y) := \inf_{\substack{d \in \mathcal{D}(d_X,d_Y) \\ \mu \in \mathcal{U}(\mu_X,\mu_Y) }} \left(\int_{X \times Y} d^{p}(x,y) \, d \mu(x,y) \right)^{1/p}. $$
For $p = \infty$, define
$$ \zeta_{\infty}(X,Y) := \inf_{\substack{d \in \mathcal{D}(d_X,d_Y) \\ \mu \in \mathcal{U}(\mu_X,\mu_Y)}} \sup_{(x,y) \in R[\mu]} d(x,y). $$
\end{definition}

It was shown in \cite{Memoli2011} that for all $X,Y \in \mathcal{M}_w$ and $1 \leq p \leq \infty$, $\zeta_p(X,Y) \geq \dgwp(X,Y)$ and $\zeta_{\infty}(X,Y) = d_{\mathrm{GW}_{\infty}}(X,Y)$.

We now prove the duality between $\sketch^S_{k,p}(X)$ and $\shatter_{k,p,p}(X)$ for metric measure spaces using a method similar to that for metric spaces. We define the Voronoi map and the Wasserstein map for metric measure spaces. Given a metric measure space $(X,d_X,\mu_X) $ and $k \in \mathbf{N}$, the Voronoi map associated to $X$ returns a partition of $X$ into $k$ clusters. The Wasserstein map, on the other hand, acts on a partition of $X$ into $k$ clusters, and returns a $k$-point metric measure space.  

\begin{definition}[\textbf{$\eps,p$-optimal metric coupling}]
Let $(X,d_X,\mu_X) \in \mathcal{M}_{w}$, $k \in \mathbf{N}$, $1 \leq p < \infty$ and $\epsilon > 0$. Given $(M_k,d_k,\mu_k) \in \mathcal{M}_{k,w}$, $d \in \mathcal{D}(d_X,d_k)$ is called $\eps,p$-optimal if 
$$\zeta_{p}(X,M_{k}) + \epsilon \geq \inf_{\mu \in \mathcal{U}(\mu_X,\mu_k)} \left(\int_{X \times M_k} d^{p}(x,i) d\mu(x \times i) \right)^{1/p}.$$ 
We denote by $\mathcal{D}_{\eps,p}(d_X,d_k)$ the set of all $\eps,p$-optimal metric couplings between $d_X$ and $d_k$.
\end{definition}

\begin{definition}[\textbf{Voronoi map for mm-spaces}]

Let $(X,d_X,\mu_X) \in \mathcal{M}_{w}$, $k \in \mathbf{N}$, $1 \leq p < \infty$ and $\epsilon > 0$. Given $(M_k,d_k,\mu_k) \in \mathcal{M}_{k,w}$, $M_k = \{1,2,\ldots,k \}$, let $d \in \mathcal{D}_{\eps,p} (d_X,d_k)$. Let $P = \{B'_i \}_{i=1}^{k}$ be a Voronoi partition of $(X \sqcup M_k, d) $ with respect to $M_k$. We define for every $i \in M_k$, $B_i =B'_i \cap X $. 
Thus, $\{B_i \}_{i=1}^k \in \partk(X) $. We now define
$$ \mathcal{V}_{X,k,M_k,\eps,p} : \mathcal{D}_{\eps,p} \rightarrow \partk(X) $$
as $\mathcal{V}_{X,k,M_k,\eps,p}(d) := \{B_i \}_{i=1}^{k} \in \partk(X)$. 
\end{definition}

\begin{Remark}
We observe that when $X$ is finite, then given $1 \leq p < \infty $, we obtain $d \in \mathcal{D}(d_X,d_k) $ such that $\zeta_p(X,M_k) = \inf_{\mu \in \mathcal{U}(\mu_X,\mu_k)} \left(\int_{X \times M_k} d^{p}(x,i) d\mu(x \times i) \right)^{1/p}$. Therefore, in the above definition, fixing an $\eps > 0 $ is not required.
\end{Remark}

\begin{definition}[\textbf{Wasserstein map for mm-spaces}] \label{def:Wasserstein-map}
Let $(X,d_X,\mu_X) \in \mathcal{M}_{w}$ and $k \in \mathbf{N}$. We define a map $\mathcal{H}_{X,k}^{w} : \partk(X) \rightarrow \mathcal{M}_{k,w}$ as follows: for any $P = \{B_i \}_{i=1}^{k} \in \partk(X)$, we define $(M_k,d_k,\mu_k)$ as 
$$d_{k}(i,j) := \dwp^{X}(\mu_{B_{i}}, \mu_{B_{j}}) ~ \forall~i,j \in [k],~i \neq j \quad \mbox{and} \quad \mu_{k}(i) = \mu_{X}(B_{i}) ~ \forall~i \in [k].$$
Here $\dwp^{X}(\mu_{B_i},\mu_{B_j})$ is the $p$-Wasserstein distance (Definition \ref{def:dW}) between $B_i$ and $B_j$ in $X$ and $\mu_{B_i} = \frac{\mu_X|_{B_i}}{\mu_X(B_i)}$ for all $i \in [k]$, i.e. $\mu_{B_i}$ arises as the renormalization of the measure $\mu_X$ when restricted to $B_i$.  We now define 
$$ \mathcal{H}_{X,k}^{w}(P) := (M_k,d_k,\mu_k). $$
\end{definition}

In the next section, we use the two maps defined above to establish the duality result for metric measure spaces.

\subsection{Relationship between $\sketch^S_{k,p}$ and $\shatter_{k,p,p}$} \label{sec:sketch-sturm-shatter}

We now use the Voronoi map and the Wasserstein map for metric measure spaces to establish a duality between $\sketch^S_{k,p} $ and $\shatter_{k,p,p}$ for all $k \in \mathbb{N}$ and $1 \leq p \leq \infty$ . We prove the stability of these maps in Lemmas \ref{lem:Voronoi-map-measure} and \ref{lem:Hausdorff-map-measure}. The first lemma, i.e.~Lemma \ref{lem:Voronoi-map-measure} states that given a metric measure space $(X,d_X)$ and a $k$-point metric measure space $M_k$, the cost of partitioning $X$ into the clusters obtained by applying the Voronoi map is arbitrarily close to twice the Gromov-Wasserstein distance between $X$ and $M_k$. This lemma is formally stated as follows.

\begin{lemma} \label{lem:Voronoi-map-measure}
For every $X \in \mathcal{M}_{w}$, $k \in \mathbf{N}$, $M_{k} = \{1,2,\ldots,k \} \in \mathcal{M}_{k,w}$, $\epsilon > 0$, $1 \leq p < \infty$ and $d \in \mathcal{D}_{\eps,p} (d_X,d_k)$, we have
$$ \Phi_{p,p}(X,\mathcal{V}_{X,k,M_k,\eps,p}(d)) \leq 2 \cdot \zeta_{p}(X,M_{k}) + 2 \epsilon.$$
\end{lemma}

We need the following claim to prove the above lemma.

\begin{claim} \label{clm:optimalcoupling}
Given $(X, d_{X}, \mu_{X}) \in \mathcal{M}_w $, $(M_{k}, d_{k}, \mu_{k}) \in \mathcal{M}_{k,w} $ and a coupling $d$ between $d_{X}$ and $d_{k}$, let $\{B'_{1}, \ldots, B'_{k} \}$ denote the Voronoi partition of $X \sqcup M_{k}$ with respect to $M_{k}$ under the metric $d$. For all $i \in [k]$, let $B_i = B'_i \setminus \{i \}$. Define $(M_{k}, d_{k}, \mu_{k}^{\mathbf{V}}) \in \mathcal{M}_{k,w} $, where for every $i \in [k]$, $\mu^{\mathbf{V}}_{k}(i) = \mu_{X}(B_{i})$. Then, there exists a coupling $\mu'$ between $\mu_{X}$ and $\mu^{\mathbf{V}}_{k}$ such that $\mathrm{supp}[\mu'] = \cup_{i \in [k]} (B_i \times \{i \})$, i.e.~$(x,i) \in \mathrm{supp}[\mu']$ if and only if $x \in B_i$, and for every coupling $\mu$ between $\mu_{X}$ and $\mu_{k}$,
$$ \left( \int_{X \times M_k} d^{p}(x,i) d \mu'(x \times i) \right)^{1/p} \leq \left( \int_{X \times M_k} d^{p}(x,i) d \mu(x \times i) \right)^{1/p}.$$
\end{claim}

\begin{proof}[Proof of Claim \ref{clm:optimalcoupling}]
Let $Z = X \sqcup M_{k}$. Let $d$ be a metric on $Z$ such that $d$ is a coupling between $d_{X}$ and $d_{k}$. Let $\mu$ be any coupling between $\mu_{X}$ and $\mu_{k}$. Since $\{ B'_{1}, B'_{2}, \ldots, B'_{k} \}$ is a Voronoi partition of $Z$ with respect to $M_{k}$, we have that for all $i,j \in [k]$, if $x \in B'_{i}$, then $d(x,i) \leq d(x,j)$ for all $j \neq i$.

Let $\mu'$ be the probability measure on $X\times M_k$  defined by $\mu'(A \times \{i\}) = \mu_{X}(A\cap B_i)$, for all $A \subseteq X$ and $i \in M_k$ . We first claim that $\mu'\in\mathcal{U}(\mu_X,\mu_k^\mathbf{V})$. Clearly, $ \mu'(X \times \{i\}) = \mu_{X}(B_i) = \mu_k^\mathbf{V}(i) ~ \forall ~ i \in M_{k}.$ Now, for all $A\subseteq X$ measurable, 
$ \mu'(A \times M_{k}) = \sum_{i\in[k]} \mu_X(A\cap B_i) = \mu_X(A)$. Thus, $\mu'$ is a coupling between $\mu_X$ and $\mu_k^\mathbf{V}$. It is straightforward to see that $\mathrm{supp}[\mu'] = \cup_{i \in [k]} (B_i \times \{i \})$.

Now, to prove the claim write:
\begin{align*}
\int_{X\times P_k} d^p(x,i)\,\mu(dx\times\{i\}) & = \sum_{i} \int_X d^p(x,i)\,\mu(dx\times \{i\})\\
& = \sum_i\sum_j\int_{B_j} d^p(x,i) \,\mu(dx\times \{i\})\\
& \geq \sum_i\sum_j \int_{B_j} d^p(x,j) \mu(dx\times \{i\})\\
&=\sum_j\int_{B_j} d^p(x,j)\mu_X(dx)\\
&= \int_{X\times P_k} d^p(x,j)\mu'(dx\times\{j\}).
\end{align*}
\end{proof}

\begin{proof}[Proof of Lemma \ref{lem:Voronoi-map-measure}]
Given $(X, d_{X}, \mu_{X}) \in \mathcal{M}_{w}$, $(M_{k}, d_{k}, \mu_{k}) \in \mathcal{M}_{k,w}$, $1 \leq p < \infty$ and $\epsilon > 0$, let $d \in \mathcal{D}_{\eps,p} (d_X,d_k)$. Then, we have that 
$$\zeta_{p}(X,M_{k}) + \epsilon \geq \inf_{\mu \in \mathcal{U}(\mu_X, \mu_{k})} \left(\int \displaylimits_{X \times M_k} d^{p}(x,i) d\mu(x \times i) \right)^{1/p}.$$ 
Let $\{B'_i \}_{i=1}^k$ be a Voronoi partition of $(X \sqcup M_k,d)$ with respect to $M_k$ and $\{B_i \}_{i=1}^k = \{B'_i \cap X \}_{i=1}^k$. By definition, $\mathcal{V}_{X,k,M_k,\eps,p}(d) = \{B_i \}_{i=1}^k $.
By Claim \ref{clm:optimalcoupling}, there exists $(M_k,d_k,\mu_k^{\mathbf{V}} ) \in \mathcal{M}_{k,w}$, and a coupling $\mu' \in \mathcal{U}(\mu_X,\mu_k^{\mathbf{V}} )$ such that $\mathrm{supp}[\mu'] = \cup_{i \in [k]} (B_i \times \{i \})$, and
$$ \inf_{\mu \in \mathcal{U}(\mu_X, \mu_k)} \left(\int \displaylimits_{X \times M_k} d^{p}(x,i) d\mu(x \times i) \right)^{1/p} \geq \left(\int \displaylimits_{X \times M_k} d^{p}(x,i) d\mu'(x \times i) \right)^{1/p}. $$

We consider the partition $P = \{B_i \}_{i=1}^k$ of $X$. Then, we have
$$ \left(\int_{X \times M_k} d^{p}(x,i)\, d\mu'(x \times i) \right)^{1/p} = \left( \sum_{i} \int_{B_{i}} d^{p}(x,i) \,d \mu_{X}(x) \right)^{1/p} \leq \zeta_{p}(X,M_{k}) + \epsilon.$$
From the above equation, we obtain
\begin{align*}
2 \cdot \zeta_{p}(X,M_{k}) + 2 \epsilon & \geq \left( \sum_{i} \int_{B_{i}} d^{p}(x,i) d \mu_{X}(x) \right)^{1/p} + \left( \sum_{i} \int_{B_{i}} d^{p}(x',i) d \mu_{X}(x') \right)^{1/p} \\
& = \left( \sum_{i} \iint \displaylimits_{B_{i} \times B_i} \frac{d^{p}(x,i) d \mu_{X}(x) d \mu_{X}(x')}{\mu_{X}(B_{i})} \right)^{1/p} + \left( \sum_{i} \iint \displaylimits_{B_{i} \times B_i} \frac{d^{p}(x',i) d \mu_{X}(x') d \mu_{X}(x)}{\mu_{X}(B_{i})} \right)^{1/p} \\
& \geq \left( \sum_{i} \iint \displaylimits_{B_{i} \times B_i} \frac{(d(x,i) + d(x',i))^{p} d \mu_{X}(x) d \mu_{X}(x')}{\mu_{X}(B_{i})} \right)^{1/p}. 
\end{align*}
The last inequality is due to Minkowski's inequality. From the triangle inequality, we have
\begin{align*}
2 \cdot \zeta_{p}(X,M_{k}) + 2 \epsilon & \geq \left( \sum_{i} \frac{1}{\mu_{X}(B_{i})} \iint \displaylimits_{B_{i} \times B_i} d^{p}(x,x') d \mu_{X}(x) d \mu_{X}(x') \right)^{1/p} \\
& = \left( \sum_{i} \mathrm{diam}_{p}^{p}(B_{i})\, \mu_{X}(B_{i}) \right)^{1/p}.
\end{align*}
Thus, we have
$$ 2 \cdot \zeta_{p}(X,M_{k}) + 2 \epsilon \geq \left( \sum_{i} \mathrm{diam}_{p}^{p}(B_{i})\, \mu_{X}(B_{i}) \right)^{1/p}. $$
Since $\mathcal{V}_{X,k,M_k,\eps,p}(d ) = \{B_{i}\}_{i=1}^{k}$ and $\Phi_{p,p}(X, \{B_i \}_{i=1}^k) = \left( \sum_{i} \mathrm{diam}_{p}^{p}(B_{i})\, \mu_{X}(B_{i}) \right)^{1/p}$, we obtain $\Phi_{p,p}(X, \mathcal{V}_{X,k,M_k,\eps,p}(d)) \leq 2 \cdot \zeta_{p}(X, M_{k}) + 2 \epsilon$.
\end{proof}

The next lemma states that given $X \in \mathcal{M}_w $ and a partition $P$ of $X$ into $k$ clusters, the Gromov-Wasserstein distance between $X$ and the $k$-point mm-space obtained by applying the Hausdorff map to $P$, is at most the cost of partitioning $X$ as $P$. This lemma is formally stated as follows.

\begin{lemma} \label{lem:Hausdorff-map-measure}
For every $X \in \mathcal{M}_{w}$, $k \in \mathbf{N}$, $P \in \partk(X)$ and $1 \leq p < \infty$, we have
$$ \zeta_{p}(X, \mathcal{H}^{w}_{X,k}(P)) \leq \Phi_{p,p}(X,P). $$
\end{lemma}

\begin{proof}
Given $(X,d_{X},\mu_{X}) \in \mathcal{M}_{w}$ and $P = \{B_{i} \}_{i=1}^{k} \in \partk(X)$, let 
$$\eta = \Phi_{p,p}(X,P) = \left( \sum_{i} \mathrm{diam}_{p}^{p}(B_{i})\, \mu_{X}(B_{i}) \right)^{1/p}.$$
Let $(M_{k}, d_{k}, \mu_{k})$ be a $k$-point metric measure space such that

$$d_{k}(i,j) = \dwp^{X} \left(\frac{\mu_{X}\vert_{B_{i}}}{\mu_{X}(B_{i})},\frac{\mu_{X}\vert_{B_{j}}}{\mu_{X}(B_{j})} \right) ~ \forall~i,j \in [k],~i \neq j \quad \mbox{and} \quad \mu_{k}(i) = \mu_{X}(B_{i}) ~ \forall~i \in [k].$$

Define $d$ on $X \sqcup M_{k}$ as follows:
$$d(x,i) = \dwp^X \left(\delta_{x}, \frac{\mu_{X}\vert_{B_{i}}}{\mu_{X}(B_{i})} \right) ~ \forall~x \in X,~i \in [k].$$ 
It follows from the definition that $d$ satisfies triangle inequality. Let $\mu$ be a probability measure on $X \times M_k$ defined by $\mu(A \times \{i\}) = \mu_{X}(A\cap B_i)$ for all $A \subseteq X$ and $i \in M_k$. Clearly, $ \mu(X \times \{i\}) = \mu_{X}(B_i) = \mu_k(i) ~ \forall ~ i \in M_{k}$. Now, for all $A\subseteq X$ measurable, $ \mu(A \times M_{k}) = \sum_{i\in[k]} \mu_X(A\cap B_i) = \mu_X(A).$ Thus, $\mu$ is a coupling between $\mu_X$ and $\mu_k$. Now, we have 
\begin{align*}
\int_{X \times P_k} d^{p}(x,i) d \mu(x,i) &= \sum_{i} \int_{B_{i}} d^{p}(x,i) \mu_{X}(dx) = \sum_{i} \int_{B_{i}} \left( \dwp^{X} \left(\delta_{x}, \frac{\mu_{X}\vert_{B_{i}}}{\mu_{X}(B_{i})} \right) \right)^p \mu_{X}(dx) \\
&= \sum_{i} \int_{B_{i}} \int_{B_{i}} d^{p}(x,y) \frac{\mu_{X}(dy)}{\mu_{X}(B_{i})} \mu_{X}(dx)  \\
&= \sum_{i} \mu_{X}(B_{i}) \iint \displaylimits_{B_{i} \times B_i} d^{p}(x,y) \frac{\mu_{X}(dx) \mu_{X}(dy)}{\mu^{2}_{X}(B_{i})} \\
&= \sum_{i} \mu_{X}(B_{i}) \,\mathrm{diam}^{p}_{p}(B_{i}) \leq \eta^{p}.
\end{align*} 
This implies that $\zeta_{p}(X, M_{k}) \leq \eta$. Since $\mathcal{H}^{w}_{X,k}(P) = (M_{k},d_k,\mu_k)$, we obtain $\zeta_{p}(X, \mathcal{H}_{X,k}^{w}(P)) \leq \Phi_{p,p}(X,P)$.
\end{proof}

Using above lemmas, we now prove the main result of this section.

\begin{proof}[Proof of Theorem \ref{thm:main2}]
Fix $(X,d_X,\mu_X) \in \mathcal{M}_{w}$, $k \in \mathbf{N}$ and $1 \leq p < \infty$. Suppose $\sketch^S_{k,p}(X) < \eta$. Then, there exists $(M_{k},d_k,\mu_k) \in \mathcal{M}_{k,w}$ such that $\zeta_{p}(X,M_k) < \eta $. Then, we have that for every $\epsilon > 0$ and $d \in \mathcal{D}_{\eps,p}(d_X,d_k)$,
$$\Phi_{p,p}(X, \mathcal{V}_{X,k,M_k,\eps,p}(d)) \leq 2 \cdot \zeta_{p}(X,M_k) + 2 \epsilon < 2 \eta + 2 \epsilon .$$ 
Since $\mathcal{V}_{X,k,M_k,\eps,p}(d ) \in \partk(X)$ and $\shatter_{k,p,p}(X) \leq \Phi_{p,p}(X, \mathcal{V}_{X,k,M_k,\eps,p}(d ))$, we have that $\shatter_{k,p,p}(X) <  2 \eta + 2 \epsilon$. This inequality is true for every $\epsilon > 0$. Therefore, we conclude that $\shatter_{k,p,p}(X) \leq  2 \eta$. We have shown that whenever $\sketch^S_{k,p}(X) < \eta$, we obtain $\shatter_{k,p,p}(X) \leq 2\eta$. Hence, we have $\shatter_{k,p,p}(X) \leq 2 \cdot \sketch^S_{k,p}(X)$. 

We now prove the second inequality. Let $\shatter_{k,p,p}(X) < \eta$. Then, there exists $P = \{B_i \}_{i=1}^{k} \in \partk(X)$ such that $\Phi_{p,p}(X,P) < \eta$. Then, from Lemma \ref{lem:Hausdorff-map-measure}, we have that
$$ \zeta_{p}(X, \mathcal{H}^{w}_{X,k}(P)) \leq \Phi_{p,p}(X,P) < \eta. $$
Since $\mathcal{H}^{w}_{X,k}(P) \in \mathcal{M}_{k,w}$ and $\sketch^{S}_{k,p}(X) \leq \zeta_{p}(X, \mathcal{H}^{w}_{X,k}(P))$, we obtain $ \sketch^S_{k,p}(X) < \eta.$ Thus, we have shown that whenever $\shatter_{k,p,p}(X) < \eta$, we have $\sketch^S_{k,p}(X) < \eta$. We conclude that $\sketch^S_{k,p}(X) \leq \shatter_{k,p,p}(X)$.
\end{proof}

A direct corollary of Theorem \ref{thm:main2} is the following. 

\begin{corollary} \label{cor:GWp}
We have that
\begin{itemize}
\item If $(M_{k}, d_{k},\mu_k) \in \mathcal{M}_{k,w}$ is such that $\sketch^{S}_{k,p}(X) = \zeta_{p}(X, M_{k})$, then for every $\epsilon > 0$ and $d \in \mathcal{D}_{\eps,p}(d_X,d_k)$, we have
$$\shatter_{k,p,p}(X) \leq \Phi_{p,p}(X, \mathcal{V}_{X,k,M_k,\eps,p}(d )) \leq 2 \cdot \shatter_{k,p,p}(X) + 2 \epsilon.$$
\item If $P = \{B_{i} \}_{i=1}^{k} \in \mathrm{Part}_{k}(X)$ is such that $\shatter_{k,p,p}(X) = \Phi_{p,p}(X,P)$, then 
$$\sketch^{S}_{k,p}(X) \leq  \zeta_{p}(X, \mathcal{H}^{w}_{X,k}(P)) \leq 2 \cdot \sketch^{S}_{k,p}(X).$$
\end{itemize}
\end{corollary}

\begin{Remark} \label{rem:GWp}
Note that Corollary \ref{cor:GWp} along with Lemmas \ref{lem:Voronoi-map-measure} and \ref{lem:Hausdorff-map-measure} tells us that for any $X \in \mathcal{M}_w$, $k \in \mathbf{N} $ and $1 \leq p < \infty$, given $(M_k,d_k,\mu_k) \in \mathcal{M}_{k,w}$ such that $\sketch^S_{k,p}(X) = \zeta_p(X,M_k)$, and $d \in \mathcal{D}(d_X,d_k)$ such that $\zeta_p(X,M_k) =  \inf_{\mu \in \mathcal{U}(\mu_X, \mu_k)} \left(\int_{X \times M_k} d^{p}(x,i) d\mu(x \times i) \right)^{1/p}$, we obtain a clustering of $X$ that is a $2$-approximation for $\shatter_{k,p,p}(X)$. On the other hand, given $P = \{B_{i} \}_{i=1}^k \in \partk(X)$ such that $\shatter_{k,p,p}(X) = \Phi_{p,p}(X,P)$, a $2$-approximation for $\sketch^S_{k,p}(X)$ is obtained by calculating $p$-Wasserstein distance between the blocks of partition $P$. 
\end{Remark}

For $p = q = \infty$, we have the following strict duality result.
\begin{theorem} \label{thm:pinf-sturm-equality}
Let $X \in \mathcal{M}_w$ and $k \in \mathbf{N}$. For $p =q= \infty$, we have 
$$\frac{1}{2} \cdot \shatter_{k,\infty,\infty}(X) = \sketch^{S}_{k,\infty}(X).$$
\end{theorem}

\begin{proof}
For $p = \infty$, we use Theorem $5.1$ from \cite{Memoli2011} that says for any $X,Y \in \mathcal{M}_{w}$, we have $d_{\mathrm{GW}_{\infty}}(X,Y) = \zeta_{\infty}(X,Y)$. This gives that
$$ \sketch^{S}_{k,\infty}(X) = \inf_{(P_{k}, d_{k}, \mu_{k})} \zeta_{\infty}(X,P_{k}) = \inf_{(P_{k}, d_{k}, \mu_{k})} d_{\mathrm{GW}_{\infty}}(X, P_{k}) = \sketch_{k,\infty}(X).$$
From Theorem \ref{thm:equalityforpinf}, we have that $\sketch_{k, \infty}(X) = \frac{1}{2} \cdot \shatter_{k, \infty, \infty}(X)$. This implies that $\sketch^{S}_{k, \infty}(X) = \frac{1}{2} \cdot \shatter_{k, \infty, \infty}(X)$.
\end{proof}

 \subsection{Computation of $\sketch^S_{k,p}$ and $\shatter_{k,p,p}$ for $\Delta_m$}
In this section, we compute the objectives $\sketch^S_{k,p}$ and $\shatter_{k,p,p}$ for the space $\Delta_m \in \mathcal{M}_w$. We note that, in general, these objectives are difficult to compute. We show the computations in the following example.

\begin{example} \label{thm:sketch-shatter-deltam}
For all natural numbers $m\geq 2$, we have
\begin{itemize}
\item $\dfrac{1}{2} \leq \dfrac{\sketch_{k,p}^{S}(\Delta_{m})}{\shatter_{k,p,p}(\Delta_{m})} \leq \dfrac{1}{2 (1-k \cdot m^{-1})^{1/p}}$ for any natural number $k < m$ and $1 \leq p \leq \infty$.
\item Furthermore, $\dfrac{\sketch_{1,p}^S(\Delta_m)}{\shatter_{1,p,p}(\Delta_m)} = \dfrac{1}{2(1-m^{-1})^{1/p}}.$ This implies that the ratio $\dfrac{\sketch_{1,p}^S(\Delta_m)}{\shatter_{1,p,p}(\Delta_m)}$ assumes infinitely many values in the interval $\left[ \frac{1}{2},1 \right]$.
\end{itemize}
\end{example}

The above statements are proved in the following three lemmas.
\begin{lemma} \label{lem:qbound}
For all $m,k \in \mathbf{N}$ with $k < m$ and every $p \geq 1$, $\shatter_{k,p,p}(\Delta_{m}) = \left( 1- \frac{k}{m} \right)^{1/p}$.
\end{lemma}
\begin{proof}

Let $\{B_{1}, \cdots, B_{k} \}$ denote a partition of $\Delta_{m}$ into $k$ blocks. For every $i \in [m]$, let $\vert B_{i} \vert = k_{i}$. Then $\sum_{i} k_{i} = m$. For any real $p \geq 1$, $\mathrm{diam}^{p}_{p}(B_{i}) = \sum_{x,y \in B_{i}} d_{m}^{p}(x,y) \cdot \frac{1}{m^{2}} \cdot \left( \frac{k_{i}}{m} \right)^{-2}$. Now, there are $k_{i}^{2}$ pairs of points $x,y$ in $B_{i}$,  out of which $k_{i}^{2} - k_{i}$ pairs satisfy $d_{m}(x,y) = 1$. This gives $\mathrm{diam}_{p}^{p}(B_{i}) = \frac{k_{i}^{2} - k_{i}}{m^{2}} \cdot \frac{m^{2}}{k_{i}^{2}} = \frac{k_{i} - 1}{k_{i}}$. Now,
$$\sum_{i} \mu_{m}(B_{i}) \mathrm{diam}_{p}^{p}(B_{i}) = \sum_{i} \frac{k_{i}}{m} \left( \frac{k_{i}- 1}{k_{i}} \right) = \frac{1}{m} \sum_{i} k_{i} - 1 = \frac{m-k}{m}. $$
The above equation holds true for any partition $\{ B_{1}, \cdots, B_{k} \}$ of $\Delta_{m}$. Thus, we conclude that for any $m,k \in \mathbf{N}$ and $k < m$, $\shatter_{k,p,p}(\Delta_{m}) = \left( 1- \frac{k}{m} \right)^{1/p}$.
\end{proof}

In the next lemma, we obtain bounds for $\sketch_{k,p}^S(\Delta_m)$.

\begin{lemma} \label{lem:etabound}
For all $m,k \in \mathbf{N}$ with $k < m$ and for every $p \geq 1$, $$ \frac{1}{2} \left( 1 - \frac{k}{m} \right)^{1/p} \leq \sketch^{S}_{k,p}(\Delta_{m}) \leq \frac{1}{2}.$$
\end{lemma}
\begin{proof}
We have $\sketch_{k,p}^{S}(\Delta_{m}) = \inf_{(M_{k}, d_{k}, \mu_{k})} \zeta_{p}(\Delta_{m}, M_{k})$. Let $P_{k} = \Delta_{k}$. Then, $\sketch^{S}_{k,p}(\Delta_{m}) \leq \zeta_{p}(\Delta_{m}, \Delta_{k})$. Now, $\zeta_{p}(\Delta_{m}, \Delta_{k}) \leq \inf_{\mu,d} \left(\int d^{p}(x, y) d \mu(x \times y) \right)^{1/p}$, where $x \in \Delta_{m},~ y \in \Delta_{k}$, $d$ varies over all metric couplings between $d_{m}$ and $d_{k}$ and $\mu$ varies over all measure couplings between $\mu_{m}$ and $\mu_{k}$. Let $d(x,y) = \frac{1}{2}$ for all $x \in \Delta_{m},~ y \in \Delta_{k}$. It can be verified that $d$ is a valid coupling. Let the coupling $\mu$ be the product measure. Then,
$$ \zeta_{p}(\Delta_{m}, \Delta_{k}) \leq \left(\sum_{x \in \Delta_{m}} \sum_{y \in \Delta_{k}} \frac{1}{2^{p}} \frac{1}{mk} \right)^{1/p} = \frac{1}{2}. $$
This establishes that $\sketch^{S}_{k,p}(\Delta_{m}) \leq \frac{1}{2}$. The lower bound follows from Theorem \ref{thm:main2} and Lemma \ref{lem:qbound}. Therefore, we obtain
$$ \frac{1}{2} \leq \frac{\sketch^{S}_{k,p}(\Delta_{m})}{\shatter_{k,p,p}^{S}(\Delta_{m})} \leq \frac{1}{2(1 - k \cdot m^{-1})^{1/p}}. $$
\end{proof}

In the next lemma, we explicitly compute $\sketch_{1,p}^S(\Delta_m)$, using the last lemma.

\begin{lemma}\label{lem:ex}
For all $m \geq 2$ and $1 \leq p < \infty $, $\sketch^{S}_{1,p}(\Delta_{m}) > \frac{1}{2} \cdot \shatter_{1,p,p}(\Delta_{m})$. Furthermore, we have:
 $$\frac{\sketch_{1,p}^S(\Delta_m)}{\shatter_{1,p,p}(\Delta_m)} = \frac{1}{2(1-m^{-1})^{1/p}}.$$
\end{lemma}

\begin{proof}
Let $m \geq 2$ and $1 \leq p < \infty$. Let $\ast $ denote the one point metric measure space $(M_1, d_1, \mu_1)$. We have that $\sketch^S_{1,p}(\Delta_m) = \inf_{(M_1, d_1, \mu_1)} \zeta_p(\Delta_m, M_1) = \zeta_p(\Delta_m, \ast) = \inf_{d \in \mathcal{D}(\Delta_m,\ast)} \left(\sum_{x \in \Delta_m} \frac{d^p(x,\ast)}{m} \right)^{1/p}$. Using Minkowski's inequality, we obtain that
\begin{align*}
(m-1) \left( \sum_{x \in \Delta_m} \frac{d^p(x, \ast) }{m} \right)^{1/p} &= 
\left( \sum_{x \in \Delta_m} \frac{d^p(x, \ast) }{m} \right)^{1/p} + \cdots + \left( \sum_{x \in \Delta_m} \frac{d^p(x, \ast) }{m} \right)^{1/p} \\
&\geq \frac{1}{m^{1/p}} \cdot \left( \sum_{x \in \Delta_m} \left(d(x, \ast) + \cdots + d(x, \ast) \right)^p \right)^{1/p}.
\end{align*}

Further analysis is divided into two cases.
\begin{itemize}
\item \textbf{Case 1} - $m$ is odd. 

Since $m-1$ is even, we combine the terms in the above inequality into pairs, such that for all $x,x' \in \Delta_m$ with $x \neq x'$ appearing in the sum, we use the triangle inequality 
$d(x,\ast) + d(x', \ast) \geq d(x,x') = 1 $, $\frac{m(m-1)}{2} $ times. Precisely, if we denote the elements of $\Delta_m$ by $x_1, x_2, \ldots, x_m$, 
then we have the following sum.
\begin{align*}
&(d(x_1, \ast) + d(x_2,\ast) + \ldots + d(x_{m-1},\ast))^p + (d(x_2, \ast) + d(x_3,\ast) + \ldots + d(x_m, \ast))^p + \\
&\ldots + (d(x_m, \ast) + d(x_1, \ast) + \ldots + d(x_{m-2}, \ast))^p.
\end{align*}

On applying the triangle inequality, we obtain the following:
$$ (m-1) \left( \sum_{x \in \Delta_m} \frac{d^p(x, \ast) }{m} \right)^{1/p} \geq \frac{1}{m^{1/p}} \left( m \left( \frac{m-1}{2} \right)^p \right)^{1/p} = \frac{m-1}{2}. $$
This implies that $\left( \sum_{x \in \Delta_m} \frac{d^p(x, \ast) }{m} \right)^{1/p} \geq \frac{1}{2} $. This proves the desired lower bound for the case when $m$ is odd.

\item \textbf{Case 2} - $m$ is even.

We know that for $x,x' \in \Delta_m$ with $x \neq x'$, we have that $d(x, \ast) + d(x', \ast) \geq d(x,x') = 1$. This implies that we cannot have both $d(x, \ast) < \frac{1}{2}$ and $d(x', \ast) < \frac{1}{2}$ simultaneously. Again, let $\Delta_m = \{x_1, x_2, \ldots, x_m \} $. Then, we conclude that out of the $m$ terms $d(x_1, \ast), d(x_2, \ast), \ldots, $ $d(x_m,\ast) $, we have $d(x_i, \ast) < \frac{1}{2}$ only for a unique $i \in [m]$. Since $m$ is even, the expression $\left( \sum_{x \in \Delta_m} \left(d(x, \ast) + \cdots + d(x, \ast) \right)^p \right)^{1/p} $ has an odd number of terms in the sum $\left(d(x, \ast) + \cdots + d(x, \ast) \right)^p $. We rearrange these terms in such a way that for $x,x' \in \Delta_m$ with $x \neq x' $, we have $\frac{m-2}{2} $ pairs $d(x,\ast), d(x',\ast)$, satisfying $d(x, \ast) + d(x',\ast) \geq d(x,x') = 1 $, and the remaining term satisfies $d(x, \ast) \geq \frac{1}{2}$. In particular, if we assume that the term $d(x_1, \ast) < \frac{1}{2} $, then we obtain the following sum.
\begin{align*}
&((d(x_1, \ast) + d(x_2, \ast) + \ldots + d(x_{m-2},\ast)) + d(x_{m-1},\ast)^p + ((d(x_2, \ast) + d(x_3, \ast) + \ldots +\\
&d(x_{m-1},\ast)) + d(x_m,\ast))^p + \ldots + ((d(x_m,\ast) + d(x_1,\ast) + \ldots + d(x_{m-3},\ast)) + d(x_{m-2},\ast))^p.
\end{align*}

By applying triangle inequality on $m-2$ terms inside the inner brackets, we obtain the following:
$$ (m-1) \left( \sum_{x \in \Delta_m} \frac{d^p(x, \ast) }{m} \right)^{1/p} \geq \frac{1}{m^{1/p}} \left(m \left( \frac{m-2}{2} + \frac{1}{2} \right)^p \right)^{1/p} = \frac{m-1}{2}. $$
Thus, we obtain that $\left( \sum_{x \in \Delta_m} \frac{d^p(x, \ast) }{m} \right)^{1/p} \geq \frac{1}{2} $. This proves the lower bound for the case when $m$ is even.
\end{itemize}
Thus, using the lower bound obtained above and the upper bound obtained in Lemma \ref{lem:etabound}, we obtain that for all $p \geq 1$, $\sketch^S_{1,p}(\Delta_m) = \frac{1}{2}$. Since $\shatter_{1,p,p}(\Delta_m) = \left( 1- \frac{1}{m} \right)^{1/p} $ (Lemma \ref{lem:qbound}), we obtain $\frac{\sketch_{1,p}^S(\Delta_m)}{\shatter_{1,p,p}(\Delta_m)} = \frac{1}{2(1-m^{-1})^{1/p}}. $
\end{proof}

\subsection{Approximation results for $\sketch^S_{k,p}$ for finite $p$} \label{sec:sketch-s-approx}

We now use the duality between $\sketch^S_{k,p}$ and $\shatter_{k,p,p}$, that we established in the last section, to obtain approximation results for $\sketch^S_{k,p}$ via the approximation results for $\shatter_{k,p,p}$. Precisely, we show the following result.

\begin{theorem}[Approximation results for $\sketch^S_{k,p}$] \label{thm:sketch-s-approx}
For any $0 < \epsilon < 1$ and $t \in \mathbb{Z}$ such that $k \geq t > 1$, there is an $f(p)$-approximation for $\sketch^S_{k,p}(X)$, where $f(p)$ is as follows:
\begin{enumerate}
\item For $p = 1$, $f(p) = 12 + \frac{8}{t} + \epsilon $.
\item For $p = 2$, $f(p) = 20 + \frac{16}{t} + \epsilon $.
\item For all reals $p > 2$, $f(p) = \left(12 + \frac{8}{t} \right)p + \epsilon $. 
\end{enumerate}
The running time of the approximation algorithm is $|X|^{O(t)} \cdot \epsilon^{-1}$.
\end{theorem}

We need the following preliminaries in order to prove the above theorem.
\begin{definition}[\textbf{$p$-radius}]
For $p\in[1,\infty)$, given $X \in \mathcal{M}_w$ and a subset $B \subseteq X$, we define the $p$-radius of $B$ as
$$ \mathrm{rad}_{p}(B) := \inf_{a \in B} \left( \int d_{X}^{p}(a,x) \frac{d \mu_{X}(x)}{\mu_{X}(B)} \right)^{1/p}. $$
For $p = \infty$, we define
$$ \mathrm{rad}_{\infty}(B) = \inf_{a \in B} \sup_{x \in B} d_{X}(a,x).$$
\end{definition}

The following theorem establishes the relationship between $p$-radius and $p$-diameter for a metric measure space.

\begin{theorem} \label{thm:radiusdiaminequality} 
For all $p\in [1,\infty]$, 
given a metric measure space $(X, d_{X}, \mu_{X})$ and subset $B \subseteq X$,
$$ 2 \, \mathrm{rad}_{p}(B) \geq \mathrm{diam}_{p}(B) \geq \mathrm{rad}_{p}(B). $$
\end{theorem}

\begin{proof} 
Let us first consider the case of finite $p$.
For $B \subseteq X$, we have $$\mathrm{diam}_{p}(B) = \left( \iint \displaylimits_{B \times B} d_{X}^{p}(x,x') \frac{d \mu_{X}(x) \, d \mu_{X}(x')}{\big(\mu_{X}(B)\big)^2} \right)^{1/p}.$$ For any $a \in X$, we have
\begin{align*}
2 \left( \int_{B} d_{X}^{p}(a,x) \frac{d \mu_{X}(x)}{\mu_{X}(B)} \right)^{1/p} &= \left( \int_{B} d_{X}^{p}(a,x) \frac{d \mu_{X}(x)}{\mu_{X}(B)} \right)^{1/p} + \left( \int_{B} d_{X}^{p}(a,x') \frac{d \mu_{X}(x')}{\mu_{X}(B)} \right)^{1/p} \\
&= \left( \iint \displaylimits_{B \times B} d_{X}^{p}(a,x) \frac{d \mu_{X}(x) \, d \mu_{X}(x')}{\big(\mu_{X}(B)\big)^2} \right)^{1/p} + \left( \iint \displaylimits_{B \times B} d_{X}^{p}(a,x') \frac{d \mu_{X}(x')\, d \mu_{X}(x)}{\big(\mu_{X}(B)\big)^2} \right)^{1/p} \\
& \geq \left( \iint \displaylimits_{B \times B} d_{X}^{p}(x,x') \frac{d \mu_{X}(x) \, d \mu_{X}(x') }{\big(\mu_{X}(B)\big)^2} \right)^{1/p} \\
&= \mathrm{diam}_{p}(B).
\end{align*}
This gives that $2 \, \mathrm{rad}_{p}(B) \geq \mathrm{diam}_{p}(B)$. For the other inequality, we observe that
\begin{align*}
\mathrm{diam}^{p}_{p}(B) &= \iint \displaylimits_{B \times B} d_{X}^{p}(x,x') \frac{d \mu_{X}(x) \, d \mu_{X}(x')}{\big(\mu_{X}(B)\big)^2} \\ 
&= \int_{B} \int_{B} d_{X}^{p}(x,x') \frac{d \mu_{X}(x) \, d \mu_{X}(x')}{\big(\mu_{X}(B)\big)^2} \\
&= \int_{B} \frac{d \mu_{X}(x')}{\mu_{X}(B)} \int_{B} d_{X}^{p}(x,x') \frac{d \mu_{X}(x)}{\mu_{X}(B)} \\
& \geq \int_{B} \mathrm{rad}^{p}_{p}(B) \frac{d \mu_{X}(x')}{\mu_{X}(B)} \\
&= \mathrm{rad}^{p}_{p}(B).
\end{align*}
Thus, we obtain $\mathrm{diam}_{p}(B) \geq \mathrm{rad}_{p}(B) $.

It remains to consider the case $p=\infty$. We recall that for any $X \in \mathcal{M}_{w}$ and $B \subseteq X$, we have
$$ \diam_{\infty}(X) = \sup_{x,x' \in X}d_{X}(x,x') \quad \mathrm{and} \quad \diam_{\infty}(B) = \sup_{x,x' \in B} d_{X}(x,x'). $$ 
For any $a \in B$, we have
$$ \diam_{\infty}(B) = \sup_{x,x' \in B}d_{X}(x,x') \leq \sup_{x,x' \in B} \left( d_{X}(x,a) + d_{X}(a,x') \right) = \sup_{x \in B}d_{X}(x,a) + \sup_{x' \in B}d_{X}(a,x'). $$
Since the above inequality holds for any $a \in B$, we have that 
$$ \diam_{\infty}(B) \leq \inf_{a \in B} \left( \sup_{x \in B}d_{X}(x,a) + \sup_{x' \in B}d_{X}(a,x') \right) = 2 \, \rad_{\infty}(B). $$
The reverse inequality holds because
$$ \rad_{\infty}(B) = \inf_{a \in B} \sup_{x \in B} d_{X}(a,x) \leq \sup_{a \in B} \sup_{x \in B} d_{X}(a,x) = \diam_{\infty}(B). $$
\end{proof}

We now define an analogue of $\shatter_{k,p,q}(X)$, replacing the diameter in its definition by radius. We denote this new objective by $\shatter^{\rad}_{k,p,q}(X) $.

\begin{definition}[$\shatter^{\rad}_{k,p,q}$]
For $X \in \mathcal{M}_{w}$, $k \in \mathbf{N}$ and $p,q \in \mathbf{R}$ with $1 \leq p < \infty$ and $1 \leq q < \infty$, define
$$ \shatter^{\rad}_{k,p,q}(X) = \min_{\{B_i\}_{i=1}^k\in\mathrm{Part}_{k}(X)} \left( \sum_{i=1}^k \mathrm{rad}_{p}^{q}(B_{i}) \,\mu_{X}(B_{i}) \right)^{1/q}. $$
For $1 \leq p \leq \infty$ and $q = \infty$, define
$$ \shatter^{\rad}_{k,p,\infty}(X) = \min_{\{B_i\}_{i=1}^k\in\mathrm{Part}_{k}(X)} \max_{i} \rad_{p}(B_{i}).$$
\end{definition} 

Now, a direct corollary of Theorem \ref{thm:radiusdiaminequality} is the following:
\begin{corollary} \label{cor:shatterrad}
For every $X \in \mathcal{M}_{w}$, for all $p\in[1,\infty]$ and for all $k \in \mathbf{N}$, we have 
$$\shatter^{\rad}_{k,p,p}(X) \leq \shatter_{k,p,p}(X) \leq 2\cdot \shatter^{\rad}_{k,p,p}(X).$$
\end{corollary}

\subsection{Proof of Theorem \ref{thm:sketch-s-approx} }
We first define some functions required for the proofs in this section. Fix $X \in \mathcal{M}_w $ with $|X| = n$. For non-empty $C \subseteq X$, define
$$ ||C||_{p} := \left(\sum_{x \in X} d^{p}(x,C) \mu_{X}(x) \right)^{1/p}. $$

For $1 \leq p < \infty $, define
$$\mathrm{opt}_{p}(X) := \min_{C \subseteq X, |C| = k} ||C||_p, $$
and 
$$ C_{p}(X) := \mathrm{argmin}_{C \subseteq X, |C| = k} ||C||_p. $$

For $p = \infty$, define
$$ \mathrm{opt}_{\infty}(X) := \min_{C \subseteq X, |C| = k} \max_{x \in X} d(x,C),  $$
and 
$$ C_{\infty}(X) := \mathrm{argmin}_{C \subseteq X, |C| = k} \max_{x \in X} d(x,C). $$

\begin{Remark} \label{rem:kmed}
The $k$-median objective \cite{har2011geometric} is the same as $\mathrm{opt}_1(X)$ and the $k$-means objective \cite{har2011geometric} is the same as $\mathrm{opt}_2(X)$.
\end{Remark}

The next lemma shows that the objective $\mathrm{opt}_p(X)$ defined above equals $\shatter^{\rad}_{k,p,p}(X)$ for all $1 \leq p \leq \infty$.

\begin{lemma} \label{lem:Qhatbounds}
For $1 \leq p \leq \infty$, any finite $X \in \mathcal{M}_{w}$ and any $k \in \mathbf{N}$, we have $$\shatter^{\rad}_{k,p,p}(X) = \mathrm{opt}_{p}(X).$$
\end{lemma}

\begin{proof}
We first assume that $1 \leq p < \infty$. For a finite metric space $X$, we have $\mathrm{rad}_{p}(X) = \mathrm{inf}_{a \in X}  \left( \sum_{x \in X} d^{p}(a,x) \mu_{X}(x) \right)^{1/p}$. Similarly for $B \subseteq X$, we have $\mathrm{rad}_{p}(B) = \mathrm{inf}_{a \in B} \left( \sum_{x \in B} d^{p}(a,x) \frac{\mu_{X}(x)}{\mu_{X}(B)} \right)^{1/p} $. 
Let $\{B_{1} \ldots, B_{k} \}$ be a partition of $X$ such that
$$ \shatter^{\rad}_{k,p,p}(X) = \left( \sum_{i=1}^{k} \rad_{p}^{p}(B_{i}) \mu_{X}(B_{i}) \right)^{1/p}. $$
From the above definitions, we have that
$$ \sum_{i=1}^{k} \mathrm{rad}^{p}_{p}(B_{i}) \mu_{X}(B_{i}) = \sum_{i=1}^{k} \sum_{x \in B_{i}} d^{p}(a_{i},x) \frac{\mu_{X}(x)}{\mu_{X}(B_{i})} \mu_{X}(B_{i}) = \sum_{i=1}^{k} \sum_{x \in B_{i}} d^{p}(a_{i},x) \mu_{X}(x).  $$
where for every $i \in [k]$, $a_{i} \in B_{i}$ is such that $ \mathrm{rad}_{p}(B_{i}) = \left( \sum_{x \in B_{i}} d^{p}(a_{i},x) \frac{\mu_{X}(x)}{\mu_{X}(B)} \right)^{1/p} $. This implies that
$$\shatter^{\rad}_{k,p,p}(X) := \left( \sum_{i=1}^{k} \sum_{x \in B_{i}} d^{p}(a_{i},x) \mu_{X}(x) \right)^{1/p}.  $$
For every $i \in [k]$, let $c_{i} = a_{i}$. Let $C = \{ c_{1}, \ldots, c_{k} \}$. Then, for any $i \in [k]$ and any $x \in B_{i}$, we have $d_{X}(x,C) \leq d_{X}(x,a_{i})$. This implies that
$$ \sum_{x \in X}d^{p}(x,C) \mu_{X}(x) = \sum_{i=1}^{k} \sum_{x \in B_{i}}d^{p}(x, C) \mu_{X}(x) \leq \sum_{i=1}^{k} \sum_{x \in B_{i}} d^{p}(a_{i}, x) \mu_{X}(x).  $$
Thus, we obtain that $\mathrm{opt}_{p}(X) \leq \shatter^{\rad}_{k,p,p}(X) $. 

We now prove that $\shatter^{\rad}_{k,p,p}(X) \geq \mathrm{opt}_{p}(X)$. Let $C^\ast = \{c_{1}, \ldots, c_{k} \}\in C_{p}(X)$ and $\{B_{1}, \ldots, B_{k} \}$ be a Voronoi partition of $X$ with respect to $C^\ast$. Then, for all $i \in [k]$, $c_{i} \in B_{i}$. For every $i \in [k]$, let $a_{i} \in B_{i}$ be such that $\mathrm{rad}_{p}(B_{i}) = \left( \sum_{x \in B_{i}} d^{p}(a_{i},x) \frac{\mu_{X}(x)}{\mu_{X}(B)} \right)^{1/p}$. Then, we have that
$$ \sum_{x \in X}d^{p}(x, C^\ast) \mu_{X}(x) = \sum_{i=1}^{k} \sum_{x \in B_{i}} d^{p}(x,c_{i}) \mu_{X}(x) \geq \sum_{i=1}^{k} \sum_{x \in B_{i}} d^{p}(x, a_{i}) \mu_{X}(x) \geq \left(\shatter^{\rad}_{k,p,p}(X) \right)^{p}. $$
Thus, we have established that $\shatter^{\rad}_{k,p,p}(X) = \mathrm{opt}_{p}(X)$. Therefore, a $c$-approximation algorithm for $\mathrm{opt}_{p}(X)$ is also a $c$-approximation for $\shatter^{\rad}_{k,p,p}(X)$. The same result holds for $p = \infty$.
\end{proof}

We use the previous lemma to prove Theorem \ref{thm:sketch-s-approx}. We first describe the approximation algorithm for $\sketch^S_{k,1} $.
\subsubsection{The case $p = 1$}

For $p = 1$, Charikar and Guha \cite{CG99} gave a polynomial time $4$-approximation algorithm for $\mathrm{opt}_{1}(X)$. For the case when $X$ has uniform measure, the best known approximation factor for $\mathrm{opt}_{1}(X)$ is $(3+ \epsilon)$ for any $\epsilon > 0$. This  approximation factor is due to Arya et al.~\cite{arya2004local} and is obtained from a local search algorithm. We describe this local search method in the next paragraph.

Let $K$ be a set of centers that is an $\alpha$-approximation for $\mathrm{opt}_{1}(X)$ and let $C \in C_1(X)$. Since $K$ is an $\alpha$-approximation, we have $||K||_{1} \leq \alpha ||C||_{1}$. We start with this set $K$ and perform a local search as follows: we choose an arbitrary $\epsilon \in (0,1)$ and set $\tau = \frac{\epsilon}{10 k}$, $k = |K|$. We fix these values of $\epsilon$ and $\tau$. We now choose an arbitrary $p \in X \setminus K$ and an arbitrary $r \in K$. We consider the new set of centers obtained by swapping $r$ by $p$. In particular we define $K_{new} = K \cup \{p \} \setminus \{r \}$. Now if $||K_{new}||_{1} \leq (1 - \tau) ||K||_{1}$, then we set $K = K_{new}$; otherwise $K$ remains the same. We then consider another swap and repeat the same procedure. There are $O(nk)$ possible swaps since $|X \setminus K| = n-k$ and $|K| = k$. We consider all such swaps. We stop when we obtain a set of centers that can no longer be improved by any swap. Let $L$ denote this locally optimal set of centers. 

We now analyze the running time of this algorithm. There are $O(nk)$ possible swaps for a fixed set of centers $K$. For every swap, computation of $||K_{new}||_{1}$ requires $O(nk)$ time, since we have to calculate the distance of every $p \in X \setminus K_{new}$ to its closest point in $K_{new}$. Since $\frac{1}{1 - \tau} \geq 1 + \tau $, we obtain that the running time of the local search algorithm is 
$$O \left((nk)^2 \log_{1/(1-\tau)} \frac{||K||_{1}}{||C||_{1}} \right) = O \left( (nk)^2 \log_{1+ \tau} \alpha \right) = O \left((nk)^{2} \frac{\log \alpha}{ \tau} \right). $$

In the above algorithm, we replace one center from a locally optimal solution $K$ with an arbitrary center from $X \setminus K$. Another method of performing local search is to simultaneously replace an arbitrary set of $t$ centers from $K$ with an arbitrary set of $t$ centers from $X \setminus K$. This method is called the \textit{$t$-swap method}. Since the algorithm involves checking all subsets of both $X$ and $K$ of size $t$, its running time is $n^{O(t)} \epsilon^{-1} $. Note that the running time is polynomial for constant values of $t$ and for polynomially small $\epsilon$. 

The $t$-swap local search algorithm has been analyzed by Arya et al.~in \cite{arya2004local} and also by Gupta and Tangwongsan in \cite{GT08}. Both papers give the same approximation factor of $\left( 3 + \frac{2}{t} + \epsilon \right)$ but for the case when measure on $X$ is uniform. Here, we consider the simpler analysis of Gupta and Tangwongsan and observe that their analysis works for any arbitrary measure on $X$. When measure on $X$ is non-uniform, we modify their cost function so as to include the measures.  

Thus, we have that for any $0 < \epsilon < 1$ and $k > t > 1$, there is a $\left( 3 + \frac{2}{t} + \epsilon \right)$-approximation algorithm for $\mathrm{opt}_{1}(X)$  with running time $|X|^{O(t)} \epsilon^{-1} $. From Lemma \ref{lem:Qhatbounds}, we obtain that this gives a $ \left( 3 + \frac{2}{t} + \epsilon \right)$-approximation for $\shatter^{\rad}_{k,1,1}(X)$. Using Corollary \ref{cor:shatterrad}, we obtain a $\left( 6 + \frac{4}{t} + \epsilon \right)$-approximation of $\shatter_{k,1,1}(X)$. Let $P$ be a partition of $X$ that achieves this $\left( 6 + \frac{4}{t} + \epsilon \right)$-approximation of $\shatter_{k,1,1}(X)$. We apply the Wasserstein map (Definition \ref{def:Wasserstein-map}) to $P$ to obtain a $k$-point metric measure space. From Remark \ref{rem:GWp}, we have that the $k$-point metric space obtained is a $\left( 12 + \frac{8}{t} + \epsilon \right)$-approximation for $\sketch^{S}_{k,1}(X)$, We note that computing the $k$-point metric measure space is polynomial time, since the computation involves calculating $1$-Wasserstein distance between blocks of partition $P$, and this requires polynomial time \cite{Orl97}.

\subsubsection{The case $2 \leq p < \infty$ }

We now describe the approximation algorithm for $2 \leq p < \infty $. We first need few definitions.  

\begin{definition}[\textbf{$\lambda$-approximate metric} \cite{MP03}]
Given $\lambda \geq 1$, a non-negative and symmetric function $d$ defined on a set $M$ is called a $\lambda$-approximate metric if, for any sequence of points $ \left\langle x_0, \ldots, x_m \right\rangle$ in $M$, we have
$ d(x_0, x_m) \leq \lambda \cdot \sum_{0 \leq i \leq m} d(x_{i}, x_{i+1}). $
\end{definition}

\begin{definition}[\textbf{Weakly $\lambda$-approximate metric} \cite{MP03}]
Given $\lambda \geq 1$, a non-negative and symmetric function $d$ defined on a set $M$ is called a weakly $\lambda$-approximate metric if, it satisfies the following inequality: for any $x,y,z \in M$, we have
$ d(x,z) \leq \lambda(d(x,y) + d(y,z)). $
The above inequality is a weaker form of the triangle inequality.
\end{definition}

For example, if we consider the distance function $d^{p}$ on our metric space $(X,d)$, where $1 < p < \infty$, then we have that for any $x,y,z \in X$, 
$$ d^{p}(x,y) \leq (d(x,z) + d(z,y))^{p} \leq 2^{p-1}(d^{p}(x,z) + d^{p}(z,y)). $$
Thus, $d^p$ is a weakly $2^{p-1}$- approximate metric. 

We refer to the work of Mettu and Plaxton \cite{MP03} in this paragraph. They provide a constant factor approximation algorithm for the $k$-median problem (Remark \ref{rem:kmed}) on spaces endowed with a weakly $\lambda$-approximate metric. Let the input metric space be $U$ and $n$ be the cardinality of $U$. Let $l$ be the ratio of the diameter of $U$ to the shortest distance between any pair of distinct points in $U$. We note that the length of any sequence $\sigma_{i}$ as described in Section $3.1$ of \cite{MP03} is at most $O(\log l)$. From Lemma $4.1$ of \cite{MP03}, we have that if $U$ is a weakly $\lambda$-approximate metric then for points $x_{0}, x_{1}, \ldots, x_{m}$ in $U$ with $m \geq 1$, $d(x_0,x_m) \leq \lambda^{\lceil \log_2 m \rceil} \sum_{0 \leq i \leq m} d(x_i, x_{i+1})$. Since all sequences used in proving Theorem $3.1$ of \cite{MP03} have length $O(\log l)$, we obtain an approximation factor of $\lambda^{O(\log_2 \log l)}$ for the $k$-median problem on a weakly $\lambda$-approximate metric. Since $d^p$ is a weakly $2^{p-1}$-approximate metric, we obtain a $(\log l)^{O(p-1)}$-approximation for $\mathrm{opt}_{p}(X)$ .

Let $K$ denote the set of centers obtained from this approximation algorithm. We now perform local search on $K$ using the $t$-swap method for $t \geq 1$, as done in the previous section for $p=1$. This method has been analyzed for the metric $d^{p}$ by Gupta and Tangwongsan in \cite{GT08}. The problem studied by Gupta and Tangwongsan in \cite{GT08} is the $l^p$-facility location problem. The objective of this problem is $\mathrm{opt}_p(X)$, but with uniform measure on $X$. For this problem, they obtain the following result:

\begin{theorem} \cite[Theorem 3.5]{GT08} \label{thm:GTapprox}
For any value of $t \in \mathbf{Z}_{+}$, the natural $t$-swap local-search algorithm for the $l_{p}$-facility location problem yields the following guarantees:

\begin{enumerate}
\item For $p = 2$, it is a $\left( 5 + \frac{4}{t} \right)$ approximation.
\item For all reals $p \geq 2$, it is a $\left( 3+\frac{2}{t} \right)p$ approximation.
\end{enumerate}
\end{theorem}

The approximation factors obtained above are for $\mathrm{opt}_p(X)$ when the measure on $X$ is uniform. However, the analysis of Theorems $3.2$ and $3.5$ of \cite{GT08} also works when $X$ has non-uniform measure. Therefore, we can perform the same analysis on the locally optimal solution obtained after performing local search on $K$ and obtain the following result: 

\begin{lemma} \label{lem:optapprox}
For any $t \in \mathbf{Z}_{+}$ and any $0 < \epsilon < 1$, there is an $f(p)$-approximation for $\mathrm{opt}_{p}(X)$ where $f(p)$ is as follows:
\begin{enumerate}
\item For $p = 2$, $f(p) = 5 + \frac{4}{t} + \epsilon$.
\item For all reals $p > 2$, $f(p) = \left( 3 + \frac{2}{t} \right)p + \epsilon$.
\end{enumerate}
The running time of the approximation algorithm is $|X|^{O(t)} \epsilon^{-1}$.
\end{lemma}

From Lemma \ref{lem:Qhatbounds} and Lemma \ref{lem:optapprox}, we obtain an $f(p)$-approximation for $\shatter^{\rad}_{k,p,p}(X)$. Using Corollary \ref{cor:shatterrad}, we obtain a $2f(p)$-approximation for $\shatter_{k,p,p}(X)$. Let $P$ be a partition of $X$ that achieves this $2f(p)$-approximation of $\shatter_{k,p,p}(X)$. We apply the Wasserstein map (Definition \ref{def:Wasserstein-map}) to $P$ to obtain a $k$-point metric measure space. From Remark \ref{rem:GWp}, we have that the $k$-point metric space obtained is a $4f(p)$-approximation of $\sketch^{S}_{k,p}(X)$, We note that computing the $k$-point metric measure space is polynomial time, since the computation involves calculating $p$-Wasserstein distance between blocks of partition $P$, and this requires polynomial time \cite{Orl97}.

\section{Relating the weak and strong sketching objectives for mm-spaces}\label{sec:doubling}
In this section, we show a relation between $\sketch_{k,p} $ and $\sketch^S_{k,p}$ for a specific class of metric measure spaces. In particular, without imposing some control on the class of mm-spaces we consider, there is no hope to obtain a comparability result between $\sketch_{k,p}$ and $\sketch^S_{k,p}$. This is demonstrated by the following example. 

\begin{example}[Blow-up of $\frac{\sketch^S_{k,1}(\Delta_m)}{\sketch_{k,1}(\Delta_m)}$] \label{eg:blowup}
 We know from the proof of Theorem \ref{thm:sketch-shatter-deltam} that for all $k,m \in \mathbf{N}$ with $k \leq m$ and $p \geq 1$, we have
$$\sketch^{S}_{k,p}(\Delta_m) \geq \frac{1}{2} \left(1- \frac{k}{m} \right)^{1/p}.$$
 Claim $5.1$ of \cite{Memoli2011} gives $d_{GW_p}(\Delta_m, \Delta_k) \leq \frac{1}{2} \left(\frac{1}{k} + \frac{1}{m} \right)^{1/p}$ which implies  $\sketch_{k,p}(\Delta_m) \leq \frac{1}{2} \left(\frac{1}{k} + \frac{1}{m} \right)^{1/p} $. Therefore, we obtain 
$$ \frac{\sketch^S_{k,1}(\Delta_m)}{\sketch_{k,1}(\Delta_m)} \geq \frac{1- \frac{k}{m}}{\frac{1}{k} + \frac{1}{m}}. $$

For $m=2^{n+1}$ and $k=2^n$, we obtain
\begin{equation}\label{eq:blowup} \frac{\sketch^S_{k,1}(\Delta_m)}{\sketch_{k,1}(\Delta_m)} \geq  \frac{m}{6}. 
\end{equation}
\end{example}

We observe that as $m \rightarrow \infty$, the right hand side of the above inequality goes to $\infty$. In what follows, we show that for a rich family of metric measure spaces, $\sketch^S_{k,p}(X)$ is bounded above by a suitable function of $\sketch_{k,p}(X)$. 

\begin{definition}[\textbf{Doubling mm-spaces}]\cite[Chapter 1]{Heinonen01} 
A metric measure space $(X, d_{X}, \mu_{X})$ is called doubling if there exists a constant $C \geq 1$ such that for all $x \in X$ and $r > 0$, we have
$$ \mu_{X}(B_{X}(x, 2r)) \leq C \cdot \mu_{X}(B_{X}(x, r)). $$
The constant $C$ is called the doubling constant of $X$.
\end{definition}

Note that for any natural number $m \geq 2$, the space $\Delta_m \in \mathcal{M}_w$ has doubling constant $m$, i.e. \facundo{it} is not bounded independently of $m$. 

We now invoke some results from Section $5$ of \cite{Memoli2011}. Let $(X, d_{X}, \mu_{X}) \in \mathcal{M}_{w}$. Given $\delta > 0$, for $\epsilon\geq 0$, define $$f_{\delta}(\epsilon) := \mu_{X}(\{x \in X~|~\mu_{X}(B_{X}(x,\epsilon)) \leq \delta\}).$$ We now define $v_{\delta}(X)$ as follows:
$v_{\delta}(X) := \inf \{\epsilon > 0~|~ f_{\delta}(\epsilon) \leq \epsilon \}. $
Note that $v_{\delta}(X)$ is an increasing function of $\delta$ i.e.~if $\delta_{1} \leq \delta_{2}$ then $v_{\delta_{1}}(X) \leq v_{\delta_{2}}(X)$. 

We now have the following result:

\begin{theorem}[\cite{Memoli2011}] \label{thm:weak-sturmrel}
Let $X,Y \in \mathcal{M}_{w}$, $p \in [1, \infty)$ and $\delta \in (0, 1/2) $. Then
$$ \zeta_{p}(X, Y) \leq (4 \cdot \min(v_{\delta}(X), v_{\delta}(Y)) + \delta)^{1/p} \cdot M,$$
whenever $\dgwp(X, Y) < \delta^{5}$, where $M = 2 \cdot \mathrm{max}(\mathrm{diam}(X), \mathrm{diam}(Y)) + 45$.
\end{theorem}

Let $\mathcal{F} \subset \mathcal{M}_{w}$ be a family for which there exists a surjective function $\rho_{\mathcal{F}}: [0, \infty) \rightarrow [0, \delta_{\mathcal{F}}]$ with $\delta_{\mathcal{F}} > 0$, satisfying the following condition: for all $\epsilon \geq 0, x \in X$ and $X \in \mathcal{F}$, $ \mu_{X}(B_{X}(x,\epsilon)) \geq \rho_{\mathcal{F}}(\epsilon).$ Then, for all $\delta \in (0, \delta_{\mathcal{F}})$, we have
$ \sup_{X \in \mathcal{F}} v_{\delta}(X) \leq \inf \{ \epsilon > 0~|~ \rho_{\mathcal{F}}(\epsilon) > \delta \}. $ \footnote{This is because for a fixed $\delta \in (0, \delta_{\mathcal{F}})$, if $\epsilon \geq 0$ is such that $\rho_{\mathcal{F}}(\epsilon) > \delta$, then for all $x \in X$, $X \in \mathcal{F}$, we have $\mu_{X}(B_{X}(x, \epsilon)) > \delta$. This implies that $f_{\delta}(\epsilon) = 0$ and $v_{\delta}(X) \leq \epsilon$. Since $\epsilon > 0$ was arbitrary, we get the above inequality.}

If $\mathcal{F} \subset \mathcal{M}_{w}$ is the set of all metric measure spaces with doubling dimension $C$, then we set $\rho_{\mathcal{F}}(\epsilon) = \left( \dfrac{\epsilon}{2D} \right)^{N} $, where $D = \diam_{\infty}(X)$ and $N = \log_{2} C$. We observe that if $\rho_{\mathcal{F}}$ is an increasing function then given $\delta > 0$, the quantity $\inf \{\epsilon>0~|~\rho_{\mathcal{F}}(\epsilon) > \delta \}$ is equal to $\rho_{\mathcal{F}}^{-1}(\delta)$. Since $\rho_{\mathcal{F}}(\epsilon) = \left( \dfrac{\epsilon}{2D} \right)^{N}$ is an increasing function, we obtain that
$ \inf \{\epsilon > 0~|~ \rho_{\mathcal{F}}(\epsilon) > \delta \} = \rho_{\mathcal{F}}^{-1}(\delta) = 2D \delta^{1/N}. $

Therefore, we conclude that for all doubling metric measure spaces $X$ with doubling constant $C>0$, $v_{\delta}(X) \leq 2 \cdot \diam(X) \cdot \delta^{1/\log_{2}C}$. We are now ready to prove Theorem \ref{thm:doubling}.

\begin{proof}[Proof of Theorem \ref{thm:doubling}]
Let $\sketch_{k,p}(X) < \delta'^{5}$, where $\delta' \in \left(0, \frac{1}{2} \right)$. Then, there exists $(M_{k}, d_{k}, \mu_{k}) \in \mathcal{M}_{k,w}$ such that $\dgwp(X, M_{k}) < \delta'^5$. Due to the following claim, we may assume that $\diam(M_{k}) \leq \diam(X)$. 

\begin{claim}
Given $X \in \mathcal{M}_{w}$, let $(M_{k}, d_{k}, \mu_{k}) \in \mathcal{M}_{k,w}$ be such that $\diam(M_k) > \diam(X)$. Then, there exists $(M'_k,d'_k,\mu'_k) \in \mathcal{M}_{k,w}$ such that $\dgwp(X,M'_k) \leq \dgwp(X,M_k)$ and $\diam(M_{k}) \leq \diam(X)$.
\end{claim}
\begin{proof}[Proof of Claim]
We consider the $k$-point space $(M'_{k}, d'_{k}, \mu'_{k})$ defined as follows: for every $i \in [k]$, $\mu'_{k}(i) = \mu_{k}(i)$ and
$$ d'_{k}(i,j) = \min (d_{k}(i,j), \diam(X)) \quad \forall~i,j \in [k] .$$
Let $S \subseteq [k] \times [k]$ be such that for all $(i,j) \in S$, $d_{k}(i,j) \neq d'_{k}(i,j)$. This means that for all $(i,j) \in S$, $d_{k}(i,j) > \diam(X)$ and therefore $d'_{k}(i,j) = \diam(X)$. Let $\overline{S} = [k] \times [k] \setminus S$. Then, for any measure coupling $\mu$ between $\mu_{X}$ and $\mu_{k}$, we have 
\begin{align*}
\iint \displaylimits_{(X \times M'_{k}) \times (X \times M'_{k})} & |d_{X}(x,x') - d'_{k}(i,j)|^{p} d \mu(x, i) d \mu(x', j)\\
 &= \iint \displaylimits_{(X \times S) \times (X \times S)} |d_{X}(x,x') - d'_{k}(i,j)|^{p} d \mu(x, i) d \mu(x', j) \\
&+ \iint \displaylimits_{(X \times \overline{S}) \times (X \times \overline{S})} |d_{X}(x,x') - d'_{k}(i,j)|^{p} d \mu(x, i) d \mu(x', j). 
\end{align*}
Note that
\begin{align*}
\iint \displaylimits_{(X \times \overline{S}) \times (X \times \overline{S})} & |d_{X}(x,x') - d'_{k}(i,j)|^{p} d \mu(x, i) d \mu(x', j)\\
 &= \iint \displaylimits_{(X \times \overline{S}) \times (X \times \overline{S})} |d_{X}(x,x') - d_{k}(i,j)|^{p} d \mu(x, i) d \mu(x', j).
\end{align*}
For the other integral, we have
\begin{align*}
\iint \displaylimits_{(X \times S) \times (X \times S)} & |d_{X}(x,x') - d'_{k}(i,j)|^{p} d \mu(x, i) d \mu(x', j) \\
&= \iint \displaylimits_{(X \times S) \times (X \times S)} |d_{X}(x,x') - \diam(X)|^{p} d \mu(x, i) d \mu(x', j) \\
&= \iint \displaylimits_{(X \times S) \times (X \times S)} \left(\diam(X) - d_{X}(x,x') \right)^{p} d \mu(x, i) d \mu(x', j) \\
&\leq \iint \displaylimits_{(X \times S) \times (X \times S)} \left(d_{k}(i,j) - d_{X}(x,x') \right)^{p} d \mu(x, i) d \mu(x', j) \\
&= \iint \displaylimits_{(X \times S) \times (X \times S)} |d_{k}(i,j) - d_{X}(x,x')|^{p} d \mu(x, i) d \mu(x', j).
\end{align*}
Therefore, we obtain that
\begin{align*}
\iint \displaylimits_{(X \times M'_{k}) \times (X \times M'_{k})} & |d_{X}(x,x') - d'_{k}(i,j)|^{p} d \mu(x, i) d \mu(x', j) \\
&\leq \iint \displaylimits_{(X \times S) \times (X \times S)} |d_{k}(i,j) - d(x,x')|^{p} d \mu(x, i) d \mu(x', j) \\
&+ \iint \displaylimits_{(X \times \overline{S}) \times (X \times \overline{S})} |d_{X}(x,x') - d_{k}(i,j)|^{p} d \mu(x, i) d \mu(x', j) \\
&= \iint \displaylimits_{(X \times M_{k}) \times (X \times M_{k})} |d_{X}(x,x') - d_{k}(i,j)|^{p} d \mu(x, i) d \mu(x', j).
\end{align*}
This implies that $\dgwp(X, M'_{k}) \leq \dgwp(X,M_{k})$ and by definition of $(M'_k,d'_k,\mu'_k)$, we have that $\diam(M'_{k}) \leq \diam(X)$.
\end{proof}
We continue with the proof of Theorem \ref{thm:doubling}. Now, we have that $\dgwp(X, M_{k}) < \delta'^{5}$ and $\diam(M_{k}) \leq \diam(X)$. We now use Theorem \ref{thm:weak-sturmrel} to obtain
$$ \zeta_{p}(X, M_{k}) \leq  (4 \cdot \min(v_{\delta'}(X), v_{\delta'}(M_k)) + \delta')^{1/p} \cdot M, $$
where $M = 2 \cdot \max (\diam(X), \diam(M_k)) + 45$. Since $X$ is a doubling metric measure space with doubling constant $C>0$, we have that $v_{\delta'}(X) \leq 2 \cdot \diam(X) \delta'^{1/\log_{2}C}$. Thus, we obtain that 
$$ \min(v_{\delta'}(X), v_{\delta'}(M_k)) \leq v_{\delta'}(X) \leq 2 \cdot \diam(X) \delta'^{1/\log_{2}C}. $$
Since $\diam(M_{k}) \leq \diam(X)$, we obtain $M = 2 \cdot \diam(X) + 45$. Therefore we have the following result:
$$ \sketch^{S}_{k,p}(X) \leq \zeta_{p}(X, M_{k}) \leq \big(8 \cdot \diam(X) \, \delta'^{1/ \log_{2} C} + \delta'\big)^{1/p} \cdot M, $$
where $M =2 \cdot \diam(X) + 45$. Let $\sketch_{k,p}(X) = \delta$. Then, for every $\epsilon > 0$, the previous inequality holds true for $\delta' = \delta^{1/5}(1+ \epsilon)$. Moreover, we know from \cite{Memoli2011} that for every $X \in \mathcal{M}_w$ and $M_k \in \mathcal{M}_{k,w}$, $\zeta_p(X,M_k) \geq \dgwp(X,M_k)$. Therefore, we obtain that
$$ \delta \leq \sketch^{S}_{k,p}(X) < \left(8 \cdot \diam(X)\cdot \delta^{1/ (5\log_{2}C)} + \delta^{1/5} \right)^{1/p} \cdot M, $$
whenever $ \sketch_{k,p}(X) = \delta < 2^{-5}$, where $M = 2 \cdot \diam(X) + 45 $. 
\end{proof}

We successfully showed that for doubling metric measure spaces, $\sketch^S_{k,p}(X)$ is bounded by a function of $\sketch_{k,p}(X)$. Now, in the next and the last section, we show that there exist clustering objectives that do not admit a dual sketching objective. 

\section{Impossibility results for sketching of metric spaces} \label{sec:impossibility}

In order to prove that there exist clustering cost functions that do not admit dual sketching cost functions, some conditions on our clustering and sketching cost functions are required. Otherwise, for any clustering cost function, we may set our sketching cost function to be equal to the clustering cost function. However, such a freedom in the choice of sketching cost function is, of course, unreasonable. We refer to our desired notions of sketching and clustering cost functions as \emph{admissible sketching} and \emph{admissible clustering}. We start by defining the notion of $k$-covering radius, which is required to define an admissible sketching.

\begin{definition}[\textbf{$k$-covering radius}]
Given $X \in \mathcal{M}$, the $k$-covering radius of $X$ is the minimum $r > 0$ for which there exist $k$ balls of radius $r$ centered at points of $X$ covering the whole space $X$. We use $\Cov_k(X)$ to denote the $k$-covering radius of $X$.
\end{definition}

For $\lambda \geq 0 $, we denote by $\lambda X$, the metric space $(X, \lambda \cdot d_X)$. 

\begin{definition}[\textbf{Admissible clustering}]
We say that a clustering cost function $\Phi$ is admissible if the following properties are satisfied for all $X \in \mathcal{M}$ and $k \in \mathbf{N}$:
\begin{enumerate}
\item $\shatter_{k}^{\Phi}(\lambda X) = \lambda \, \shatter_{k}^{\Phi}(X)$ for all $\lambda \geq 0$.
\item $\shatter_{k}^{\Phi}(X) \leq \shatter_{k-1}^{\Phi}(X)$.
\end{enumerate}
\end{definition}

\begin{definition}[\textbf{Admissible sketching}]
We say that a sketching cost function $\Psi$ is admissible if the following properties are satisfied for all $X \in \mathcal{M}$ and $k \in \mathbf{N}$:
\begin{enumerate}
\item $\sketch_{k}^{\Psi}(\lambda X) = \lambda \,\sketch_{k}^{\Psi}(X)$ for all $\lambda \geq 0$.
\item $\sketch_{k}^{\Psi}(X) \leq \sketch_{k-1}^{\Psi}(X)$.
\item $\sketch_{k}^{\Psi}(X) \geq \alpha_k \cdot \Cov_{k}(X)$, where $\alpha_k > 0$ is a constant. 
\end{enumerate}
\end{definition}
In the following section, we provide examples of admissible clustering cost functions that do not admit dual admissible sketching cost functions.

\subsection{Specific examples}

We now provide three examples of admissible clustering cost functions for which dual admissible sketching cost functions do not exist.

\begin{example}
Given $(X,d_X) \in \mathcal{M}$, we consider $u_X$, the maximal sub-dominant ultrametric on $X$ \cite{RTV86}, which is defined as follows: for every $x,x' \in X$, we first let $S_{x,x'}$ denote the collection of all finite sequences of points in $X$ starting at $x$ and ending at $x'$. Then, we define
$$ u_X(x,x') := \inf \big\{\max_i d_X(x_i, x_{i+1})~|~(x_0,x_1, \ldots, x_n) \in S_{x,x'}\big\}. $$
Let $U(X) = (X, u_X)$. Given $k \leq |X|$ and $P \in \partk(X)$, we consider the clustering cost function:

$$\Phi_U(X,P) := \Phi_{\mathcal{M}} (U(X),P) = \max_{B \in P} \diam(B) .$$ 
It is straightforward to see that $\Phi_U$ is admissible. We have that
$$\shatter_{k}^{\Phi_U}(X) = \inf_{P \in \mathrm{Part}_k(X)} \Phi_U(X,P),$$ 
and we observe that this corresponds to finding a partition of $X$ into $k$ blocks that maximizes the minimum inter-cluster distance. We now show that $\Phi_U $ does not admit a dual admissible sketching cost function.

\begin{proposition}[$\Phi_U$ does not admit a dual admissible sketching cost function] \label{thm:ultrametric}
For every $k \in \mathbf{N}$, there exists a sequence of spaces $\{Y_{n,k} \}_{n \in \mathbf{N}} \in \mathcal{M}$ such that for any admissible sketching cost function $\Psi$, we have $\sketch^{\Psi}_{k}(Y_{n,k}) \geq \Theta(1)$ but as $n \rightarrow \infty$, $\shatter_{k}^{\Phi_U}(Y_{n,k}) \rightarrow 0$. 
\end{proposition}

\begin{proof}
For $k=1$, consider the space $Y_{n,1} = \left\lbrace 0, \frac{1}{n}, \frac{2}{n}, \ldots, \frac{n-1}{n}, 1 \right\rbrace \subset \mathbf{R}$. For an arbitrary $k \in \mathbf{N}$, we define $Y_{n,k} = \cup_{i=0}^{k-1} (2k+Y_{n,1})$, where $\alpha + Y_{n,1} = \left\lbrace \alpha, \alpha + \frac{1}{n}, \alpha + \frac{2}{n}, \ldots, \alpha + \frac{n-1}{n}, \alpha + 1 \right\rbrace$. In particular, for every $k \in \mathbf{N}$, we obtain a sequence of spaces $\{Y_{n,k} \}_{n \in \mathbf{N}} \in \mathbf{R}$. We observe that for any $n,k \in \mathbf{N}$, $\shatter_{k}^{\Phi_U}(Y_{n,k}) = \frac{1}{n}$. In addition, for all $n,k \in \mathbf{N}$ we have $\Cov_k(Y_{n,k}) \geq \frac{1}{2}$. Therefore, for any admissible clustering $\Psi$, we have $\sketch_{k}^{\Psi}(Y_{n,k}) \geq \frac{\alpha_k}{2} > 0$. However, as $n \rightarrow \infty$, $\shatter_{k}^{\Phi_U}(Y_{n,k}) = \frac{1}{n} \rightarrow 0$. Thus, we obtain that for every $k \in \mathbf{N}$, $\{Y_{n,k} \}_{n \in \mathbf{N}} \in \mathcal{M}$ is such that as $n \rightarrow \infty$, $\shatter_{k}^{\Phi_U}(Y_{n,k}) \rightarrow 0$ but any admissible sketching cost function $\Psi$ satisfies $\sketch_{k}^{\Psi}(Y_{n,k}) \geq \Theta(1)$. This implies that the admissible clustering cost function $\Phi_{U}$ does not admit a dual admissible sketching cost function.   
\end{proof}
\end{example}

\begin{example}
The second example is given by the clustering cost function $\overline{\Phi}$ defined as follows: for a partition $P = \{B_{i} \}_{i=1}^{k}$ of $X$, we consider
$$ \overline{\Phi}(X,P) := \max_{i} (\diam(X) - \diam(B_{i})). $$

It is straightforward to see that $\overline{\Phi}$ is an admissible clustering cost function.
\begin{proposition} \label{thm:phi-bar-example}
The admissible clustering cost function $\overline{\Phi}$ does not admit a dual admissible sketching cost function.
\end{proposition}

\begin{proof}
Let $X \in \mathcal{M}$ be arbitrary and let $k = 1$. Then, we have $\mathrm{Part}_{1}(X) = \{X \}$ and this implies that $\shatter^{\overline{\Phi}}_{1}(X) = \overline{\Phi}(X, \{X \}) = 0$. However, for any admissible clustering cost function $\Psi$, we have $\sketch^{\Psi}_{1}(X) \geq \alpha_1 \cdot \diam(X) > 0$ since $\Cov_{1}(X) \geq \frac{\diam(X)}{2}$. Here, $\alpha_1 > 0$ is a constant. This shows that the inequality 
$\sketch^{\Psi}_{1}(X) \leq C_{2} \cdot \shatter^{\overline{\Phi}}_{1}(X)$
does not hold for any constant $C_2 >  0$. Since $X \in \mathcal{M}$ was arbitrary, we conclude that the admissible clustering cost function $\overline{\Phi}$ does not admit a dual admissible sketching cost function.
\end{proof}
\end{example}

\begin{example}
The third clustering cost function is defined using metric transforms \cite{Deza2013}. Given $1 < p < \infty$, a metric transform $M_{p} : \mathcal{M} \rightarrow \mathcal{M}$ is given by $M_{p}((X, d_{X})) = (X, m_p)$, where the metric $m_p$ is defined as follows: for any $x, x' \in X$, let $S_{x,x'}$ denote the collection of all sequences of points in $X$ starting at $x$ and ending at $x'$. 
Then, we define
$$ m_p(x,x') := \min_{(x_{0}, \ldots, x_{m}) \in S_{x,x'}} \left( \sum_{i=0}^{m-1} \left(d_{X}(x_{i}, x_{i+1})\right)^p \right)^{1/p}.$$
For any $k \in \mathbf{N}$ and $P \in \mathrm{Part}_{k}(X)$, define $\Phi_{p}(X,P) = \Phi_{\mathcal{M}} (M_{p}(X), P)$. It is straightforward to see that $\Phi_{p}(X,P)$ is admissible. We also note that the clustering cost function $\Phi_U $ that was defined as the first example is a special case of $\Phi_p $. In particular, $\Phi_U = \Phi_\infty $. We now have the following theorem.

\begin{proposition} \label{thm:metric-transform-example}
For any $1 < p < \infty$, the admissible clustering cost function $\Phi_{p}$ does not admit a dual admissible sketching cost function.
\end{proposition}
\begin{figure}
\begin{center}
\scalebox{0.6}{\includegraphics{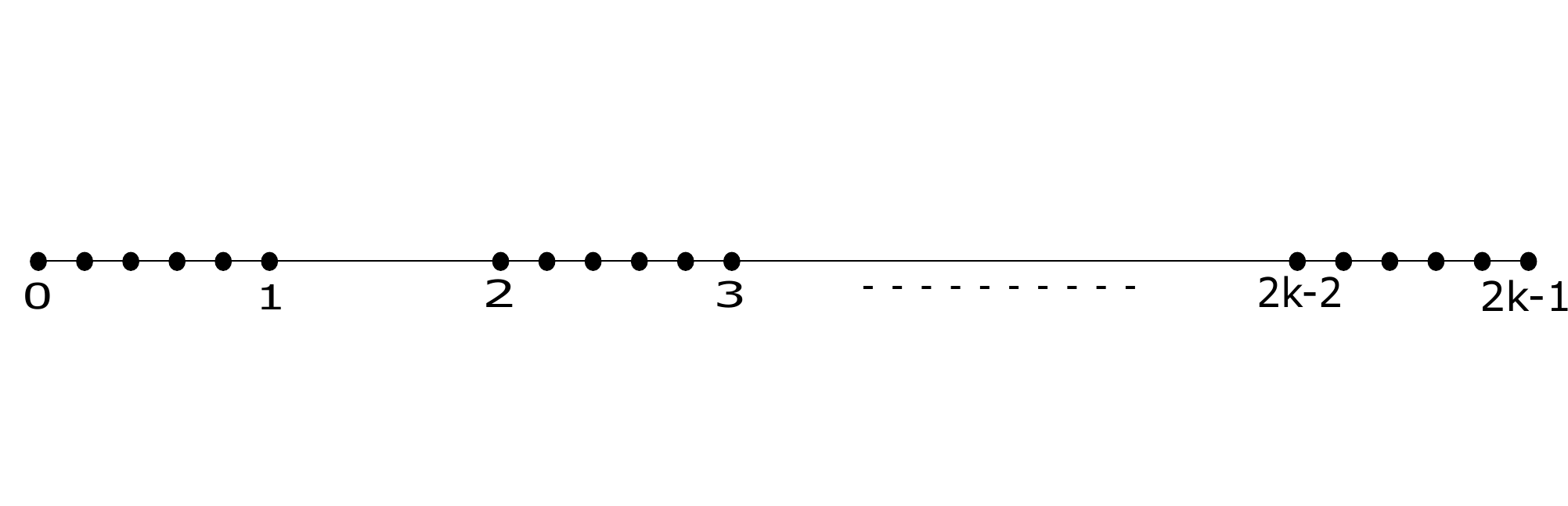}}
\caption{The set $Y_{n,k}$. Points within an interval of length $1$ are at distance $\frac{1}{n}$ from each other. }
\end{center}
\end{figure}
\begin{proof}
We use construction similar to one in the proof of Proposition \ref{thm:ultrametric}. Let $p \in (1, \infty)$ be fixed. For $k =1$, consider the space $Y_{n,1} = \left\lbrace 0, \frac{1}{n}, \frac{2}{n}, \ldots, \frac{n-1}{n}, 1 \right\rbrace \subset \mathbf{R}$. For an arbitrary $k \in \mathbf{N}$, we define $Y_{n,k} = \cup_{i=0}^{k-1} \{2k+Y_{n,1}\}$, where for $\alpha\in\mathbf{R}$, $\alpha + Y_{n,1} = \left\lbrace \alpha, \alpha + \frac{1}{n}, \alpha + \frac{2}{n}, \ldots, \alpha + \frac{n-1}{n}, \alpha + 1 \right\rbrace$. In particular, for every $k \in \mathbf{N}$, we obtain a sequence of spaces $\{Y_{n,k} \}_{n \in \mathbf{N}} \in \mathbf{R}$. We observe that for all $n,k \in \mathbf{N} $, $\diam(Y_{n,k}) = 2k-1$. For any $n \in \mathbf{N}$, we have 
$$\shatter^{\Phi_{p}}_{1}(Y_{n,1}) = \diam(M_{p}(Y_{n,1})) = \left( \frac{n}{n^p} \right)^{1/p} = \frac{1}{n^{1-1/p}}. $$
Thus, we obtain that as $n \rightarrow \infty$, $\shatter^{\Phi_p}_{1}(Y_{n,1}) \rightarrow 0$. However $\Cov_{1}(Y_{n,1}) \geq \frac{\diam(Y_{n,1})}{2} = \frac{1}{2}$. This implies that for any admissible sketching cost function $\Psi$, $\sketch^{\Psi}_{1}(Y_{n,1}) \geq \frac{\alpha_1}{2} > 0$ for all $n \in N$ and some constant $\alpha_1 > 0$. Thus, the inequality $\sketch^{\Psi}_1(Y_{n,1}) \leq C_2 \cdot \shatter^{\Phi_p}_1(Y_{n,1}) $ does not hold for any constant $C_2 > 0$.

For all $k > 1$, we consider $Y_{n,k}$. We again have $\Cov_{k}(Y_{n,k}) \geq \frac{1}{2}$. This implies that for any admissible sketching cost function $\Psi$, $\sketch^{\Psi}_k(Y_{n,k}) \geq \frac{\alpha_k}{2} > 0$ for all $n \in N$ and some constant $\alpha_k > 0$. We have $\shatter^{\Phi_p}_k(Y_{n,k}) = \frac{1}{n^{1-1/p}} $ since $Y_{n,k}$ consists of $k$ blocks of $Y_{n,1}$ arranged in a path, with the distance between adjacent blocks equal to $1$. Thus, we obtain an equipartition of $Y_{n,k}$ into $k$ blocks. We again have that $\shatter^{\Phi_p}_k(Y_{n,k}) \rightarrow 0$ as $n \rightarrow \infty$. Thus, the inequality $\sketch^{\Psi}_k(Y_{n,k}) \leq C_2 \cdot \shatter^{\Phi_p}_k(Y_{n,k})$ does not hold for any constant $C_2 > 0$. We conclude that the admissible sketching cost function $\Phi_p$ does not admit a dual admissible clustering cost function. Since $p \in (1, \infty)$ was arbitrary, the statement of the theorem holds.
\end{proof}
\end{example}

%%%%%%%%%%%%%%%%%%%%%%%%%%%%%%%
\section{Discussion}
In this paper, we identified a relationship between the two fundamental data analysis tasks of sketching and clustering. By sketching, we mean optimally approximating a metric space with a $k$-point metric space, for some $k \in \mathbf{N}$, under the Gromov-Hausdorff distance. We define sketching similarly for metric measure spaces, but by using the Gromov-Wasserstein distance instead of the Gromov-Hausdorff distance. The sketching objectives introduced are novel, while for the clustering objectives, several theoretical and computational results were known. We show that for both metric spaces and metric measure spaces, their respective sketching and clustering objectives are within constant factors of each other. This phenomenon is referred to as "duality". In addition, for $k \in \mathbf{N}$, approximation algorithms for computing $k$-sketch for both metric spaces and metric measure spaces are presented. Two sketching objectives are defined for metric measure spaces, out of which only one is dual to the respective clustering objective. A relation between these two sketching objectives is shown for a specific class of metric measure spaces. In the end, it is shown that the duality between sketching and clustering objectives is non-trivial. This is done by means of several examples of clustering objectives that do not admit a dual sketching objective. 

We remark that there are clustering problems that involve a different objective than the ones considered here.
One such example is partitioning a metric space into clusters of diameter at most $R$, for some given $R>0$, minimizing the number of clusters.
This is known as the $R$-dominating set problem \cite{lund1994hardness,feige1998threshold}.
It is not known whether similar duality results can be obtained in this case.

There are also several clustering objectives that do not fit within our duality framework. 
For example, in some clustering problems, such as general Facility Location \cite{Li11}, the number of centers, $k$, is not given.
Moreover, in some problems such as hierarchical clustering, the solution can be thought of as a family of clusterings.
Exploring duality and impossibility results for these clustering problems is left as an interesting research direction.

\smallskip
\noindent
\textbf{Acknowledgments.} The first and third authors were supported by NSF grants DMS-1723003 and CCF-1740761. 
The second author was supported by NSF CAREER award 1453472, and NSF grants CCF-1815145 and CCF-1423230.

\bibliographystyle{alpha}
\bibliography{references}

\appendix

\section{Proofs}

\begin{proof}[Proof of Proposition \ref{thm:tree-metric-spaces}]
Let $m=1$. Let $X = \{x_{1}, \ldots, x_{6} \}$ denote the nodes of a rooted tree with three branches, as shown in Figure \ref{fig:tree}. Let $r$ denote the root. For $i= \{1,2,3 \}$, the $i$-th branch contains $x_{2i-1}, x_{2i}$ with $d_{X}(r, x_{2i-1}) = 1$ and $d_{X}(x_{2i-1}, x_{2i}) = 1$. Then we have
$$ d_{X}(x_{1}, x_{2}) = d_{X}(x_{3}, x_{4}) = d_{X}(x_{5}, x_{6}) = 1, $$
$$ d_{X}(x_{1}, x_{3}) = d_{X}(x_{1}, x_{5}) = d_{X}(x_{3}, x_{5}) = 2, $$
$$ d_{X}(x_{2}, x_{4}) = d_{X}(x_{2}, x_{6}) = d_{X}(x_{4}, x_{6}) = 4, $$
$$ d_{X}(x_{1}, x_{4}) = d_{X}(x_{1}, x_{6}) = d_{X}(x_{3}, x_{2}) = d_{X}(x_{3}, x_{6}) = d_{X}(x_{5}, x_{2}) = d_{X}(x_{5}, x_{4}) = 3. $$ 
Let $Y = \{y_{1}, y_{2}, y_{3} \}$ be such that for $i \in \{1,2,3 \} $, $y_{i}$ lies on the $i$-th branch of $X$, and $d_{Y}(r, y_{i}) = \frac{3}{2}$. 

We now show that for every $K \subset X$ with $|K| \leq 3$, $\dgh(X, K) > \dgh(X,Y)$. We first calculate $\dgh(X,Y)$. Let $R = \{(x_1,y_1),(x_2,y_1),(x_3,y_2),(x_4,y_2),(x_5,y_3),(x_6,y_3) \} $ be a correspondence between $X$ and $Y$. Then, we have $\dis(R) \geq d_X(x_1,x_2) = 1$. This implies that $\dgh(X,Y) \leq \frac{1}{2}$. Moreover, we have $\dgh(X,Y) \geq \frac{1}{2} |\mathrm{diam}(X) - \mathrm{diam}(Y)| \geq \frac{1}{2}$. Thus, we conclude that $\dgh(X,Y) = \frac{1}{2}$. Now, in order to prove the proposition, it suffices to show that for every $K \subset X$ with $|K| \leq 3$, $\dgh(X,K) > \frac{1}{2}$. We use the fact that for metric spaces $X,K$, if $\phi:X \rightarrow K$ is a surjective map, then $R(\phi) = \{(x,\phi(x))~|~x \in X \} $ is a correspondence between $X$ and $K$. We have the following cases:

\begin{enumerate}
\item $K = \{x_{1}, x_{2}, x_{3} \}$.

Let $\phi: X \rightarrow K$ be a surjective map such that $\phi|_{K} = \mathrm{id}$. Then, if $\phi(x_{6}) = x_{1}$, we obtain $\mathrm{dis}(R(\phi)) \geq 3$; if $\phi(x_{6}) = x_{2}$, we obtain $\mathrm{dis}(R(\phi)) \geq 4$ and if $\phi(x_{6}) = x_{3}$, we obtain $\mathrm{dis}(R(\phi)) \geq 3$. This implies that $\dgh(X, K) \geq \frac{3}{2}$.

\item $K = \{ x_{1}, x_{2}, x_{4} \}$

Let $\phi: X \rightarrow K$ be a surjective map such that $\phi|_{K} = \mathrm{id}$. Then, if $\phi(x_{6}) = x_{1}$, we obtain $\mathrm{dis}(R(\phi)) \geq 3$; if $\phi(x_{6}) = x_{2}$, we obtain $\mathrm{dis}(R(\phi)) \geq 4$ and if $\phi(x_{6}) = x_{4}$, we obtain $\mathrm{dis}(R(\phi)) \geq 4$. This implies that $\dgh(X, K) \geq \frac{3}{2}$.

\item $K = \{x_{1}, x_{3}, x_{5} \}$

We use that $d_{GH}(X, K) \geq \frac{1}{2} |\mathrm{diam}(X) - \mathrm{diam}(K)|$. This implies $\dgh(X,K) \geq 1$.

\item $K = \{ x_{2}, x_{4}, x_{6} \}$

Let $\phi: X \rightarrow K$ be a surjective map such that $\phi|_{K} = \mathrm{id}$. Then, it is trivial to see that $\mathrm{dis}(R(\phi))$ is minimized if $\phi(x_{1}) = x_{2}, \phi(x_{3}) = x_{4}$ and $\phi(x_{5}) = x_{6}$. This implies that $\mathrm{dis}(R(\phi)) = 2$. Thus, $\dgh(X, K) = 1 $.

\item $K = \{ x_{1}, x_{3}, x_{6} \}$

Let $\phi: X \rightarrow K$ be a surjective map such that $\phi|_{K} = \mathrm{id}$. Then, it is trivial to see that $\mathrm{dis}(R(\phi))$ is minimized if $\phi(x_{2}) = x_{1}, \phi(x_{4}) = x_{3}$ and $\phi(x_{5}) = x_{6}$. This implies that $\mathrm{dis}(R(\phi)) = 2$. Thus, $\dgh(X, K) = 1 $.

\item $K = \{ x_{2}, x_{4}, x_{5} \}$

Let $\phi: X \rightarrow K$ be a surjective map such that $\phi|_{K} = \mathrm{id}$. Then, it is trivial to see that $\mathrm{dis}(R(\phi))$ is minimized if $\phi(x_{1}) = x_{2}, \phi(x_{3}) = x_{4}$ and $\phi(x_{6}) = x_{5}$. This implies that $\mathrm{dis}(R(\phi)) = 2$. Thus, $\dgh(X, K) = 1 $.

\item $K = \{x_{1}, x_{2} \}$

Let $\phi: X \rightarrow K$ be a surjective map such that $\phi|_{K} = \mathrm{id}$. Then, if $\phi(x_{6}) = x_{1}$, we obtain $\mathrm{dis}(R(\phi)) \geq 3$ and if $\phi(x_{6}) = x_{2}$, we obtain $\mathrm{dis}(R(\phi)) \geq 4$. Thus, $\dgh(X, K) \geq \frac{3}{2}$.

\item $K = \{x_{1}, x_{3} \}$

Let $\phi: X \rightarrow K$ be a surjective map such that $\phi|_{K} = \mathrm{id}$. Then, if $\phi(x_{6}) = x_{1}$, we obtain $\mathrm{dis}(R(\phi)) \geq 3$ and if $\phi(x_{6}) = x_{3}$, we again obtain $\mathrm{dis}(R(\phi)) \geq 3$. Thus, $\dgh(X, K) \geq \frac{3}{2}$.

\item $K = \{x_{1}, x_{4} \}$

Let $\phi: X \rightarrow K$ be a surjective map such that $\phi|_{K} = \mathrm{id}$. Then, if $\phi(x_{6}) = x_{1}$, we obtain $\mathrm{dis}(R(\phi)) \geq 3$ and if $\phi(x_{6}) = x_{4}$, we obtain $\mathrm{dis}(R(\phi)) \geq 4$. Thus, $\dgh(X, K) \geq \frac{3}{2}$.

\item $K = \{x_{2}, x_{4} \}$

Let $\phi: X \rightarrow K$ be a surjective map such that $\phi|_{K} = \mathrm{id}$. Then, if $\phi(x_{6}) = x_{2}$, we obtain $\mathrm{dis}(R(\phi)) \geq 4$ and if $\phi(x_{6}) = x_{4}$, we again obtain $\mathrm{dis}(R(\phi)) \geq 4$. Thus, $\dgh(X, K) \geq 2$.

\item $K \subset X$ with $|K| = 1$

In this case, $\dgh(X, K) = \frac{1}{2} \mathrm{diam}(X) = 2$.
\end{enumerate}
The analysis for the remaining subsets $K$ of $X$ is symmetric to the cases above. This proves the proposition for $n=3$. In general, for any $m \in \mathbf{N}$ and $n=3m$, we create $m$ copies of $X$ that are connected at the root $r$ and prove the proposition similarly.
\end{proof}

\begin{proof}[Proof of Proposition \ref{thm:euclidean-metric-spaces}]
Let $m=1$. We construct the set $X$ on lines similar to those in Proposition \ref{thm:tree-metric-spaces}, see Figure \ref{fig:tree}. Let $x_{1} = (0,1), x_{2} = (0,2), x_{3} = (-\frac{\sqrt{3}}{2}, -\frac{1}{2}), x_{4} = (-\sqrt{3}, -1), x_{5} = (\frac{\sqrt{3}}{2}, -\frac{1}{2}), x_{6} = (\sqrt{3}, -1) $. Then, we have
$$ d(x_{1}, x_{2}) = d(x_{3}, x_{4}) = d(x_{5}, x_{6}) = 1, $$
$$ d(x_{1}, x_{3}) = d(x_{1}, x_{5}) = d(x_{3}, x_{5}) = \sqrt{3}, $$
$$ d(x_{2}, x_{4}) = d(x_{2}, x_{6}) = d(x_{4}, x_{6}) = 2 \sqrt{3}, $$
$$ d(x_{1}, x_{4}) = d(x_{1}, x_{6}) = d(x_{3}, x_{2}) = d(x_{3}, x_{6}) = d(x_{5}, x_{2}) = d(x_{5}, x_{4}) = \sqrt{7}. $$

Let $X = \{ x_{1}, x_{2}, x_{3}, x_{4}, x_{5}, x_{6} \}$. Then, $\mathrm{diam}(X) = 2 \sqrt{3}$. Let $y_{1} = \big(0, \frac{3}{2} \big), y_{2} = \big(\frac{-3 \sqrt{3}}{4}, -\frac{3}{4} \big)$ and $y_{3} = \big(\frac{3 \sqrt{3}}{4}, -\frac{3}{4} \big)$. Let $Y = \{ y_{1}, y_{2}, y_{3} \}$. We now show that for every $K \subset X$ with $|K| = 3$, $\dgh(X,K) > \dgh(X,Y)$.

We first calculate $\dgh(X,Y)$. Let $R = \{(x_1,y_1),(x_2,y_1),(x_3,y_2),(x_4,y_2),(x_5,y_3),(x_6,y_3) \} $ be a correspondence between $X$ and $Y$. Then, we have $\dis(R) \geq d_X(x_1,x_2) = 1$. This implies that $d_{GH}(X,Y) \leq \frac{1}{2}$. Now, in order to prove the proposition, it suffices to show that for every $K \subset X$ with $|K| \leq 3$, $\dgh(X,K) > \frac{1}{2}$. We use the fact that for metric spaces $X,K$, if $\phi:X \rightarrow K$ is a surjective map, then $R(\phi) = \{(x,\phi(x))~|~x \in X \} $ is a correspondence between $X$ and $K$. We have the following cases:
\begin{enumerate}
\item $K = \{x_{1}, x_{2}, x_{3} \}$.

Let $\phi: X \rightarrow K$ be a surjective map such that $\phi|_{K} = \mathrm{id}$. Then, if $\phi(x_{6}) = x_{1}$, we obtain $\mathrm{dis}(R(\phi)) \geq \sqrt{7} $; if $\phi(x_{6}) = x_{2}$, we obtain $\mathrm{dis}(R(\phi)) \geq 2 \sqrt{3} $ and if $\phi(x_{6}) = x_{3}$, we obtain $\mathrm{dis}(R(\phi)) \geq \sqrt{7} $. This implies that $\dgh(X, K) \geq \frac{\sqrt{7}}{2}$.

\item $K = \{ x_{1}, x_{2}, x_{4} \}$

Let $\phi: X \rightarrow K$ be a surjective map such that $\phi|_{K} = \mathrm{id}$. Then, if $\phi(x_{6}) = x_{1}$, we obtain $\mathrm{dis}(R(\phi)) \geq \sqrt{7}$; if $\phi(x_{6}) = x_{2}$, we obtain $\mathrm{dis}(R(\phi)) \geq 2 \sqrt{3} $ and if $\phi(x_{6}) = x_{4}$, we obtain $\mathrm{dis}(R(\phi)) \geq 2 \sqrt{3} $. This implies that $\dgh(X, K) \geq \frac{\sqrt{7}}{2}$.

\item $K = \{x_{1}, x_{3}, x_{5} \}$

We use that $\dgh(X, K) \geq \frac{1}{2} |\mathrm{diam}(X) - \mathrm{diam}(K)|$. This implies that $\dgh(X,K) \geq \frac{\sqrt{3}}{2}$.

\item $K = \{ x_{2}, x_{4}, x_{6} \}$

Let $\phi: X \rightarrow K$ be a surjective map such that $\phi|_{K} = \mathrm{id}$. Then, it is trivial to see that $\mathrm{dis}(R(\phi))$ is minimized if $\phi(x_{1}) = x_{2}, \phi(x_{3}) = x_{4}$ and $\phi(x_{5}) = x_{6}$. This implies that $\mathrm{dis}(R(\phi)) = \sqrt{3}$. Thus, $\dgh(X, K) = \frac{\sqrt{3}}{2} $.

\item $K = \{ x_{1}, x_{3}, x_{6} \}$

Let $\phi: X \rightarrow K$ be a surjective map such that $\phi|_{K} = \mathrm{id}$. Then, it is trivial to see that $\mathrm{dis}(R(\phi))$ is minimized if $\phi(x_{2}) = x_{1}, \phi(x_{4}) = x_{3}$ and $\phi(x_{5}) = x_{6}$. This implies that $\mathrm{dis}(R(\phi)) = \sqrt{3}$. Thus, $\dgh(X, K) = \frac{\sqrt{3}}{2} $.

\item $K = \{ x_{2}, x_{4}, x_{5} \}$

Let $\phi: X \rightarrow K$ be a surjective map such that $\phi|_{K} = \mathrm{id}$. Then, it is trivial to see that $\mathrm{dis}(R(\phi))$ is minimized if $\phi(x_{1}) = x_{2}, \phi(x_{3}) = x_{4}$ and $\phi(x_{6}) = x_{5}$. This implies that $\mathrm{dis}(R(\phi)) = \sqrt{3} $. Thus, $\dgh(X, K) = \frac{\sqrt{3}}{2} $.

\item $K = \{x_{1}, x_{2} \}$

Let $\phi: X \rightarrow K$ be a surjective map such that $\phi|_{K} = \mathrm{id}$. Then, if $\phi(x_{6}) = x_{1}$, we obtain $\mathrm{dis}(R(\phi)) \geq \sqrt{7} $ and if $\phi(x_{6}) = x_{2}$, we obtain $\mathrm{dis}(R(\phi)) \geq 2 \sqrt{3} $. Thus, $\dgh(X, K) \geq \frac{\sqrt{7}}{2}$.

\item $K = \{x_{1}, x_{3} \}$

Let $\phi: X \rightarrow K$ be a surjective map such that $\phi|_{K} = \mathrm{id}$. Then, if $\phi(x_{6}) = x_{1}$, we obtain $\mathrm{dis}(R(\phi)) \geq \sqrt{7} $ and if $\phi(x_{6}) = x_{3}$, we again obtain  $\mathrm{dis}(R(\phi)) \geq \sqrt{7} $. Thus, $\dgh(X, K) \geq \frac{\sqrt{7}}{2}$.

\item $K = \{x_{1}, x_{4} \}$

Let $\phi: X \rightarrow K$ be a surjective map such that $\phi|_{K} = \mathrm{id}$. Then, if $\phi(x_{6}) = x_{1}$, we obtain $\mathrm{dis}(R(\phi)) \geq \sqrt{7} $ and if $\phi(x_{6}) = x_{4}$, we obtain $\mathrm{dis}(R(\phi)) \geq 2 \sqrt{3} $. Thus, $\dgh(X, K) \geq \frac{\sqrt{7}}{2}$.

\item $K = \{x_{2}, x_{4} \}$

Let $\phi: X \rightarrow K$ be a surjective map such that $\phi|_{K} = \mathrm{id}$. Then, if $\phi(x_{6}) = x_{2}$, we obtain $\mathrm{dis}(R(\phi)) \geq 2 \sqrt{3}$ and if $\phi(x_{6}) = x_{4}$, we again obtain $\mathrm{dis}(R(\phi)) \geq 2 \sqrt{3}$. Thus, $\dgh(X, K) \geq \sqrt{3}$.

\item $K \subset X$ with $|K| = 1$

In this case, $\dgh(X, K) = \frac{1}{2} \mathrm{diam}(X) = \sqrt{3}$.
\end{enumerate}
The analysis for the remaining subsets $K$ of $X$ is similar to the cases above. This proves the proposition for $n=3$. In general, for any $m \in \mathbf{N}$ and $n=3m$, we create $m$ disjoint copies of $X$ and prove the proposition similarly.
\end{proof}

\end{document}